\documentclass[graybox,color]{svmult}

\usepackage{amssymb}
\usepackage{mathptmx}       
\usepackage{helvet}         
\usepackage{courier}        
\usepackage{type1cm}        
\usepackage{makeidx}         
\usepackage{graphicx}        
\usepackage{multicol}        
\usepackage[bottom]{footmisc}
\usepackage[normalem]{ulem}
\usepackage{cancel}

\makeindex             

\begin{document}

\title*{Magnetic Reconnection in Astrophysical Environments}
\titlerunning{Magnetic Reconnection}

\author{Alex Lazarian, Gregory L. Eyink, Ethan T. Vishniac and Grzegorz Kowal}

\institute{A. Lazarian \at Department of Astronomy, University of Wisconsin, 475 North Charter Street, Madison, Wisconsin 53706, USA \email{lazarian@astro.wisc.edu}
\and G. L. Eyink \at Department of Applied Mathematics and Statistics, The Johns Hopkins University, Baltimore, Maryland 21218, USA \email{eyink@jhu.edu}
\and E. T. Vishniac \at Department of Physics and Astronomy, McMaster University, 1280 Main Street West, Hamilton, Ontario L8S 4M1, Canada \email{ethan@mcmaster.ca}
\and G. Kowal \at Escola de Artes, Ci\^{e}ncias e Humanidades, Universidade de
S\~{a}o Paulo, Av. Arlindo B\'{e}ttio, 1000 -- Ermelino Matarazzo, CEP
03828-000, S\~{a}o Paulo, SP, Brazil \email{g.kowal@iag.usp.br}}

\maketitle

\abstract*{Magnetic reconnection is a process that changes magnetic field
topology in highly conducting fluids.  Traditionally, magnetic reconnection was
associated mostly with solar flares.  In reality, the process must be ubiquitous
as astrophysical fluids are magnetized and motions of fluid elements necessarily
entail crossing of magnetic frozen in field lines and magnetic reconnection.  We
consider magnetic reconnection in realistic 3D geometry in the presence of
turbulence.  This turbulence in most astrophysical settings is of pre-existing
nature, but it also can be induced by magnetic reconnection itself.  In this
situation turbulent magnetic field wandering opens up reconnection outflow
regions, making reconnection fast.  We discuss Lazarian \& Vishniac (1999) model
of turbulent reconnection, its numerical and observational testings, as well as
its connection to the modern understanding of the Lagrangian properties of
turbulent fluids.  We show that the predicted dependences of the reconnection
rates on the level of MHD turbulence make the generally accepted Goldreich \&
Sridhar (1995) model of turbulence self-consistent.  Similarly, we argue that
the well-known Alfv\'{e}n theorem on flux freezing is not valid for the
turbulent fluids and therefore magnetic fields diffuse within turbulent volumes.
 This is an element of magnetic field dynamics that was not accounted by earlier
theories.  For instance, the theory of star formation that was developing
assuming that it is only the drift of neutrals that can violate the otherwise
perfect flux freezing, is affected and we discuss the consequences of the
turbulent diffusion of magnetic fields mediated by reconnection.  Finally, we
briefly address the first order Fermi acceleration induced by magnetic
reconnection in turbulent fluids which is discussed in detail in the chapter by
de Gouveia Dal Pino and Kowal in this volume.}

\abstract{Magnetic reconnection is a process that changes magnetic field
topology in highly conducting fluids.  Traditionally, magnetic reconnection was
associated mostly with solar flares.  In reality, the process must be ubiquitous
as astrophysical fluids are magnetized and motions of fluid elements necessarily
entail crossing of magnetic frozen in field lines and magnetic reconnection.  We
consider magnetic reconnection in realistic 3D geometry in the presence of
turbulence.  This turbulence in most astrophysical settings is of pre-existing
nature, but it also can be induced by magnetic reconnection itself.  In this
situation turbulent magnetic field wandering opens up reconnection outflow
regions, making reconnection fast.  We discuss Lazarian \& Vishniac (1999) model
of turbulent reconnection, its numerical and observational testings, as well as
its connection to the modern understanding of the Lagrangian properties of
turbulent fluids.  We show that the predicted dependences of the reconnection
rates on the level of MHD turbulence make the generally accepted Goldreich \&
Sridhar (1995) model of turbulence self-consistent.  Similarly, we argue that
the well-known Alfv\'{e}n theorem on flux freezing is not valid for the
turbulent fluids and therefore magnetic fields diffuse within turbulent volumes.
This is an element of magnetic field dynamics that was not accounted by earlier
theories.  For instance, the theory of star formation that was developing
assuming that it is only the drift of neutrals that can violate the otherwise
perfect flux freezing, is affected and we discuss the consequences of the
turbulent diffusion of magnetic fields mediated by reconnection.  Finally, we
briefly address the first order Fermi acceleration induced by magnetic
reconnection in turbulent fluids which is discussed in detail in the chapter by
de Gouveia Dal Pino and Kowal in this volume.}

\section{Introduction}
\label{sec:introduction}

Magnetic fields modify fluid dynamics and it is generally believed that magnetic
fields embedded in a highly conductive fluid retain their topology for all time
due to the magnetic fields being frozen-in \cite{Alfven42, Parker79}.
Nevertheless, highly conducting ionized astrophysical objects, like stars and
galactic disks, show evidence of changes in topology, i.e. ``magnetic
reconnection'', on dynamical time scales \cite{Parker70, Lovelace76,
PriestForbes02}.  Historically, magnetic reconnection research was motivated by
observations of the solar corona \cite{Innesetal97, YokoyamaShibata95,
Masudaetal94} and this influenced attempts to find peculiar conditions conducive
for flux conservation violation, e.g. special magnetic field configurations or
special plasma conditions.  For instance, much  work has concentrated on showing
how reconnection can be rapid in plasmas with very small collision rates
\cite{Shayetal98, Drake01, Drakeetal06, Daughtonetal06}.  However, it is clear
that reconnection is a ubiquitous process taking place in various astrophysical
environments, e.g. magnetic reconnection can be inferred from the existence of
large-scale dynamo activity inside stellar interiors \cite{Parker93,
Ossendrijver03}, as well as from the eddy-type motions in magnetohydrodynamic
turbulence.  Without fast magnetic reconnection magnetized fluids would behave
like Jello or felt, rather than as a fluid.

In fact, solar flares \cite{Sturrock66} are just one vivid example of
reconnection activity.  Some other reconnection events, e.g. $\gamma$-ray bursts
\cite{ZhangYan11, Lazarianetal04, Foxetal05, Galamaetal98} also occur in
collisionless media, while others take place in collisional media.  Thus
attempts to explain only collisionless reconnection substantially limits
astrophysical applications of the corresponding reconnection models.  We also
note that magnetic reconnection occurs rapidly in computer simulations due to
the high values of resistivity (or numerical resistivity) that are employed at
the resolutions currently achievable.  Therefore, if there are situations where
magnetic fields reconnect slowly, numerical simulations do not adequately
reproduce astrophysical reality.  This means that if collisionless reconnection
is the only way to make reconnection rapid, then numerical simulations of many
astrophysical processes, including those of the interstellar medium (ISM), which
is collisional, are in error.  Fortunately, this scary option is not realistic,
as recent observations of the collisional parts of the solar atmosphere indicate
fast reconnection \cite{ShibataMagara11}.

What makes reconnection enigmatic is that it is not possible to claim that
reconnection must always be rapid empirically, as solar flares require periods
of flux accumulation time, which correspond to slow reconnection.  Thus magnetic
reconnection should have some sort of trigger, which should not depend on the
parameters of the local plasma.  In this review we argue that the trigger is
turbulence.

We may add that some recent reviews dealing with turbulent magnetic reconnection
include \cite{BrowningLazarian13} and \cite{KarimabadiLazarian13}.  The first
one analyzes the reconnection in relation to solar flares, the other provides
the comparison of the PIC simulations of the reconnection in collisionless
plasmas with the reconnection in turbulent MHD regime.

In the review below we provide a simple description of the basics of magnetic
reconnection and astrophysical turbulence in \S\ref{sec:basics}, present the
theory of magnetic reconnection in the presence of turbulence and its testing in
\S\ref{sec:reconnection} and \S\ref{sec:numerical_testing}, respectively.
Observational tests of the magnetic reconnection are described in
\S\ref{sec:obs_cons_tests} while the extensions of the reconnection theory are
discussed in \S\ref{sec:lv99_extension} and its astrophysical implications are
summarized in \S\ref{sec:implications}. In \S\ref{sec:discussion} we present a
discussion and summary of the review.

\section{Basics of Magnetic Reconnection and Astrophysical Turbulence}
\label{sec:basics}

\subsection{Models of laminar reconnection}
\label{ssec:lam_rec}

Turbulence is usually not a welcome ingredient in theoretical modeling.
Turbulence carries an aura of mystery, especially magnetic turbulence, which is
still a subject of ongoing debates.  Thus, it is not surprising that researchers
prefer to consider laminar models whenever possible.

\begin{figure}
\centering
\includegraphics[width=0.9\textwidth]{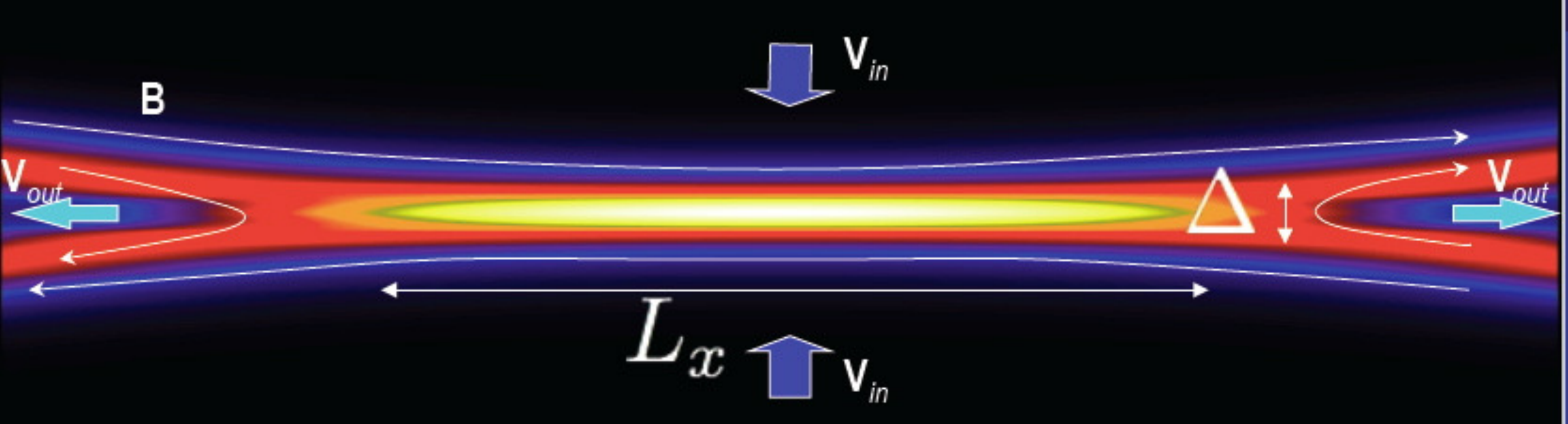}
\caption{Sweet-Parker reconnection.  Simulations of laminar reconnection from
\cite{Kowaletal09} are used.  The current sheet has $L_x$ extension, while the
ejection of matter and shared component of magnetic field happens through
$\Delta$.  The cross-section of the reconnection is shown.  Generically, the
shared component of magnetic field is directed perpendicular to the picture
plane.  This component should be also ejected through $\Delta$. \label{fig:sp_recon}}
\end{figure}

The classical Sweet-Parker model, the first analytical model for magnetic
reconnection, was proposed by Parker \cite{Parker57} and Sweet
\cite{Sweet58}\footnote{The basic idea of the model was first discussed by Sweet
and the corresponding paper by Parker refers to the model as ``Sweet model''.}.
Sweet-Parker reconnection has the virtue that it relies on a robust and
straightforward geometry (see Figure~\ref{fig:sp_recon}).  Two regions with uniform {\it
laminar} magnetic fields are separated by thin current sheet.  The speed of
reconnection is given roughly by the resistivity divided by the sheet thickness,
i.e.
\begin{equation}
V_{rec1}\approx \eta/\Delta.
\label{eq.1}
\end{equation}
One might incorrectly assume that by decreasing the current sheet thickness one
can increase the reconnection rate.  In fact, for {\it steady state
reconnection} the plasma in the current sheet must be ejected from the edge of
the current sheet at the Alfv\'{e}n speed, $V_A$.  Thus the reconnection speed
is
\begin{equation}
V_{rec2}\approx V_A \Delta/L,
\label{eq.2}
\end{equation}
where $L$ is the length of the current sheet, which requires $\Delta$ to be
large for a large reconnection speed.

In other words, we face two contradictory requirements on the outflow thickness,
namely, $\Delta$ should be large so as to not constrain the outflow of plasma
and $\Delta$ should be small for the Ohmic diffusivity to do its job of
dissipating magnetic field lines.  As a result, the steady state Sweet-Parker
reconnection rate is a compromise between the two contradictory requirements.
If $\Delta$ becomes small, the reconnection rate $V_{rec1}$ increases, but the
insufficient outflow of plasma from the current sheet will lead to an increase
in $\Delta$ and slow down the reconnection process.  If $\Delta$ increases, the
outflow will speed up but the oppositely directed magnetic field lines get
further apart and $V_{rec1}$ drops.  The slow reconnection rate limits the
supply of plasma into the outflow and decreases $\Delta$.  This self regulation
ensures that in the steady state $V_{rec1}=V_{rec2}$ which determines both the
steady state reconnection rate and the steady state $\Delta$.  As a result, the
overall reconnection speed is reduced from the Alfv\'{e}n speed by the square root
of the Lundquist number, $S\equiv L_xV_A/\eta$, i.e.
\begin{equation}
V_{rec, SP}=V_A S^{-1/2}.
\label{SP}
\end{equation}

For astrophysical conditions the Lundquist number $S$ may easily be $10^{16}$
and larger.  The corresponding Sweet-Parker reconnection speed is negligible.
If this sets the actual reconnection speed then we should expect magnetic field
lines in the fluid not to change their topology, which in the presence of
chaotic motions should result in a messy magnetic structure with the properties
of Jello.  On the contrary, the fast reconnection suggested by solar flares,
dynamo operation etc. requires that the dependence on $S$ be erased.

A few lessons can be learned from the analysis of the Sweet-Parker reconnection.
 First of all, it is a self-regulated process.  Second, even with the
Sweet-Parker scheme the instantaneous rates of reconnection are not restricted.
Indeed, under the external forcing the Ohmic annihilation rate given by
$V_{rec1}$ can be arbitrary large, which, nevertheless does not mean that the
time averaged rate of reconnection is also large.  This should be taken into
account when the probability distribution functions of currents are interpreted
in terms of magnetic reconnection (see \S\ref{ssec:rec_reg_struct}).

\begin{figure}
\centering
\includegraphics[width=0.8\textwidth]{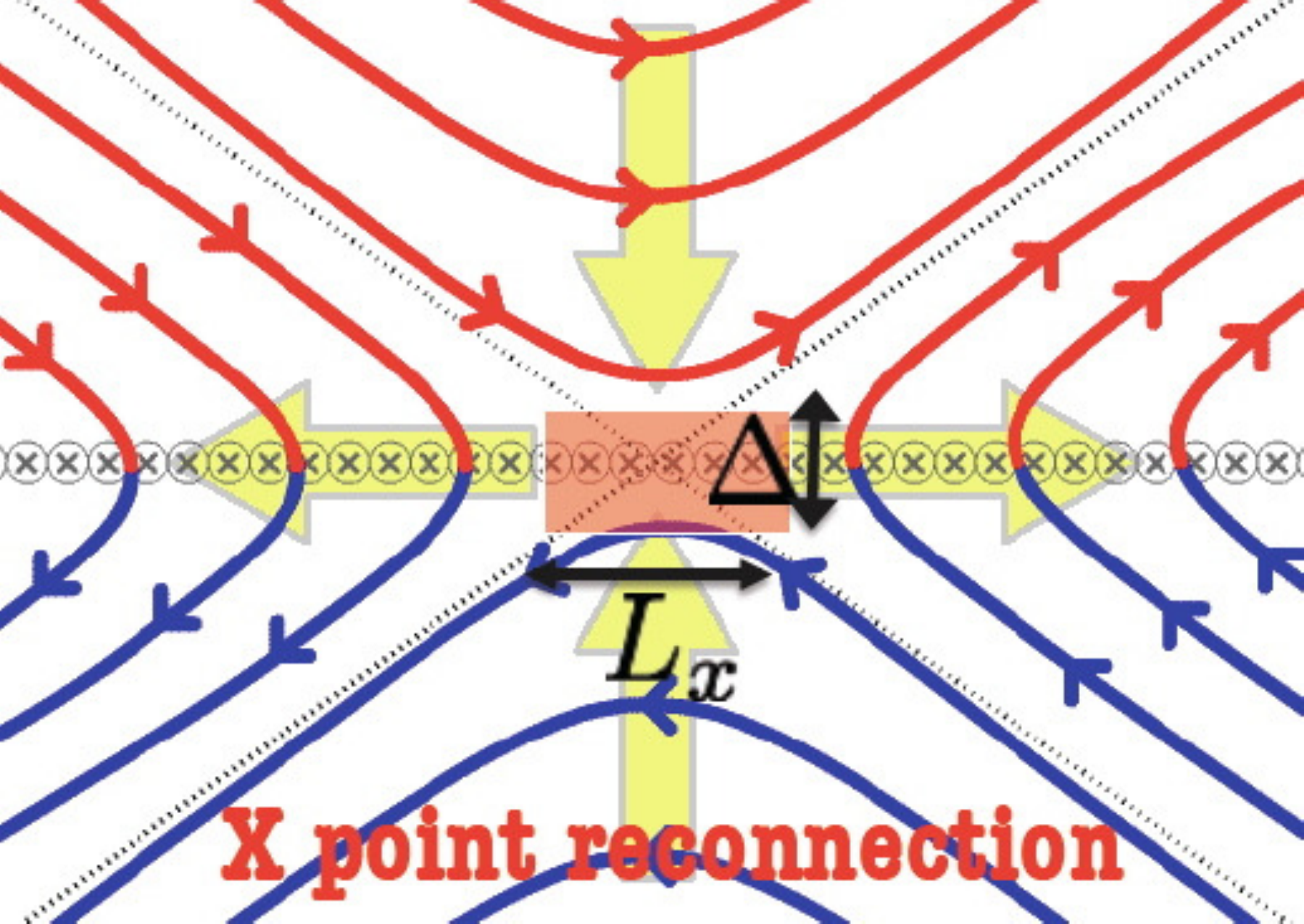}
\caption{Petschek reconnection is an X-point reconnection where due to the
formation of shocks the magnetic field lines are bent sharply towards the
reconnection ``point'' with $L_x\sim \Delta$. \label{fig:petschek}}
\end{figure}

The low efficiency of the Sweet-Parker reconnection arises from the disparity of
the scales of $\Delta$, which is determined by microphysics, i.e. depends on
$\eta$, and $L_x$ that has a huge, i.e. astronomical, size.  The introduction of
plasma effects does not change this problem as in this case $\Delta$ should be
of the order of the ion Larmor radius, which is $\ll L_x$.  There are two ways
to make the reconnection speed faster.  One way is to reduce $L_x$, by changing
the geometry of reconnection region, e.g. making magnetic field lines come at a
sharp angle rather than in a natural Sweet-Parker way.  This is called X-point
reconnection.  The most famous example of this is Petschek reconnection
\cite{Petschek64} (see Figure~\ref{fig:petschek}).  The other way is to extend
$\Delta$ and make it comparable to $L_x$.  Obviously, a factor different from
resistivity should be involved.  In this review we provide evidence that
turbulence can do the job of increasing $\Delta$.  However, before focusing on
this process, we shall first discuss very briefly the Petschek reconnection
model, which for a few decades served as the default model of fast reconnection.

Figure~\ref{fig:petschek} illustrates the Petschek model of reconnection.  The model
suggests that extended magnetic bundles come into contact over a tiny area
determined by the Ohmic diffusivity.  This configuration differs dramatically
from the expected generic configuration when magnetic bundles try to press their
way through each other.  Thus the first introduction of this model raised
questions of dynamical self-consistency.  An X-point configuration has to
persist in the face of compressive bulk forces.  However, numerical simulations
have shown that an initial X-point configuration of magnetic field reconnection
is unstable in the MHD limit for small values of the Ohmic diffusivity
\cite{Biskamp96} and the magnetic field will relax to a Sweet-Parker
configuration.  The physical explanation for this effect is simple.  In the
Petschek model shocks are required in order to maintain the geometry of the
X-point.  These shocks must persist and be supported by the flows driven by fast
reconnection.  The simulations showed that the shocks fade away and the contact
region spontaneously increases.

X-point reconnection can be stabilized when the plasma is collisionless.
Numerical simulations \cite{Shayetal98, Shayetal04} have been encouraging in
this respect and created the hope that there was at last the solution of the
long-standing problem of magnetic reconnection.  However, there are several
important issues that remain unresolved.  First, it is not clear that this kind
of fast reconnection persists on scales greater than the ion inertial scale
\cite{Bhattacharjeeetal03}.  Several numerical studies \cite{Wangetal01,
Smithetal04, Fitzpatrick04} have found large scale reconnection speeds which are
not fast in the sense that they show dependence on resistivity.  There are
countervailing analytical studies \cite{Malyshkin08, Shivamoggi11} which suggest
that Hall X-point reconnection rates are independent of resistivity or other
microscopic plasma mechanisms of line slippage, but the rates determined in
these studies become small when the ion inertial scale is much less than $L_x$.
Second, in many circumstances the magnetic field geometry does not allow the
formation of X-point reconnection. For example, a saddle-shaped current sheet
cannot be spontaneously replaced by an X-point.  The energy required to do so is
comparable to the magnetic energy liberated by reconnection, and must be
available beforehand.  Third, the stability of the X-point is questionable in
the presence of the external random forcing, which is common, as we discuss
later, for most of the astrophysical environments.  Finally, the requirement
that reconnection occurs in a collisionless plasma restricts this model to a
small fraction of astrophysical applications.  For example, while reconnection
in stellar coronae might be described in this way, stellar chromospheres can
not.  This despite the fact that we observe fast reconnection in those
environments \cite{ShibataMagara11}.  More generally, Yamada \cite{Yamada07}
estimated that the scale of the reconnection sheet should not exceed about 40
times the electron mean free path.  This condition is not satisfied in many
environments which one might naively consider to be collisionless, among them
the interstellar medium.  The conclusion that stellar interiors and atmospheres,
accretion disks, and the interstellar medium in general does not allow fast
reconnection is drastic and unpalatable.

Petschek reconnection requires an extended X-point configuration of reconnected
magnetic fluxes and Ohmic dissipation concentrated within a microscopic region.
As we discuss in this review (see \S\ref{sec:obs_cons_tests}), neither of these
predictions were supported by solar flare observations.  This suggests that
neither Sweet-Parker nor Petschek models present a universally applicable
mechanism of astrophysical magnetic reconnection.  This does not preclude that
these processes are important in particular special situations.  In what follows
we argue that Petschek-type reconnection may be applicable for magnetospheric
current sheets or any collisionless plasma systems, while Sweet-Parker can be
important for reconnection at small scales in partially ionized gas.

\subsection{Turbulence in Astrophysical fluids}
\label{sec1.2}

Neither of these models take into account turbulence, which is ubiquitous in
astrophysical environments.  Indeed, plasma flows at high Reynolds numbers are
generically turbulent, since laminar flows are then prey to numerous linear and
finite-amplitude instabilities.  This is sometimes driven turbulence due to an
external energy source, such as supernova in the ISM \cite{NormanFerrara96,
Ferriere01}, merger events and AGN outflows in the intercluster medium (ICM)
\cite{Subramanianetal06, EnsslinVogt06, Chandran05}, and baroclinic forcing
behind shock waves in interstellar clouds.  In other cases, the turbulence is
spontaneous, with available energy released by a rich array of instabilities,
such as the MRI in accretion disks \cite{BalbusHawley98}, the kink instability
of twisted flux tubes in the solar corona \cite{GalsgaardNordlund97a,
GerrardHood03}, etc.  Whatever its origin, observational signatures of
astrophysical turbulence are seen throughout the universe.  The turbulent
cascade of energy leads to long ``inertial ranges'' with power-law spectra that
are widely observed, e.g. in the solar wind \cite{Leamonetal98, Baleetal05}, and
in the ICM \cite{Schueckeretal04, VogtEnsslin05}.

\begin{figure}
\centering
\includegraphics[width=0.9\textwidth]{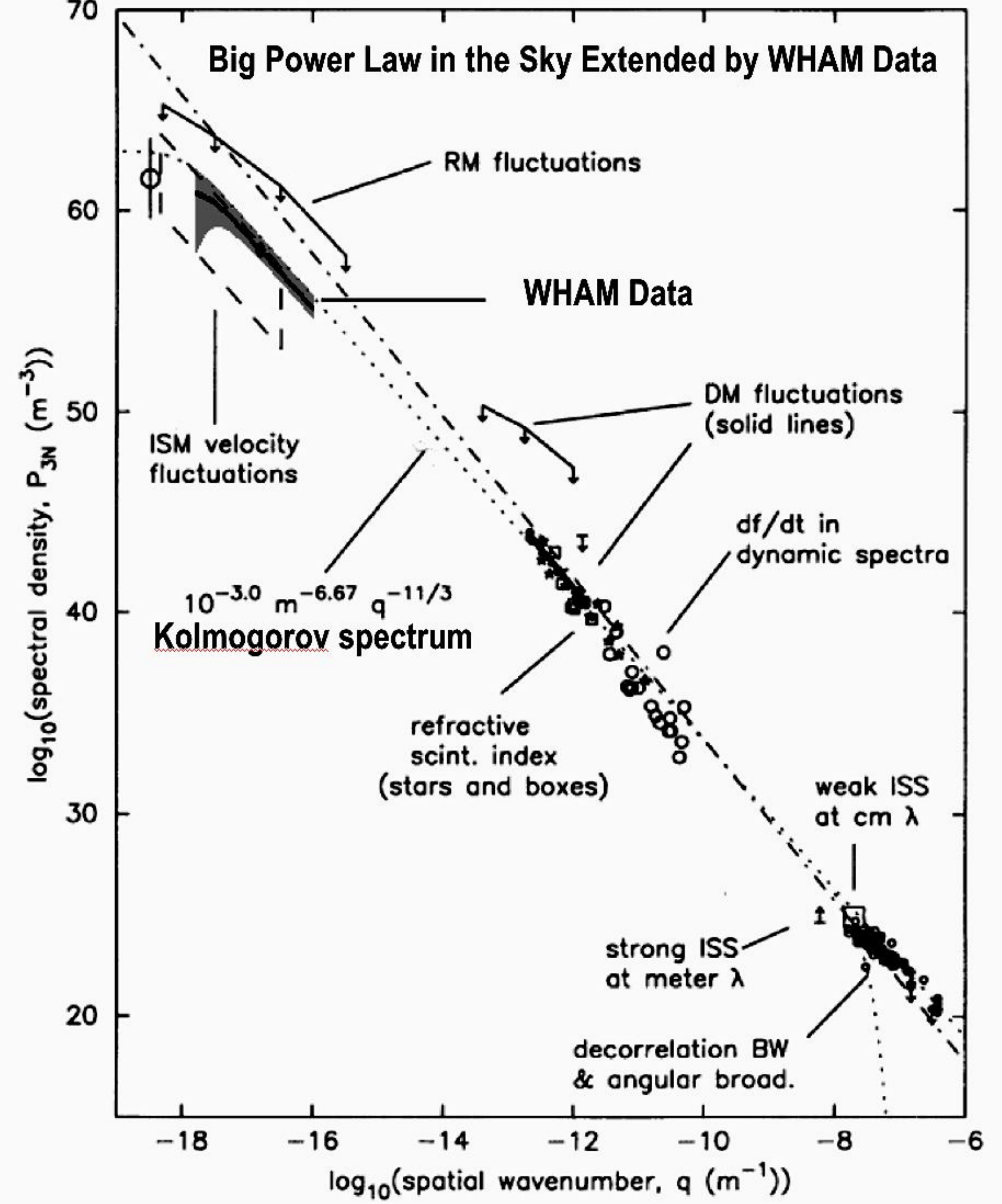}
\caption{Turbulence in the interstellar gas of the Milky Way as revealed by
electron density fluctuations.  ``Big Power Law in the Sky''
\cite{Armstrongetal95} extended using WHAM data.  The slope corresponds to that
of Kolmogorov turbulence. From \cite{ChepurnovLazarian10}. \label{fig:power}}
\end{figure}

Figure~\ref{fig:power} illustrates the so-called ``Big Power Law in the Sky'' of the
electron density fluctuations.  The original version of the law was presented by
Armstrong et al. \cite{Armstrongetal95} for electron scattering and
scintillation data.  It was later extended by Chepurnov et al.
\cite{ChepurnovLazarian10} who used Wisconsin H$\alpha$ Mapper (WHAM) electron
density data.  We clearly see the power law extending over many orders of of
spatial scales and suggesting the existence of turbulence in the interstellar
medium. With more surveys, with more developed techniques we are getting more
evidence of the turbulent nature of astrophysical fluids.  For instance, for
many years non-thermal line Doppler broadening  of the spectral lines was used
as an evidence of turbulence\footnote{The power-law ranges that are universal
features of high-Reynolds-number turbulence can be inferred to be present from
enhanced rates of dissipation and mixing \cite{Eyink08} even when they are not
seen.}. The development of new techniques, namely, Velocity Channel Analysis
(VCA) and Velocity Correlation Spectrum (VCS) in a series of papers by Lazarian
\& Pogosyan \cite{LazarianPogosyan00, LazarianPogosyan04, LazarianPogosyan06,
LazarianPogosyan08} enabled researchers to use HI and CO spectral lines to
obtain the power spectra of turbulent velocities (see \cite{Lazarian09} for a
review and references therein).

As turbulence is known to change dramatically many processes, in particular,
diffusion and transport processes, it is natural to pose the question to what
extent the theory of astrophysical reconnection must take into account the
pre-existing turbulent environment.  We note that even if the plasma flow is
initially laminar, kinetic energy release by reconnection due to some slower
plasma process is expected to generate vigorous turbulent motion in high
Reynolds number fluids.

\subsection{MHD description of plasma motions}

Turbulence in plasma happens at many scales, from the largest to those below the
proton Larmor radius.  The effect of turbulence on magnetic reconnection is
different for different types of turbulence.  For instance, micro turbulence can
change the microscopic resistivity of plasmas and induce anomalous resistivity
effects (see \cite{Vekshteinetal70}).  In this review we advocate the idea that
for solving the problem of magnetic reconnection in most astrophysical important
cases the approach invoking MHD rather than plasma turbulence is adequate.  To
provide an initial support for this point, we shall reiterate a few known facts
about the applicability of MHD approximation (\cite{Kulsrud83},
\cite{Eyinketal11}).  Below we argue that MHD description is applicable to many
settings that include both collisional and collisionless plasmas, provided that
we deal with plasmas at sufficiently large scales.  To describe magnetized
plasma dynamics one should deal with three characteristic length-scales: the ion
gyroradius $\rho_i,$ the ion mean-free-path length $\ell_{mfp,i}$ arising from
Coulomb collisions, and the scale $L$ of large-scale variation of magnetic and
velocity fields.

One case of reconnection that is clearly not dealt with by the popular models of
collisionless reconnection (see above) is the ``strongly collisional'' plasma
with $\ell_{mfp,i}\ll \rho_i$.  This is the case e.g. of star interiors and
most accretion disk systems. For such ``strongly collisional'' plasmas a standard
Chapman-Enskog expansion provides a fluid description of the plasma
\cite{Braginsky65},  with a two-fluid model for scales between $\ell_{mfp,i}$
and the ion skin-depth $\delta_i= \rho_i/\sqrt{\beta_i}$ and an MHD description
at scales much larger than $\delta_i$.  This is the most obvious case of MHD
description for plasmas.

Hot and rarefied astrophysical plasmas are often ``weakly collisional'' with
$\ell_{mfp,i}\gg \rho_i$.  Indeed, the relation that follows from the standard
formula for the Coulomb collision frequency (e.g. see \cite{Fitzpatrick11},
Eq.~1.25) is
\begin{equation}
\frac{\ell_{mfp,i}}{\rho_i}\propto \frac{\Lambda}{\ln\Lambda}\frac{V_A}{c}, \label{lmfp-rho}
\end{equation}
where  $\Lambda=4\pi n\lambda_D^3$ is the plasma parameter, or the number of
particles within the Debye screening sphere, which indicates that  $\Lambda$ can
be very large.  Typical values for some weakly coupled cases are shown in
Table~\ref{tab:parameters} \cite{Eyinketal11}.

\begin{table}
\caption{Representative Parameters for Some Weakly-Coupled Astrophysical Plasmas (from \cite{Eyinketal11}) \label{tab:params}}
\begin{tabular}{llll}
\hline
\hline
Parameter & warm ionized & post-CME & solar wind at  \\
& ISM${\,\!}^a$ & current sheets${\,\!}^b$ & magnetosphere${\,\!}^c$ \\
\hline
density $n,\,cm^{-3}$ & .5 & $7\times 10^7$ & 10 \\
temperature $T,\, eV$ & .7 & $10^3$ &10 \\
plasma parameter $\Lambda$ & $4\times 10^9$ & $2\times 10^{10}$ & $5\times 10^{10}$ \\
ion thermal velocity $v_{th,i},\,cm/s$ & $10^6$ & $3\times 10^7$ & $5\times 10^6$ \\
ion mean-free-path $\ell_{mfp,i},\,cm$ & $6\times 10^{11}$ & $10^{10}$ & $7\times 10^{12}$ \\
magnetic diffusivity $\lambda,\, cm^2/s$ & $10^7$ & $8\times 10^2$ & $6\times 10^5$ \\
\hline
magnetic field $B,\, G$ & $10^{-6}$ & 1 & $10^{-4}$ \\
plasma beta $\beta$ & 14 & 3 & 1 \\
Alfv\'{e}n speed $V_A,\,cm/s$ & $3\times 10^5$ & $3\times 10^7$ & $7\times 10^6$ \\
ion gyroradius $\rho_i,\,cm$ & $10^8$ & $3\times 10^3$ & $6\times 10^6$ \\
\hline
large-scale velocity $U,\,cm/s$ & $10^6$ & $4\times 10^6$ & $5\times 10^6$ \\
large length scale $L,\,cm$ & $10^{20}$ & $5\times 10^{10}$ & $10^8$ \\
Lundquist number $S_L=\frac{V_A L}{\lambda}$ & $3\times 10^{18}$ & $2\times 10^{15}$ & $10^9$ \\
resistive length${\,\!}^*$ $\ell_\eta^\perp,\, cm$ & $5\times 10^5$ & 1 & 20\\
\hline
& & & \\
\multicolumn{4}{l}{\footnotesize{${\,\!}^a$\cite{NormanFerrara96, Ferriere01} \,\,\,\, ${\,\!}^b$\cite{Bemporad08}
\,\,\,\, ${\,\!}^c$\cite{Zimbardoetal10}}}\\
\multicolumn{4}{l}{\footnotesize{*This nominal resistive scale is calculated from $\ell_\eta^\perp \simeq L (V_A/U)
S_{L}^{-3/4}$, assuming GS95 turbulence holds}}\\
\multicolumn{4}{l}{\footnotesize{down to that scale, and should not be taken literally
when $\ell_\eta^\perp<\rho_i.$}}\\
 \label{tab:parameters}
\end{tabular}
\end{table}

For the ``weakly collisional'' but well magnetized plasmas one can invoke the
expansion over the small ion gyroradius.  This results in the ``kinetic MHD
equations'' for lengths much larger than $\rho_i$.  The difference between these
equations and the MHD ones is that the pressure tensor in the momentum equation
is anisotropic, with the two components $p_\|$ and $p_\perp$ of the pressure
parallel and perpendicular to the local magnetic field direction
\cite{Kulsrud83}.  ``Weakly collisional'', i.e.  $L\gg \ell_{mfp,i}.$, and
collisionless, i.e.  $\ell_{mfp,i}\gg L$ systems have been studied recently
\cite{Kowaletal11, SantosLimaetal13}.  While the direct collisions are
infrequent, compressions of the magnetic field induces anisotropies, as a
consequence of the adiabatic invariant conservation, in the phase space particle
distribution.  This induces instabilities that act upon plasma causing particle scattering \cite{SchekochihinCowley06,
LazarianBeresnyak06}.  Thus instead of Coulomb collisional frequency a new
frequency of scattering is invoked.  In other words, particles do not interact
between each other, but each particle interacts with the ensemble of small scale
perturbations induced by instabilities in the compressed magnetized plasmas.  By
adopting the in-situ measured distribution of particles in the collisionless
solar wind Santos-Lima et al. \cite{SantosLimaetal13} showed numerically that
the dynamics of such plasmas is identical to that of MHD.

Even without invoking instabilities, one can approach ``weakly collisional''
plasmas solving for the magnetic field using an ideal induction equation, if one
ignores all collisional effects. In many cases, e.g. in the ISM and the
magnetosphere (see Table~\ref{tab:parameters}) the resistive length-scale
$\ell_\eta^\perp$ is much smaller than both $\rho_i$ and $\rho_e\approx
\frac{1}{43}\rho_i$. Magnetic field-lines are, at least formally, well
``frozen-in'' on these scales\footnote{In \S\ref{ssec:flux_freez} we discuss the
modification of the frozen in concept in the presence of turbulence. This is not
important for the present discussion, however.}. In the ``weakly collisional''
case the``kinetic MHD'' description can be simplified at  scales greater than
$\ell_{mfp,i}$ by including the Coulomb collision operator and making a
Chapman-Enskog expansion. This reproduces a fully MHD description
at those large scales.  The idealized warm ionized phase of ISM represents
``weakly collisional'' plasmas in Table~\ref{tab:parameters}.

We can also note that additional simplifications that justify the MHD approach
occur if the  turbulent fluctuations are small compared to the mean magnetic
field, and having length-scales parallel to the mean field much larger than
perpendicular length-scales.  Treating wave frequencies that are low compared to
the ion cyclotron frequency we enter the domain of ``gyrokinetic approximation''
which is commonly used in fusion plasmas. This approximation was advocated for
application in astrophysics by \cite{Schekochihinetal07, Schekochihinetal09}.

For the ``gyrokinetic approximation'' at length-scales larger than the ion
gyroradius $\rho_i$ the incompressible shear-Alfv\'{e}n wave modes get decoupled
from the compressive modes and can be described by the simple ``reduced MHD''
(RMHD) equations.  As we argue later in the review, the shear-Alfv\'{e}n modes
are the modes that induce fast magnetic reconnection, while the other modes are
of auxiliary importance for the process.

All in all, our considerations in this part of the review support the generally
accepted notion that the MHD approximation is adequate for most astrophysical
fluids at sufficiently large scales.  A lot of work on reconnection is
concentrated on the small scale dynamics, but if magnetic reconnection is
determined by large scale motions, as we argue in this review, then the MHD
description of magnetic reconnection is appropriate.

\subsection{Modern understanding of MHD turbulence}
\label{ssec:modern_turb}

Within this volume MHD turbulence is described in the chapter by Beresnyak \&
Lazarian (see also a description of MHD turbulence in the star formation context
in the chapter by H. Vazquez-Semadeni).  Therefore in presenting the major MHD
turbulence results that are essential for our further derivation in the review,
we shall be very brief. We will concentrate on Alfv\'{e}nic modes, while
disregarding the slow and fast magnetosonic modes that in principle contribute
to MHD turbulence \cite{ChoLazarian02, ChoLazarian03, KowalLazarian10}.  The
interaction between the modes is in many cases not significant, which allows the
separate treatment of Alfv\'{e}n modes \cite{ChoLazarian02, GoldreichSridhar95,
LithwickGoldreich01}.

While having a long history of competing ideas, the theory of MHD turbulence has
become testable recently due to the advent of numerical simulations (see
\cite{Biskamp03}) which confirmed the prediction of magnetized Alfv\'{e}nic
eddies being elongated in the direction of the local magnetic field (see
\cite{Shebalinetal83, Higdon84}) and provided results consistent with the
quantitative relations for the degree of eddy elongation obtained  in the
fundamental study by \cite{GoldreichSridhar95} (henceforth GS95).

The relation between the parallel and perpendicular dimensions of eddies in GS95
picture are presented by the so called critical balance condition, namely,
\begin{equation}
\ell_{\|}^{-1}V_A\sim \ell_{\bot}^{-1}\delta u_\ell,
\label{crit}
\end{equation}
where $\delta u_\ell$ is the eddy velocity, while $\ell_{\|}$ and $\ell_{\bot}$
are, respectively, eddy scales parallel and perpendicular to the {\it local}
direction of magnetic field.  The {\it local} system of reference is that
determined by the direction of magnetic field at the scale in the vicinity of
the eddy.  It should be definitely distinguished from the mean magnetic field
reference frame \cite{LithwickGoldreich01, LazarianVishniac99, ChoVishniac00,
MaronGoldreich01, Choetal02}, where no universal relations between the eddy
scale exist.  This is very natural, as small scale turnover eddies can be
influenced only by the magnetic field around these eddies.

The motions perpendicular to the local magnetic field are essentially
hydrodynamic.  Therefore, combining (\ref{crit}) with the Kolmogorov cascade
notion, i.e. that the energy transfer rate is $\delta
u^2_{\ell}/(\ell_{\bot}/\delta u_{\ell})=const$ one gets $\delta u_\ell\sim
\ell_\bot^{1/3}$, which coincides with the known Kolmogorov relation between the
turbulent velocity and the scale. For the relation between the parallel and
perpendicular scales one gets
\begin{equation}
\ell_{\|}\propto L_{i}^{1/3}\ell_{\bot}^{2/3},
\label{anis}
\end{equation}
where $L_i$ is the turbulence injection scale. Note that recent measurements of
anisotropy in the solar wind are consistent with Eq. (\ref{anis})
\cite{Podesta10, Wicksetal10, Wicksetal11}.

In its original form the GS95 model was proposed for energy injected
isotropically with velocity amplitude $u_L=V_A$.  If the turbulence is injected
at velocities $u_L\ll V_A$ (or anisotropically with $L_{i,\|}\ll L_{i,\bot}$),
then the turbulent cascade is weak and $\ell_{\bot}$ decreases while
$\ell_{\|}=L_i$ stays the same \cite{LazarianVishniac99, MontgomeryMatthaeus95,
Galtieretal00, NgBhattacharjee96}.  In other words, as a result of the weak
cascade the eddies become thinner, but preserve the same length along the local
magnetic field.  It is possible to show that the interactions within weak
turbulence increase and transit to the regime of the strong MHD turbulence
at the scale
\begin{equation}
l_{trans}\sim L_i(u_L/V_A)^2\equiv L_i M_A^2~~~M_A<1
\label{trans}
\end{equation}
and the velocity at this scale is $v_{trans}=u_L M_A$, with $M_A=u_L/V_A\ll 1$
beeing the Alfv\'{e}nic Mach number of the turbulence \cite{LazarianVishniac99,
Lazarian06}.  Thus, weak turbulence has a limited,  i.e. $[L_i, L_i M_A^2]$
inertial range and at small scales it transits into the regime of strong
turbulence\footnote{We should stress that weak and strong are not the
characteristics of the amplitude of turbulent perturbations, but the strength of
non-linear interactions (see more discussion in \cite{Choetal03}) and small
scale Alfv\'{e}nic perturbations can correspond to a strong Alfv\'{e}nic
cascade.}.

Table~\ref{tab:regimes} illustrates different regimes of MHD turbulence both
when it is injected isotropically at superAlfv\'{e}nic and subAlfv\'{e}nic
velocities. Naturally, superAlfv\'{e}nic turbulence at large scales is similar
to the ordinary hydrodynamic turbulence, as weak magnetic fields cannot strongly
affect turbulent motions.  However, at the scale
\begin{equation}
l_A=L_i(V_A/u_L)^3=L_iM_A^{-3}~~~M_A>1
\label{alf}
\end{equation}
the motions become Alfv\'{e}nic.

\begin{table*}[t]
\caption{Regimes and ranges of MHD turbulence. \label{tab:regimes}}
\centering
\begin{tabular}{lllll}
\hline
\hline
Type                        & Injection                                                 &  Range   & Motion & Ways\\
of MHD turbulence  & velocity                                                   & of scales & type         & of study\\
\hline
Weak                       & $u_L<V_A$ & $[L_i, L_i M_A^2]$          & wave-like & analytical\\
\hline
Strong                      &                      &                                        &                 &                \\
subAlfv\'{e}nic            &  $u_L<V_A$ & $[L_i M_A^2, l_{min}]$ & eddy-like & numerical \\
\hline
Strong                    &                        &                                          &                 &                   \\
superAlfv\'{e}nic       & $u_L > V_A$ & $[l_A, l_{min}]$                    & eddy-like & numerical \\
\hline
& & & \\
\multicolumn{5}{l}{\footnotesize{$L_i$ and $l_{min}$ are injection and dissipation scales, respectively}}\\
\multicolumn{5}{l}{\footnotesize{$M_A\equiv u_L/V_A$. }}\\
\end{tabular}
\end{table*}

In this review we address the reconnection mediated by turbulence.  For this the
regime of weak, i.e. wave-like, perturbations can be an important part of the
dynamics.  A description of MHD turbulence that incorporates both weak and
strong regimes was presented in \cite{LazarianVishniac99} (henceforth LV99). In
the range of length-scales where turbulence is strong, this theory implies that
\begin{equation}
\ell_{\|}\approx L_i \left(\frac{\ell_{\bot}}{L_i}\right)^{2/3} M_A^{-4/3}
\label{Lambda}
\end{equation}
\begin{equation}
\delta u_{\ell}\approx u_{L} \left(\frac{\ell_{\bot}}{L_i}\right)^{1/3} M_A^{1/3},
\label{vl}
\end{equation}
when the turbulence is driven isotropically on a scale $L_i$ with an amplitude
$u_L$.  These are equations that we will use further to derive the magnetic
reconnection rate.

Here we do not discuss attempts to modify GS95 theory by adding concepts like
``dynamical alignment'', ``polarization'', ``non-locality'' \cite{Boldyrev06,
BeresnyakLazarian06, BeresnyakLazarian09, Gogoberidze07}.  First of all, those
do not change the nature of turbulence to affect the reconnection of the weakly
turbulent magnetic field.  Indeed, in LV99 the calculations were provided for a
wide range of possible models of anisotropic Alfv\'{e}nic turbulence and
provided fast reconnection.  Moreover, more recent studies
\cite{BeresnyakLazarian10, Beresnyak11, Beresnyak12} support the GS95 model.  A
more detailed discussion of MHD turbulence can be found in the recent review
(e.g. \cite{BrandenburgLazarian13}) and in Beresnyak and Lazarian's Chapter in
this volume.

GS95 presents a model of 3D MHD turbulence that exists in our 3D world.
Historically, due to computational reasons, many MHD related studies were done
in 2D.  The problem of such studies in application to magnetic turbulence is
that shear Alfv\'{e}n waves that play the dominant role for 3D MHD turbulence
are entirely lacking in 2D. Furthermore, all magnetized turbulence in 2D is
transient, because the dynamo mechanism required to sustain magnetic fields is
lacking in 2D \cite{Zeldovich57}.  Thus the relation of 2D numerical studies
invoking MHD turbulence, e.g. magnetic reconnection in 2D turbulence, and the
processes in the actual 3D geometry is not clear.  A more detailed discussion of
this point can be found in \cite{Eyinketal11}.

\section{Magnetic reconnection in the presence of turbulence}
\label{sec:reconnection}

\subsection{Initial attempts to invoke turbulence to accelerate magnetic reconnection}

The first attempts to appeal to turbulence in order to enhance the reconnection
rate were made more than 40 years ago.  For instance, some papers have
concentrated on the effects that turbulence induces on the microphysical level.
In particular, Speiser \cite{Speiser70} showed that in collisionless plasmas the
electron collision time should be replaced with the electron retention time in
the current sheet.  Also Jacobson \cite{Jacobson84} proposed that the current
diffusivity should be modified to include the diffusion of electrons across the
mean field due to small scale stochasticity.  However, these effects are
insufficient to produce reconnection speeds comparable to the Alfv\'{e}n speed
in most astrophysical environments.

``Hyper-resistivity'' \cite{Strauss86, BhattacharjeeHameiri86,
HameiriBhattacharjee87, DiamondMalkov03} is a more subtle attempt to derive fast
reconnection from turbulence within the context of mean-field resistive MHD.
The form of the parallel electric field can be derived from magnetic helicity
conservation.  Integrating by parts one obtains a term which looks like an
effective resistivity proportional to the magnetic helicity current.  There are
several assumptions implicit in this derivation.  The most important objection
to this approach is that by adopting a mean-field approximation, one is already
assuming some sort of small-scale smearing effect, equivalent to fast
reconnection.  Furthermore, the integration by parts involves assuming a large
scale magnetic helicity flux through the boundaries of the exact form required
to drive fast reconnection. The problems of the hyper-resistivity approach are
discussed in detail in \cite{Eyinketal11}.

A more productive development was related to studies of instabilities of the
reconnection layer.  Strauss \cite{Strauss88} examined the enhancement of
reconnection through the effect of tearing mode instabilities within current
sheets. However, the resulting reconnection speed enhancement is roughly what
one would expect based simply on the broadening of the current sheets due to
internal mixing\footnote{In a more recent work Shibata \& Tanuma
\cite{ShibataTanuma01} extended the concept suggesting that tearing may result
in fractal reconnection taking place on very small scales.}.  Waelbroeck
\cite{Waelbroeck89} considered not the tearing mode, but the resistive kink mode
to accelerate reconnection. The numerical studies of tearing have become an
important avenue for more recent reconnection research \cite{Loureiroetal09,
Bhattacharjeeetal09}.  As we discuss later in realistic 3D settings tearing
instability develops turbulence \cite{Karimabadietal13, Beresnyak13b}) which
induces a transfer from laminar to turbulent reconnection\footnote{Also earlier
works suggest such a transfer \cite{Dahlburgetal92, DahlburgKarpen94,
Dahlburg97, FerraroRogers04}.}.

Finally, a study of 2D magnetic reconnection in the presence of external
turbulence was done by \cite{MatthaeusLamkin85, MatthaeusLamkin86}.  An enhancement of the reconnection rate was
reported, but the numerical setup precluded the calculation of a long term
average reconnection rate.  As we discussed in \S\ref{ssec:lam_rec} bringing
in the Sweet-Parker model of reconnection magnetic field lines closer to each
other one can enhance the instantaneous reconnection rate, but this does not
mean that averaged long term reconnection rate increases.  This, combined with
the absence of the theoretical predictions of the expected reconnection rates
makes it difficult to make definitive conclusions from the study.  Note that, as
we discussed in \S\ref{ssec:modern_turb}, the nature of turbulence is different in
2D and 3D. Therefore, the effects accelerating magnetic reconnection mentioned in
the study, i.e. formation of X-points, compressions, may be relevant for 2D set
ups, but not relevant for the 3D astrophysical reconnection.  These effects are
not invoked in the model of the turbulent reconnection that we discuss below. We
also may note that a more recent study along the approach in
\cite{MatthaeusLamkin85} is one in \cite{Watsonetal07}, where the effects of
small scale turbulence on 2D reconnection were studied and no significant
effects of turbulence on reconnection were reported for the setup chosen by the
authors.

In a sense, the above study is the closest predecessor of
LV99 work that we deal below.  However, there are very
substantial differences between the approach of LV99 and
\cite{MatthaeusLamkin85}.  For instance, LV99, as is clear from the text below,
uses an analytical approach and, unlike \cite{MatthaeusLamkin85}, (a) provides
analytical expressions for the reconnection rates; (b) identifies the broadening
arising from magnetic field wandering as the mechanism for inducing fast
reconnection; (c) deals with 3D turbulence and identifies incompressible
Alfv\'{e}nic motions as the driver of fast reconnection.

\subsection{Model of magnetic reconnection in weakly turbulent media}
\label{ssec:lv99_model}

\begin{figure}
\centering
\includegraphics[width=0.65\textwidth]{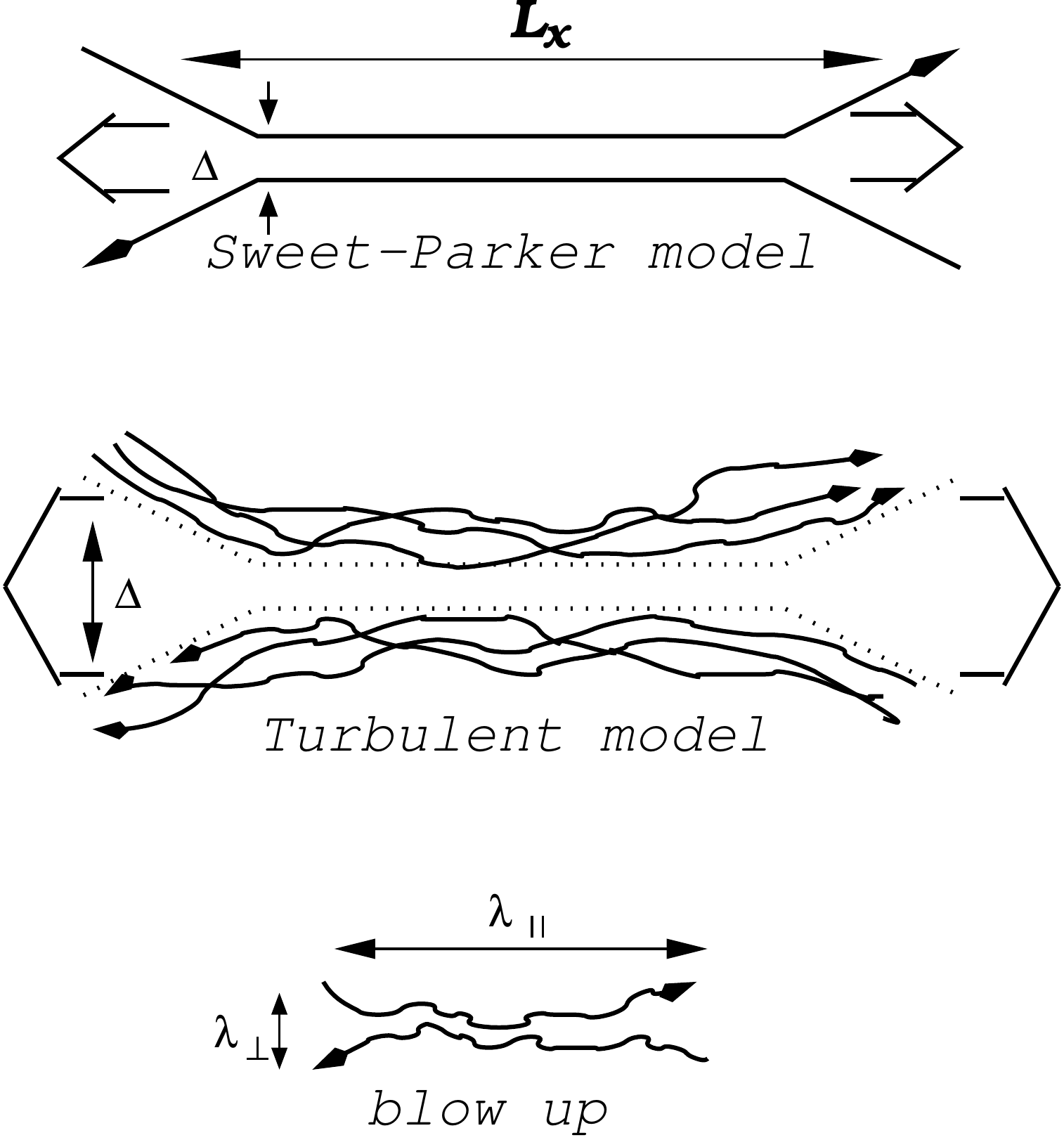}
\caption{{\it Upper plot}: Sweet-Parker model of reconnection.  The outflow is
limited to a thin width $\delta$, which is determined by Ohmic diffusivity.  The
other scale is an astrophysical scale $L \gg \delta$.  Magnetic field lines are
assumed to be laminar.
{\it Middle plot}: Turbulent reconnection model that accounts for the
stochasticity of magnetic field lines.  The stochasticity introduced by
turbulence is weak and the direction of the mean field is clearly defined.  The
outflow is limited by the diffusion of magnetic field lines, which depends on
macroscopic field line wandering rather than on microscales determined by
resistivity.
{\it Low plot}: An individual small scale reconnection region.  The reconnection
over small patches of magnetic field determines the local reconnection rate. The
global reconnection rate is substantially larger as many independent patches
reconnect simultaneously.  Conservatively, the LV99 model assumes that the small
scale events happen at a slow Sweet-Parker rate.  Following
\cite{Lazarianetal04}.}
\label{recon}
\end{figure}

As we discussed earlier, considering astrophysical reconnection in laminar
environments is not normally realistic.  As a natural generalization of the
Sweet-Parker model it is appropriate to consider 3D magnetic field wandering
induced by turbulence as in LV99.  The corresponding model of magnetic
reconnection is illustrated by Figure~\ref{recon}.

Like the Sweet-Parker model, the LV99 model deals with a generic configuration,
which should arise naturally as magnetic flux tubes try to make their way one
through another.  This avoids the problems related to the preservation of wide
outflow which plagues attempts to explain magnetic reconnection via
Petscheck-type solutions.  In this model if the outflow of reconnected flux and
entrained matter is temporarily slowed down, reconnection will also slow down,
but, unlike Petscheck solution, will not change the nature of the solution.

The major difference between the Sweet-Parker model and the LV99 model is that
while in the former the outflow is limited by microphysical Ohmic diffusivity,
in the latter model the large-scale magnetic field wandering determines the
thickness of outflow.  Thus LV99 model does not depend on resistivity and,
depending on the level of turbulence, can provide both fast and slow
reconnection rates.  This is a very important property for explaining
observational data related to reconnection flares.

For extremely weak turbulence, when the range of magnetic field wandering
becomes smaller than the width of the Sweet-Parker layer $L S^{-1/2}$, the
reconnection rate reduces to the Sweet-Parker rate, which is the ultimate
slowest rate of reconnection.  As a matter of fact, this slow rate holds only
for Lundquist numbers less than $S_c$, the critical value for tearing mode
instability of the Sweet-Parker solution.  At higher Lundquist numbers,
self-generated turbulence will be the inevitable outcome of unstable breakdown
of the Sweet-Parker current sheet and this will yield the minimal reconnection
rate in an otherwise quiet environment (see, in particular,
\cite{Beresnyak13b}).

We note that LV99 does not appeal to a chaotic field created within a
hydrodynamic weakly magnetized turbulent flow.  On the contrary, the model
considers the case of a large scale, well-ordered magnetic field, of the kind
that is normally used as a starting point for discussions of reconnection.  In
the presence of turbulence one expects that the field will have some small scale
`wandering' and this effect changes the nature of magnetic reconnection.

Ultimately, the magnetic field lines will dissipate due to microphysical
effects, e.g. Ohmic resistivity.  However, it is important to understand that in
the LV99 model only a small fraction of any magnetic field line is subject to
direct Ohmic annihilation.  The fraction of magnetic energy that goes directly
into heating the fluid approaches zero as the fluid resistivity vanishes.  In
addition, 3D Alfv\'{e}nic turbulence enables many magnetic field lines to enter the
reconnection zone simultaneously, which is another difference between 2D and 3D
reconnection.

\subsection{Opening up of the outflow region via magnetic field wandering}

To get the reconnection speed one should calculate the thickness of the outflow
$\Delta$ that is determined by the magnetic field wandering.  This was done in
LV99, where the scaling relations for the wandering field lines were
established.

The scaling relations for Alfv\'{e}nic turbulence discussed in
\S\ref{ssec:modern_turb} allow us to calculate the rate of magnetic field
spreading.  A bundle of field lines confined within a region of width $y$ at
some particular point spreads out perpendicular to the mean magnetic field
direction as one moves in either direction following the local magnetic field
lines.  The rate of field line diffusion is given by
\begin{equation}
{d\langle y^2\rangle\over dx}\sim {\langle y^2\rangle\over \lambda_{\|}},
\end{equation}
where $\lambda_{\|}^{-1}\approx \ell_{\|}^{-1}$, $\ell_{\|}$ is the parallel
scale and the corresponding transversal scale, $\ell_{\perp}$, is $\sim \langle
y^2\rangle^{1/2}$, and $x$ is the distance along an axis parallel to the
magnetic field. Therefore, using equation (\ref{Lambda}) one gets
\begin{equation}
{d\langle y^2\rangle\over dx}\sim L_i\left({\langle y^2\rangle\over L_i^2}\right)^{2/3}
\left({u_L\over V_A}\right)^{4/3}
\label{eq:diffuse}
\end{equation}
where we have substituted $\langle y^2\rangle ^{1/2}$ for $\ell_{\perp}$.  This
expression for the diffusion coefficient will only apply when $y$ is small
enough for us to use the strong turbulence scaling relations, or in other words
when $\langle y^2\rangle < L_i^2(u_L/V_A)^4$.  Larger bundles will diffuse at
the rate of $L_i^2(u_L/V_A)^4$, which is the maximal rate.  For $\langle
y^2\rangle$ small, equation (\ref{eq:diffuse}) implies that a given field line
will wander perpendicular to the mean field line direction by an average amount
\begin{equation}
\langle y^2\rangle^{1/2}\approx {x^{3/2}\over L_i^{1/2}} \left({u_L\over V_A}\right)^{2}~~~x<L_i
\label{eq:diffuse2}
\end{equation}
in a distance $x$.  The fact that the rms perpendicular displacement grows
faster than $x$ is significant.  It implies that if we consider a reconnection
zone, a given magnetic flux element that wanders out of the zone has only a
small probability of wandering back into it.  We also note that $y$ proportional
to $x^{3/2}$ is a consequence of the process of Richardson diffusion that we
discuss below.

When the turbulence injection scale is less than the extent of the reconnection
layer, i.e. $Lx\gg L_i$ magnetic field wandering obeys the usual random walk
scaling with $L_x/L_i$ steps and the  mean squared displacement per step equal
to $L_i^2(u_L/V_A)^4$. Therefore
\begin{equation}
\langle y^2\rangle^{1/2}\approx (L_i x)^{1/2} (u_L/V_A)^2~~~x>L_i
\label{eq:diffuse3}
\end{equation}

Using Eqs. (\ref{eq:diffuse2}) and (\ref{eq:diffuse3}) one can derive the
thickness of the outflow $\Delta$ (see Figure~\ref{fig:sp_recon}) and obtain (LV99):
\begin{equation}
V_{rec}\approx V_A\min\left[\left({L_x\over L_i}\right)^{1/2},
\left({L_i\over L_x}\right)^{1/2}\right]
M_A^2,
\label{eq:lim2a}
\end{equation}
where $V_AM_A^2$ is proportional to the turbulent eddy speed.  This limit on the reconnection
speed is fast, both in the sense that it does not depend on the resistivity, and
in the sense that it represents a large fraction of the Alfv\'{e}n speed when
$L_i$ and $L_x$ are not too different and $M_A$ is not too small.  At the same
time, Eq. (\ref{eq:lim2a}) can lead to rather slow reconnection velocities for
extreme geometries or small turbulent velocities.  This, in fact, is an
advantage, as this provides a natural explanation for flares of reconnection,
i.e. processes which combine both periods of slow and fast magnetic
reconnection.  The parameters in Eq. (\ref{eq:lim2a}) can change in the process
of magnetic reconnection, as the energy injected by the reconnection will
produce changes in $M_A$ and $L_i$.  In fact, we claim that in the process of
magnetic reconnection and the energy injection that this entails for
magnetically dominated plasmas, one can expect both $L_i \rightarrow L_x$ and
$M_A\rightarrow 1$, which will induce efficient reconnection with $V_{rec}\sim
V_A$.

\subsection{Richardson diffusion and LV99 model}
\label{ssec:richardson_diff}

It is well known that at scales larger than the turbulence injection scale the
fluid exhibits diffusive properties.  At the same time, at scales less than the
turbulence injection scale the properties of diffusion are different.  Since the
velocity difference increases with separation, one expects that accelerated
diffusion, or super diffusion should take place.  This process was first
described by Richardson for hydrodynamic turbulence.  A similar effect is
present for MHD turbulence (see \cite{EyinkBenveniste13} and references
therein).

Richardson diffusion can be illustrated with a simple model.  Consider the
growth of the separation between two particles $dl(t)/dt\sim v(l),$ which for
Kolmogorov turbulence is $\sim \alpha_t l^{1/3}$, where $\alpha_t$ is proportional
to the energy cascading rate, i.e. $\alpha_t\approx V_L^3/L$ for
turbulence injected with superAlv\'{e}nic velocity $V_L$ at the scale $L$.  The
solution of this equation is
\begin{equation}
l(t)=[l_0^{2/3}+\alpha_t (t-t_0)]^{3/2},
\label{sol}
\end{equation}
which at late times leads to Richardson diffusion or $l^2\sim t^3$ compared with
$l^2\sim t$ for ordinary diffusion.

Richardson diffusion provides explosive separation of magnetic field lines.  It
is clear from Eq. (\ref{sol}) that the separation of magnetic field lines does
not depend on the initial separation $l_0$ after sufficiently long intervals of
time $t$.  Potentially, one can make $l_0$ very small, but, realistically, $l_0$
should not be smaller than the scale of the marginally damped eddies $l_{damp}$,
as the derivation of the Richardson diffusion assumes the existence of
inertial-range turbulence at the scales under study.  At scales less than
$l_{damp}$  diffusion is determined by the shearing by the marginally damped
eddies.  This is known to result in Lagrangian chaos and Lyapunov exponential
separation of the points.  Separation at long times in this regime does depend
on the initial separation of points.  In other words, in realistic turbulence up
to the scale of $l_{damp}$ the distance between the points preserves the memory
of the initial separation of points, while at scales larger than $l_{damp}$ this
dependence is washed out.

Richardson diffusion is important in terms of spreading magnetic fields.  In
fact, the magnetic field line spread as a function of the distance measured
along magnetic field lines, which we discussed in the previous subsection, is
also a manifestation of Richardson diffusion, but in space rather than in time.
Below, we, however, use the time dependence of Richardson diffusion to re-derive
the LV99 results.

Sweet-Parker reconnection can serve again as our guide. There we deal with Ohmic
diffusion.  The latter induces the mean-square distance across the reconnection
layer that a magnetic field-line can diffuse by resistivity in a time $t$ given
by
\begin{equation}
\langle y^2(t)\rangle \sim \lambda t.
\label{diff-dist}
\end{equation}
where $\lambda=c^2/4\pi\sigma$ is the magnetic diffusivity.  The field lines are
advected out of the sides of the reconnection layer of length $L_x$ at a
velocity of order $V_A$.  Therefore, the time that the lines can spend in the
resistive layer is the Alfv\'{e}n crossing time $t_A=L_x/V_A$. Thus, field lines
that can  reconnect are separated by a distance
\begin{equation}
\Delta = \sqrt{\langle y^2(t_A)\rangle} \sim \sqrt{\lambda t_A} = L_x/\sqrt{S},
\label{Delta}
\end{equation}
where $S$ is Lundquist number.  Combining Eqs. (\ref{eq.2}) and
(\ref{Delta}) one gets again the well-known Sweet-Parker result,
$v_{rec}=V_A/\sqrt{S}$.

Below, following  \cite{Eyinketal11} (henceforth ELV11) we provide a different
derivation of the reconnection rate within the LV99 model.  We make use of the
fact that in Richardson diffusion \cite{Kupiainen03} the mean squared
separation of particles $\langle |x_1(t)-x_2(t)|^2 \rangle\approx \epsilon t^3$,
where $t$ is time, $\epsilon$ is the energy cascading rate and
$\langle...\rangle$ denote an ensemble averaging.  For subAlfv\'{e}nic
turbulence $\epsilon\approx u_L^4/(V_A L_i)$ (see LV99) and therefore
analogously to Eq. (\ref{Delta}) one can write
\begin{equation}
\Delta\approx \sqrt{\epsilon t_A^3}\approx L(L/L_i)^{1/2}M_A^2
\label{D2}
\end{equation}
where it is assumed that $L<L_i$.  Combining Eqs. (\ref{eq.2})
and (\ref{D2}) one gets
\begin{equation}
v_{rec, LV99}\approx V_A (L/L_i)^{1/2}M_A^2.
\label{LV99}
\end{equation}
in the limit of $L<L_i$. Similar considerations allow to recover the LV99
expression for $L>L_i$, which differs from Eq.~(\ref{LV99}) by the change of the
power $1/2$ to $-1/2$.  These results coincide with those given by
Eq.~(\ref{eq:lim2a}).

\subsection{Role of plasma effects for magnetic reconnection}
\label{ssec:role_plasma_effects}

In the LV99 model the outflow is determined by turbulent motions that are
determined by the motions on the small scales.  The small scale physics in this
situation gets irrelevant if the level of turbulence is fixed.  Following
\cite{Eyinketal11} it is possible to define the criterion for the Hall effect to
be important within the LV99 reconnection model.

Using the GS95 model one can estimate the pointwise ratio of the Hall electric
field to the MHD motional field as
\begin{equation}
\frac{J/en}{u_L} \simeq \frac{c\delta B(\ell_\eta^\perp)/4\pi ne\ell_\eta^\perp}{u_L}
  \simeq \frac{\delta_i}{L_i} M_A S_L^{1/2}
\label{A3}
\end{equation}
where $S_L=V_AL_i/\lambda$ is the Lundquist number based on the forcing
length-scale of the turbulence and $M_A=u_L/V_A$ is the Alfv\'{e}nic Mach
number, $\ell_\eta^\perp$ is the resistive cutoff length, $J$ current density,
and $n$ electron density.  This can be expressed as a ratio $(J/en)/u_L\simeq
\delta_i/\delta_{T}$ of ion skin depth to the turbulent Taylor scale
\begin{equation}
\delta_{T} =L_i M_A^{-1} S_L^{-1/2},
\end{equation}
which can be interpreted heuristically as the current sheet thickness of
small-scale Sweet-Parker reconnection layers.  If the magnetic diffusivity in
the definition of the Lundquist number is assumed to be that based on the
Spitzer resistivity, given by $\lambda=\delta_e^2 v_{th,e}/\ell_{ei}$ where
$\delta_e$ is the electron skin depth, $v_{th,e}$ is the electron thermal
velocity, and $\ell_{ei}$ is the electron mean-free-path length for collisions
with ions, then $ S_L=\left(\frac{m_e}{m_i}\right)^{1/2} \beta^{-1/2}
\left(\frac{\ell_{ei}}{\delta_e}\right)^2 \left(\frac{L_i}{\ell_{ei}}\right), $
with $\beta=v_{th,i}^2/V_A^2$ the plasma beta.  Substituting into (\ref{A3})
provides
\begin{equation}
\frac{\delta_i}{\delta_{SP}}\simeq \left(\frac{m_i}{m_e}\right)^{1/4} (v_{th,i}/u_*)
               \beta^{1/4}  \left(\frac{\ell_{ei}}{L_i}\right)^{1/2},
\label{A6}
\end{equation}
which coincides precisely with the ratio defined by \cite{Yamadaetal06} (see their Eq.~(6)),
who proposed a ratio $\delta_i/\delta_{SP} > 1$ as the applicability criterion for Hall
reconnection rather than Sweet-Parker. However, satisfaction of this criterion does
not imply that the LV99 model is inapplicable!  Eq.~(\ref{A6}) states only
that small scale reconnection occurs via collisionless effects
and the structure of local, small-scale reconnection events should be strongly
modified by Hall or other collisionless effects, possibly with an $X$-type
structure, an ion layer thickness $\sim \delta_i,$ quadrupolar magnetic fields,
etc.  However, these local effects do not alter the resulting reconnection
velocity. See \cite{Eyinketal11}, Appendix B, for a more detailed discussion.

The LV99 model assumes that the thickness $\Delta$ of the reconnection layer is
set by turbulent MHD dynamics (line-wandering and Richardson diffusion).  Thus,
self-consistency requires that the length-scale $\Delta$ must be within the
range of scales where shear-Alfv\'{e}n modes are correctly described by
incompressible MHD.  This implies a criterion for collisionless reconnection in
the presence of turbulence
\begin{equation}
\rho_i \geq \Delta
\label{LV99-break}
\end{equation}
with $\Delta$ calculated from Eq.~(\ref{D2}) and $\rho_i$ the ion cyclotron
radius.  Since $\Delta \propto L_x,$ the large length-scale of the reconnecting
flux structures, this criterion is far from being satisfied in most
astrophysical settings.  For example, in the three cases of Table~\ref{tab:parameters},
one finds using $\Delta = LM_A^2$ that $\rho_i/\Delta\simeq 10^{-13}$ for the
warm ISM, $\simeq 10^{-6}$ for post-CME sheets, and $\simeq .1$ for the
magnetosphere.  In the latter case the criterion (\ref{LV99-break}) implies that
the effect of collisonless plasmas are important.  This is not a typical
situation, however.  To what extent turbulence below the Larmor radius should be
accounted for is an interesting open issue that we address only very briefly in
\S\ref{sec:obs_cons_tests}.

\section{Numerical testing of theory predictions}
\label{sec:numerical_testing}

\subsection{Approach to numerical testing}
\label{ssec:numerical_approach}

Numerical studies have proven to be a very powerful tool of the modern
astrophysical research.  However, one must admit their limits.  The
dimensionless ratios that determine the importance of  Ohmic resistivity are the
Lundquist and magnetic Reynolds numbers.  The difference between the two numbers
is not big and they are usually of the same order. Indeed, the magnetic Reynolds
number, which is the ratio of the magnetic field decay time to the eddy turnover
time, is defined using the injection velocity $v_l$ as a characteristic speed
instead of the Alfv\'{e}n speed $V_A$, as in the Lundquist number.  Therefore
for the sake of simplicity we shall be talking only about the Lundquist number.

As we discussed in \S\ref{ssec:lam_rec} because of the very large astrophysical
length-scales $L_x$ involved, astrophysical Lundquist numbers are huge, e.g. for
the ISM they are about $10^{16}$, while present-day MHD simulations correspond
to $S<10^4$.  As the numerical resource requirements scale as $N^4$, where $N$
is the ratio between the maximum and minimum scales resolved in a computational
model, it is feasible neither at present nor in the foreseeable future to have
simulations with realistically large Lundquist numbers.  In this situation,
numerical results involving magnetic reconnection cannot be directly related to
astrophysical situation and a brute force approach is fruitless.

Fortunately, numerical approach is still useful for testing theories and the
LV99 theory presents clear predictions to be tested for the moderate Lundquist
numbers available  with present-day computational facilities.  Below we present
the results of theory testing using this approach.

\subsection{Numerical simulations}

To simulate reconnection a code that uses a higher-order shock-capturing
Godunov-type scheme based on the essentially non oscillatory (ENO) spatial
reconstruction and Runge-Kutta (RK) time integration was used to solve
isothermal non-ideal MHD equations.  For selected simulations plasma effects
were simulated using accepted procedures \cite{Kowaletal09}.

\begin{figure*}
\centering
\raisebox{-0.5\height}{\includegraphics[trim = 20mm -10mm 20mm 0mm, clip, width=0.48\textwidth]{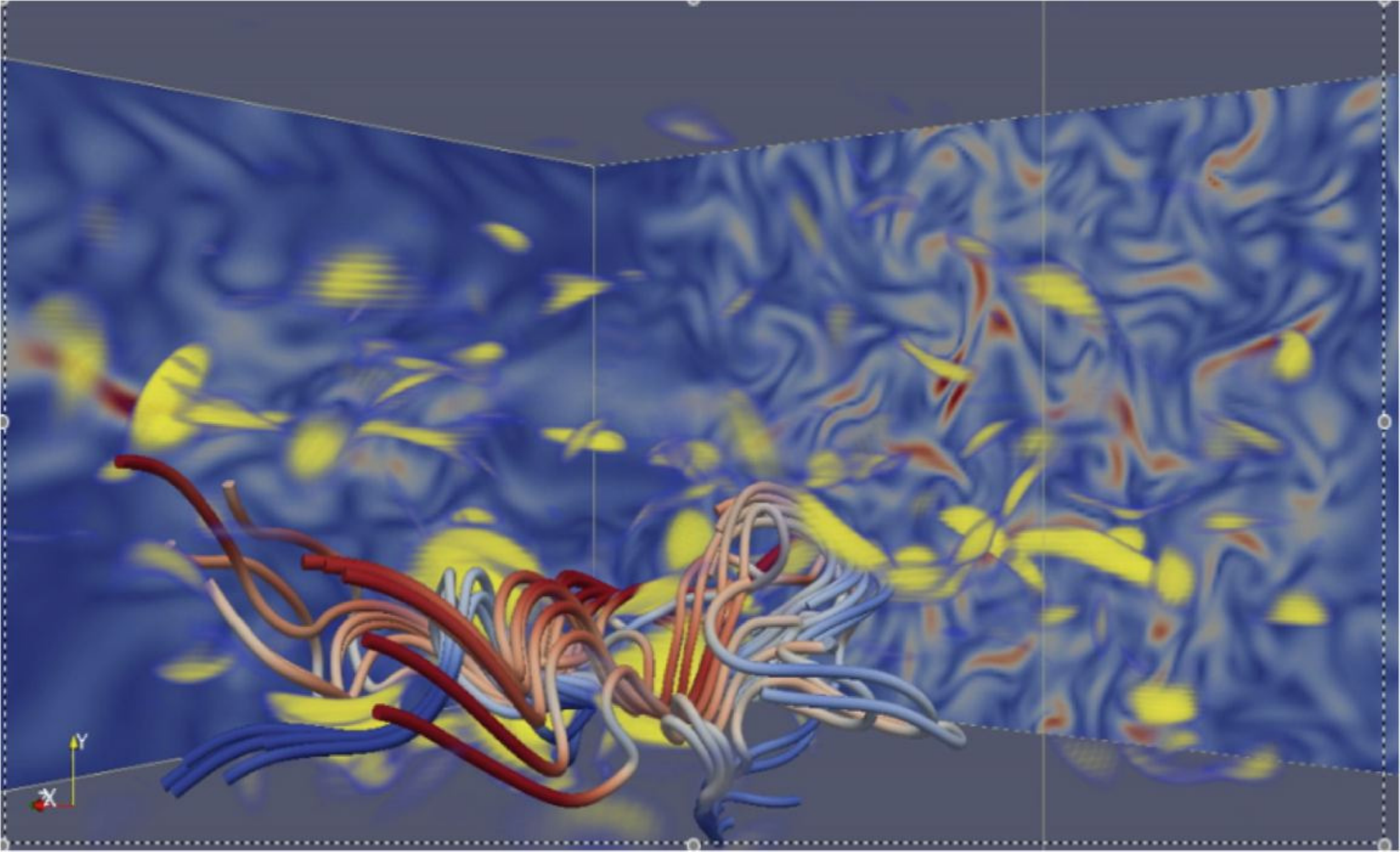}}
\raisebox{-0.5\height}{\includegraphics[width=0.25\textwidth]{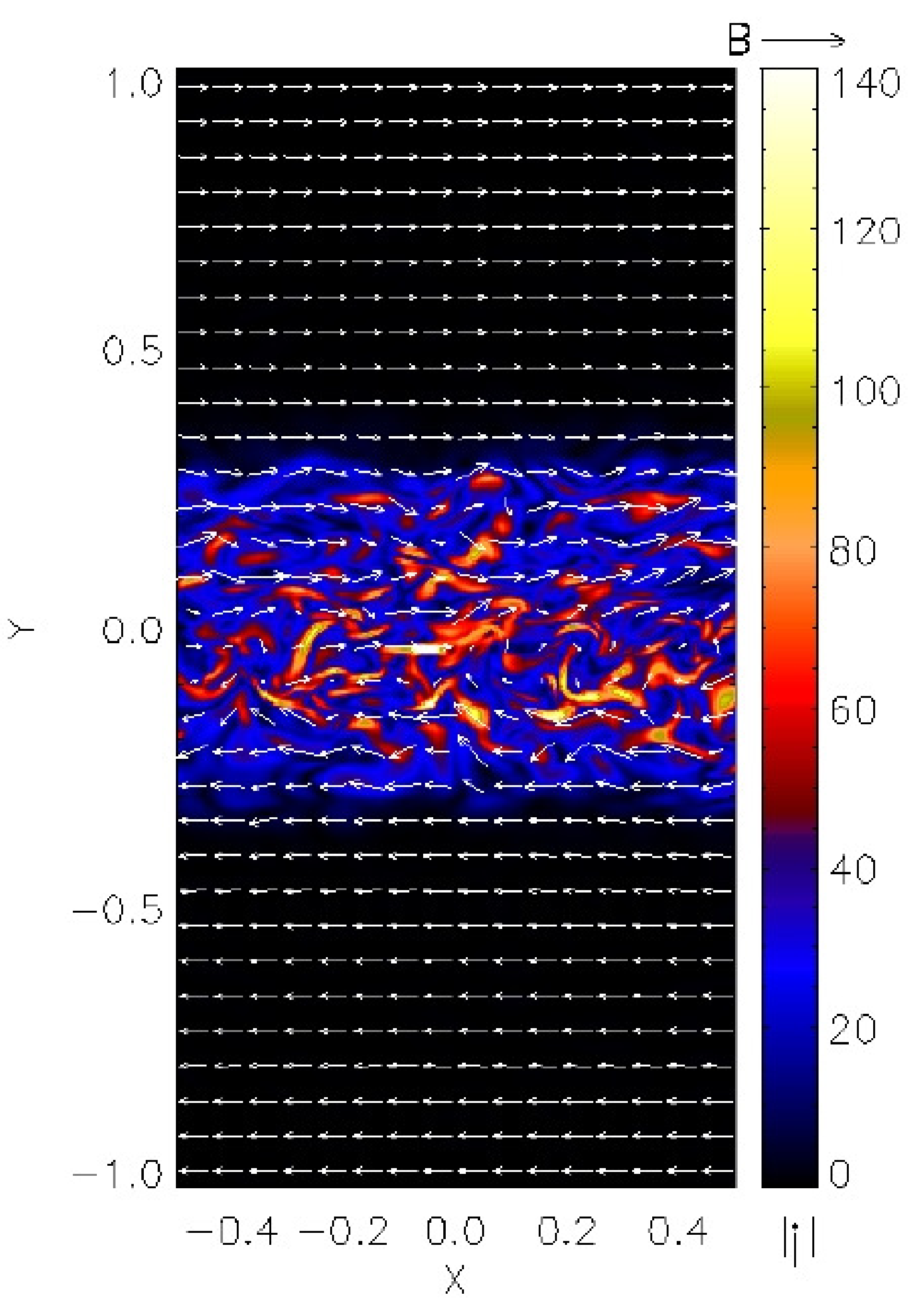}}
\raisebox{-0.5\height}{\includegraphics[width=0.25\textwidth]{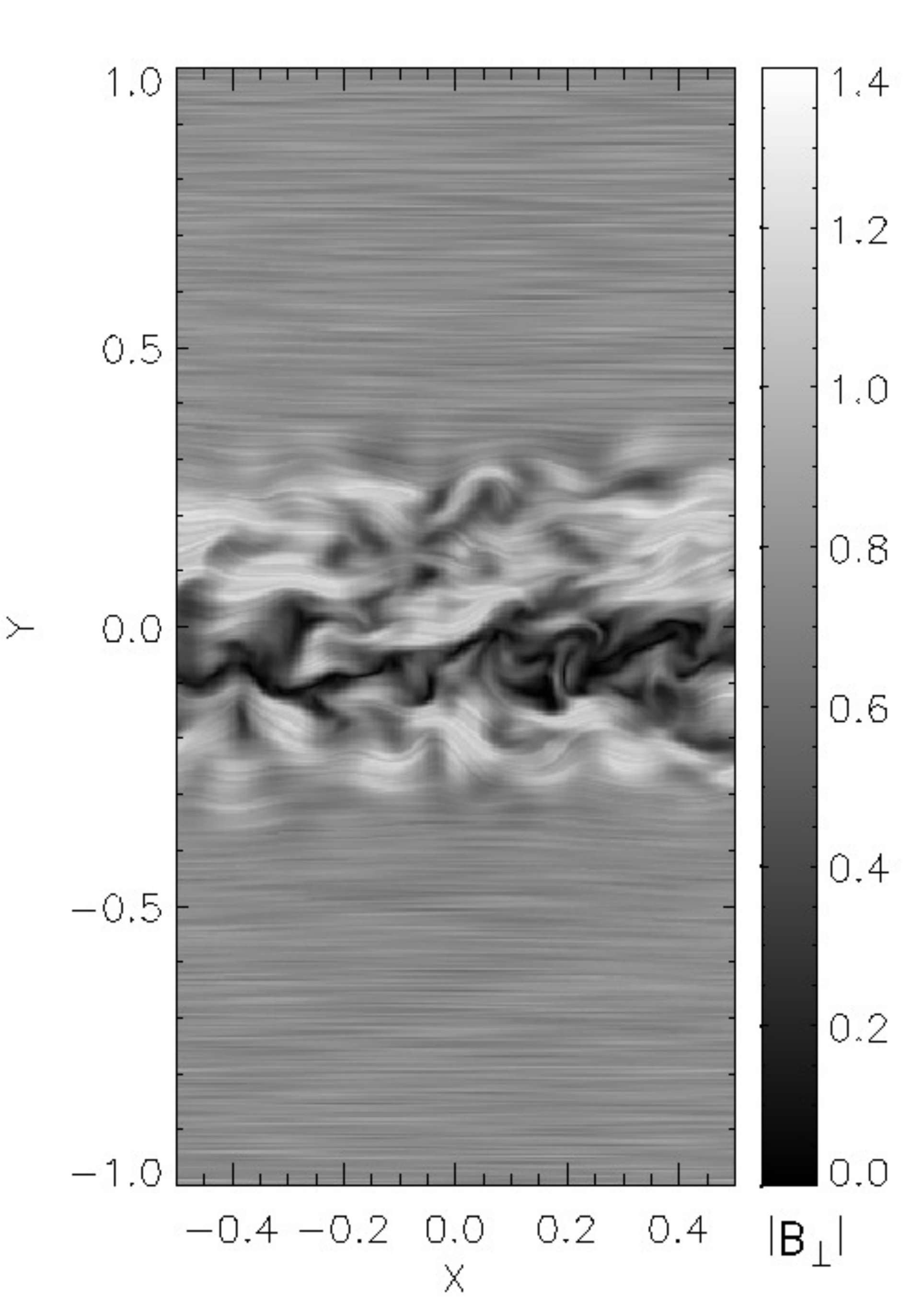}}
\caption{ Visualization of reconnection simulations in \cite{Kowaletal09, Kowaletal12}.
{\it Left panel}: Magnetic field in the reconnection region.  Large
perturbations of magnetic field lines arise from reconnection rather than
driving; the latter is subAlfv\'{e}nic.  The color corresponds to the polarization
of magnetic component $B_x$.
{\it Central panel}: Current intensity and magnetic field configuration during
stochastic reconnection.  We show a slice through the middle of the
computational box in the xy plane after twelve dynamical times for a typical
run.  The guide field is perpendicular to the page. The
intensity and direction of the magnetic field is represented by the length and
direction of the arrows.  The color bar gives the intensity of the current. The
reversal in $B_x$  is confined to the vicinity of y=0 but the current sheet is
strongly disordered with features that extend far from the zone of reversal.
{\it Right panel}: Representation of the magnetic field in the reconnection zone
with textures.
\label{visual}}
\end{figure*}

The driving of turbulence was performed using wavelets in \cite{Kowaletal09} and
in real space in \cite{Kowaletal12}.  In both cases the driving was supposed to
simulate pre-existing turbulence.  The visualization of simulations is provided
in Figure~\ref{visual}.

\subsection{Dependence on resistivity, turbulence injection power and turbulence scale}

\begin{figure}
\centering
\includegraphics[height=.45\textheight]{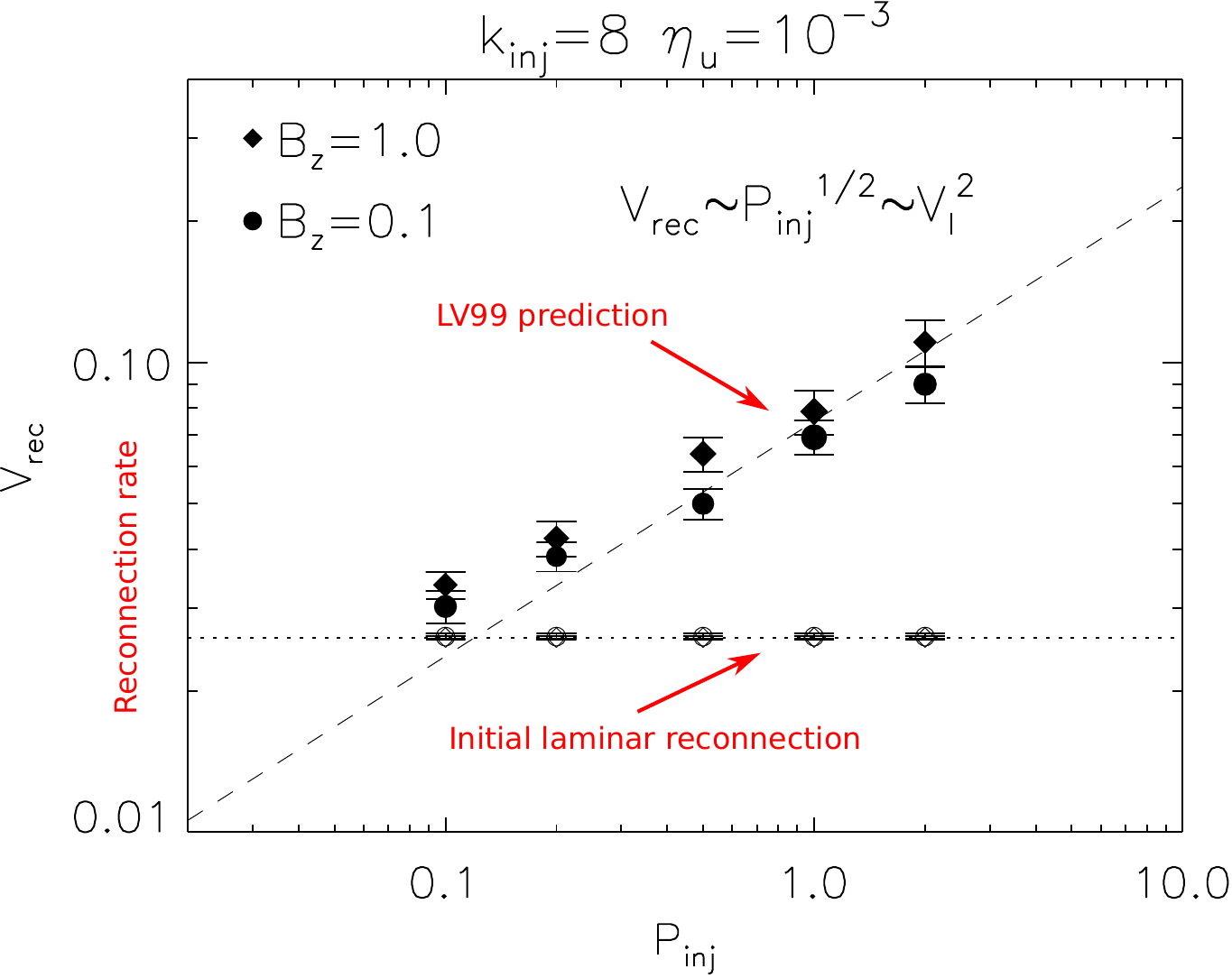}
\caption{The dependence of the reconnection velocity on the injection power for
different simulations with different drivings. The predicted LV99 dependence is
also shown. $P_{inj}$ and $k_{inj}$ are the injection power and scale,
respectively, $B_z$ is the guide field strength, and $\eta_u$ the value of
unifor resistivity coefficient. From \cite{Kowaletal12}. \label{figure6}}
\end{figure}

As we show below, simulations in \cite{Kowaletal09, Kowaletal12} provided very
good correspondence to the LV99 analytical predictions for the dependence on
resistivity, i.e. no dependence on resistivity for sufficiently strong
turbulence driving, and the injection power, i.e. $V_{rec}\sim P_{inj}^{1/2}$.
The corresponding dependence is shown in Figure~\ref{figure6}.

\begin{figure}
\centering
\includegraphics[height=.45\textheight]{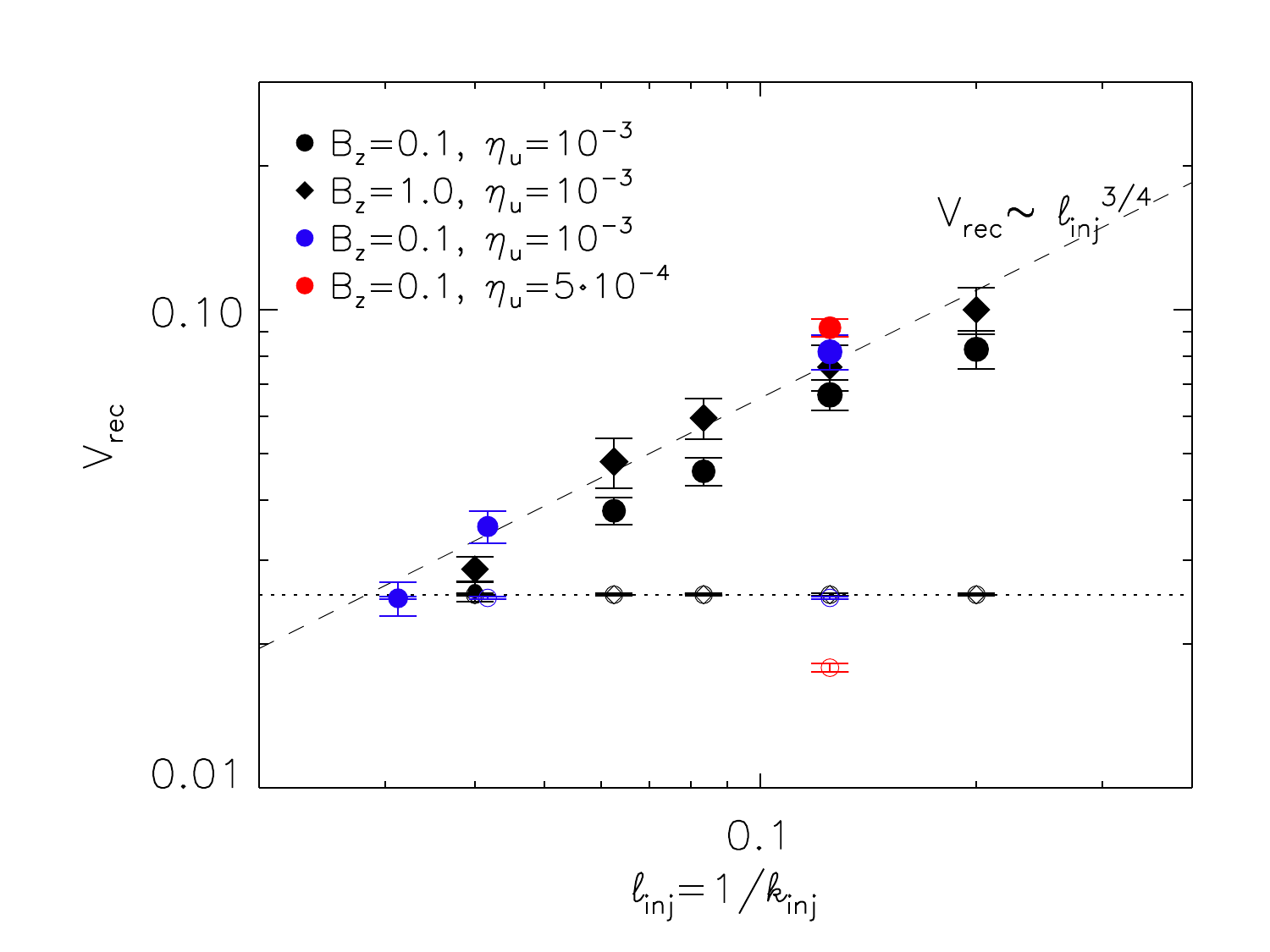}
\caption{The dependence of the reconnection velocity on the injection scale.
From \cite{Kowaletal12}. \label{figure7}}
\end{figure}

The measured dependence on the turbulence scale was a bit more shallow compared
to the LV99 predictions (see Figure~\ref{figure7}).  This may be due to the
existence of an inverse cascade that changes the driving from the idealized
assumptions in LV99 theory.

\subsection{Dependence on guide field strength, anomalous resistivity and viscosity}

The simulations did not reveal any dependence on the strength of the guide field $B_z$
(see Figure~\ref{figure6}).  This raises an interesting question.  In the limit
where the parallel wavelength of the strong turbulent eddies is less than the
length of the current sheet, we can rewrite the reconnection speed as
\begin{equation}
V_{rec} \approx \left({P L_x\over  V_{Ax}}\right)^{1/2} {1 \over k_\| V_A}.
\label{eddy}
\end{equation}
Here $P$ is the power in the strong turbulent cascade, $L_x$ and $V_{Ax}$ are
the length scale and Alfv\'{e}n velocity in the direction of the reconnecting
field, and $V_A$ is the total Alfv\'{e}n velocity, including the guide field.
The parallel wavenumber, $k_\|$, is characteristic of the large scale strongly
turbulent eddies.  We have assumed that such eddies are smaller than the size of
the current sheet. The point of rewriting the reconnection speed in this way is
that it is insensitive to assumptions about the connection between the input
power and driving scale and the parameters of the strongly turbulent cascade.

In a physically realistic situation, the dynamics that drive the turbulence,
whatever they are, provide a characteristic frequency and input power.  Since
the guide field enters only in the combination $k_\| V_A$, i.e. through the eddy
turn over rate, this implies that varying the guide field will not change the
reconnection speed.  However, in the numerical simulations cited above the
driving forces are independent of time scale, and sensitive to length scale, so
getting the physically realistic scaling is unexpected.  Further complicating
matters, we note that the dependence on length scale, described in the previous
section, is roughly what we expect if $k_\|$ is given by the forcing wavenumber.

This is the only clear discrepancy between the simulations and our predictions.
It is clearly important to understand its nature. One possibility is that the
transfer of energy from the weak turbulence driven by isotropic forcing to the
strongly turbulent eddies does not proceed in the expected manner.  This may be
due to the effect of the strong magnetic shear when a guide field is present.
Alternatively, the periodicity of the box, or the possibility that some wave
modes may leave the computational box faster than the nonlinear decay rate, may
skew the weakly turbulent spectrum.  The latter possibilities can be tested by
simulating strong turbulence and comparing the results with equation
(\ref{eddy}).  The former will require a more detailed theoretical and
computational study of the nature of the strong turbulence in the presence of
strong magnetic shear.

\begin{figure}[t]
\centering
\raisebox{-0.5\height}{\includegraphics[width=0.48\textwidth]{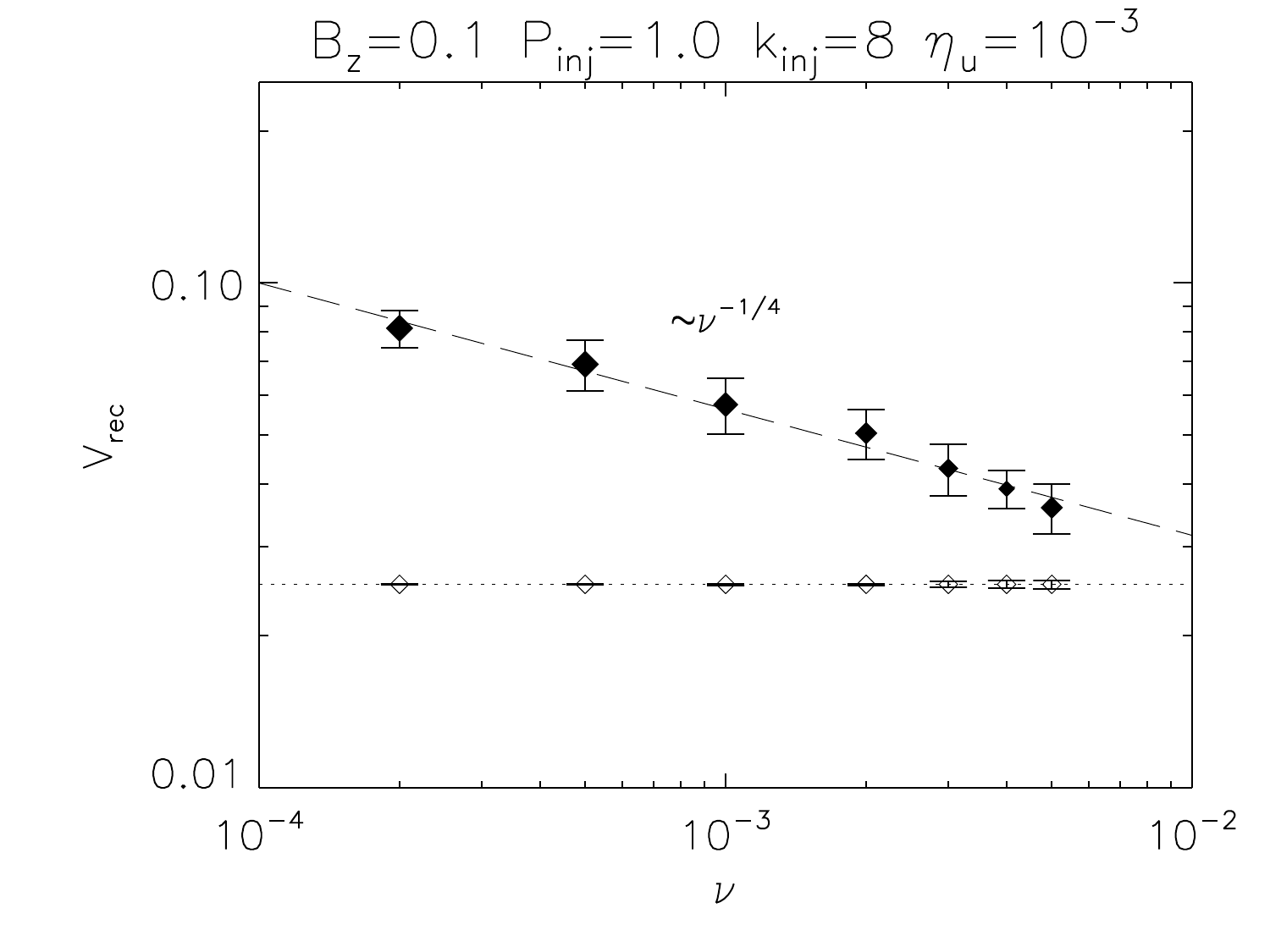}}
\raisebox{-0.5\height}{\includegraphics[width=0.48\textwidth]{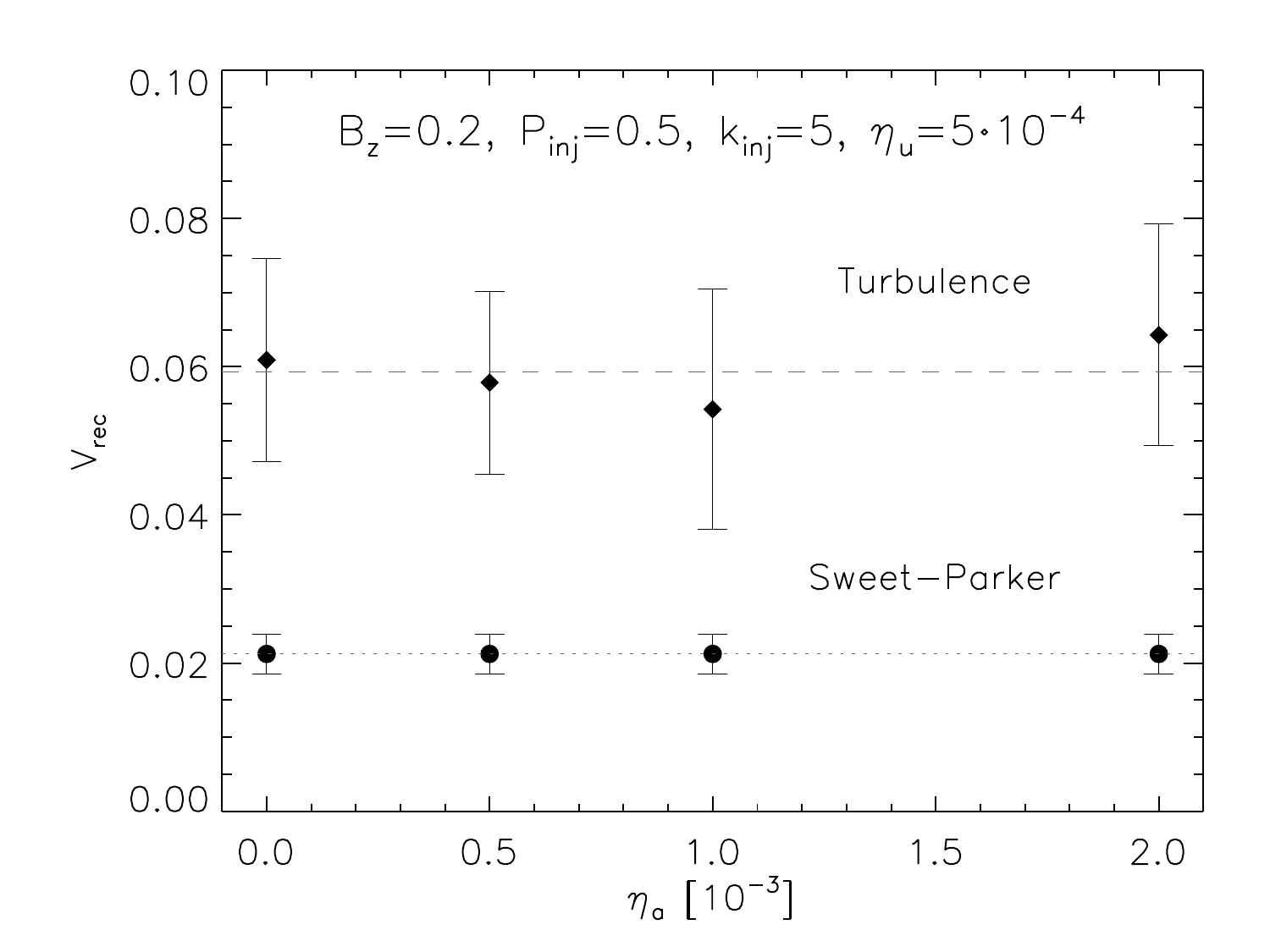}}
\caption{{\it Left panel}. The dependence of the reconnection velocity on
viscosity.  From \cite{Kowaletal12}.  {\it Right panel.} The dependence of
reconnection velocity on anomalous resistivity.  From \cite{Kowaletal09}.
\label{figure8}}
\end{figure}

The left panel of Figure~\ref{figure8} shows the dependence of the reconnection
rate on viscosity.  This can be explained as the effect of the finite inertial
range of turbulence.  For an extended range of motions, LV99 does not predict
any viscosity dependence.  However, for numerical simulations the range of
turbulent motions is very limited and any additional viscosity decreases the
resulting velocity dispersion and therefore the field wandering.

LV99 predicted that in the presence of sufficiently strong turbulence, plasma
effects should not play a role.  The accepted way to simulate plasma effects
within MHD code is to use anomalous resistivity.  The results of the
corresponding simulations are shown in the right panel of Figure~\ref{figure8}
and they confirm that the change of the anomalous resistivity does not change
the reconnection rate.

\subsection{Structure of the reconnection region}
\label{ssec:rec_reg_struct}

The internal structure of the reconnection region is important, both for the
role it plays in determining the overall reconnection speed, and for what it
tells us about the nature of local electric currents.  We can imagine two
extreme pictures.  First, the magnetic shear might be concentrated in a narrow,
albeit highly distorted sheet, whose width is determined by microphysics.  In
this case the outflow region would be much broader than the current sheet and
particle acceleration would take place in a nearly two dimensional, and highly
singular, region. The electric field in the current sheet would be very large,
much larger than one would be able to simulate directly.   At the other extreme,
the current sheet and the outflow zone may roughly coincide.  In this case the
current sheet is broad and the  currents are distributed widely within a three
dimensional volume. The electric fields would be roughly similar to what we
expect in homogeneous turbulence.   In the former case the turbulence within the
current sheet is difficult to estimate.  In the latter case, it would be similar
to the turbulence within a statistically homogeneous volume, of the sort that we
can simulate.  This would imply that the basic derivation of reconnection speeds
in LV99 is valid and particle acceleration takes place in a broad volume.
While both of these models are caricatures, they give a good sense of the basic
issues at stake.

\begin{figure*}
\centering
\includegraphics[width=0.45\textwidth]{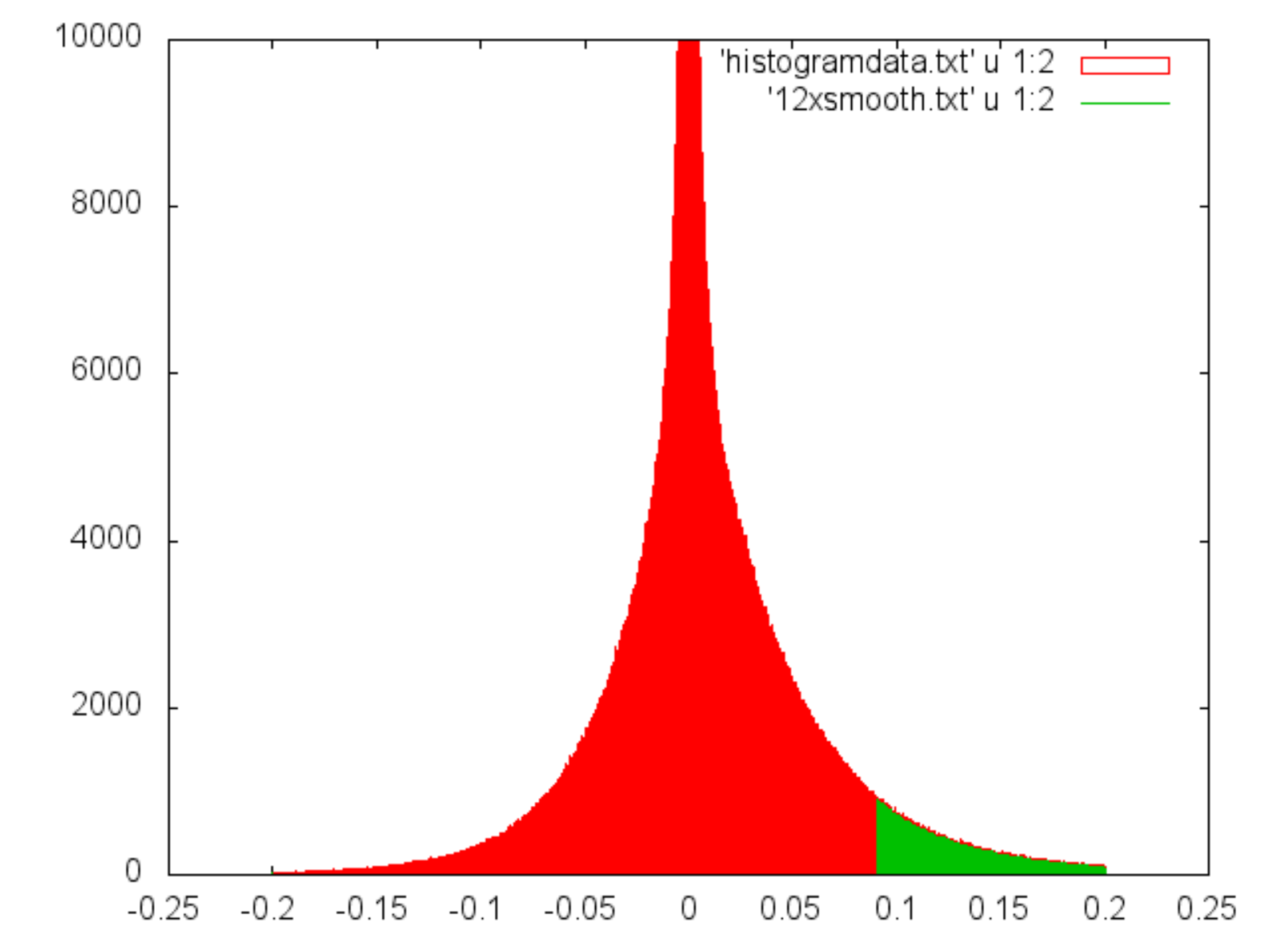}
\includegraphics[width=0.45\textwidth]{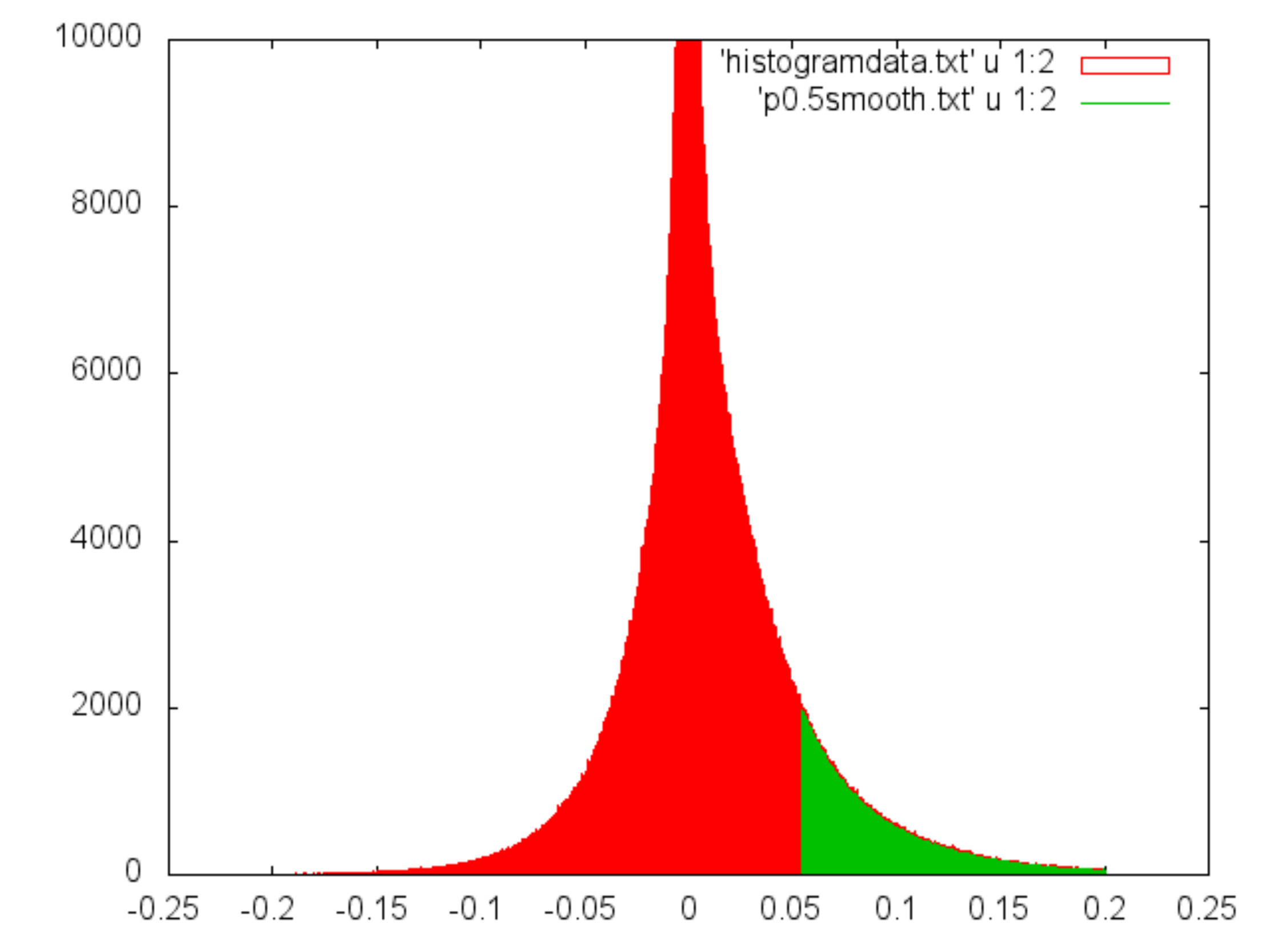}
\includegraphics[width=0.45\textwidth]{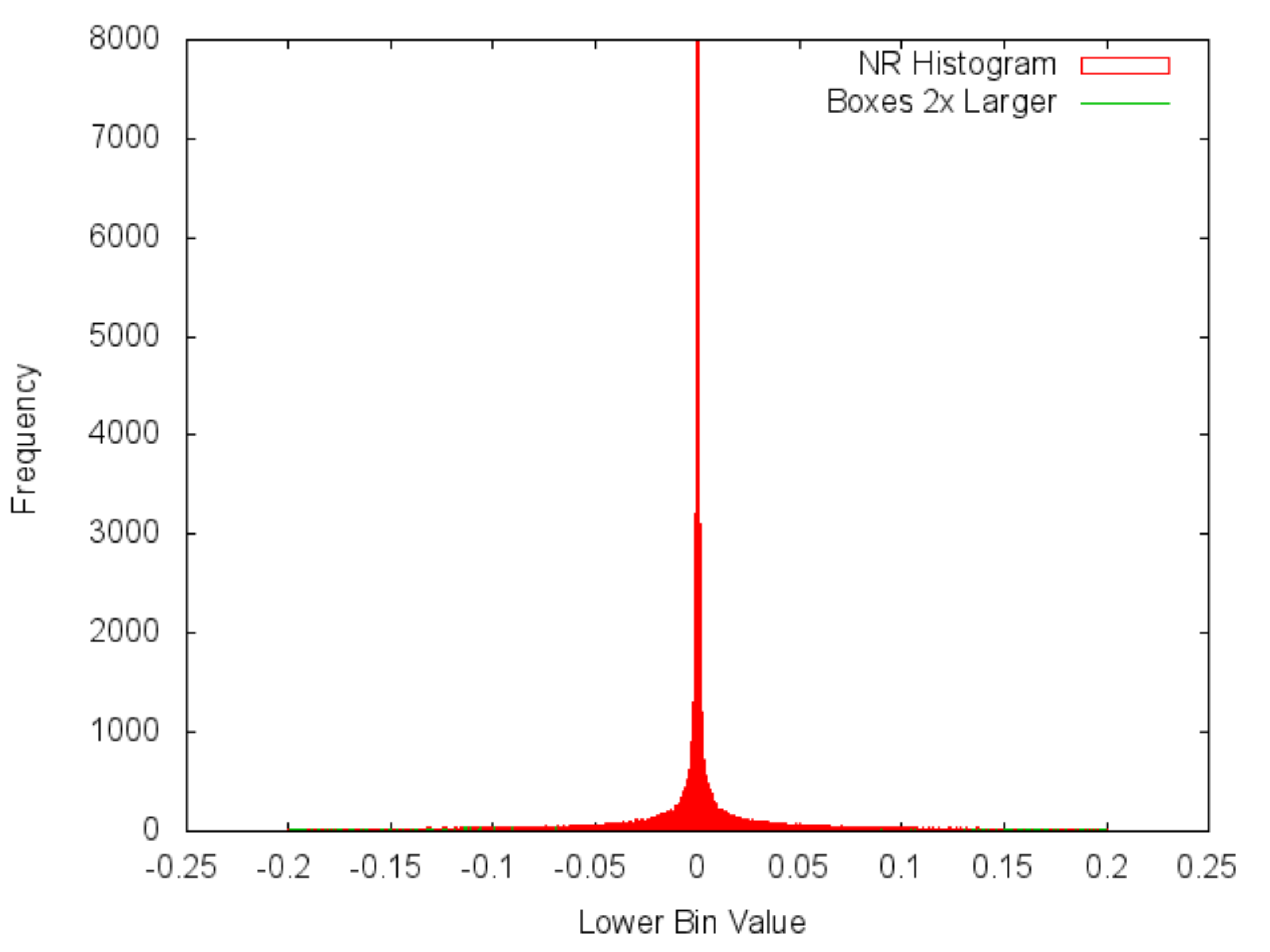}
\includegraphics[width=0.45\textwidth]{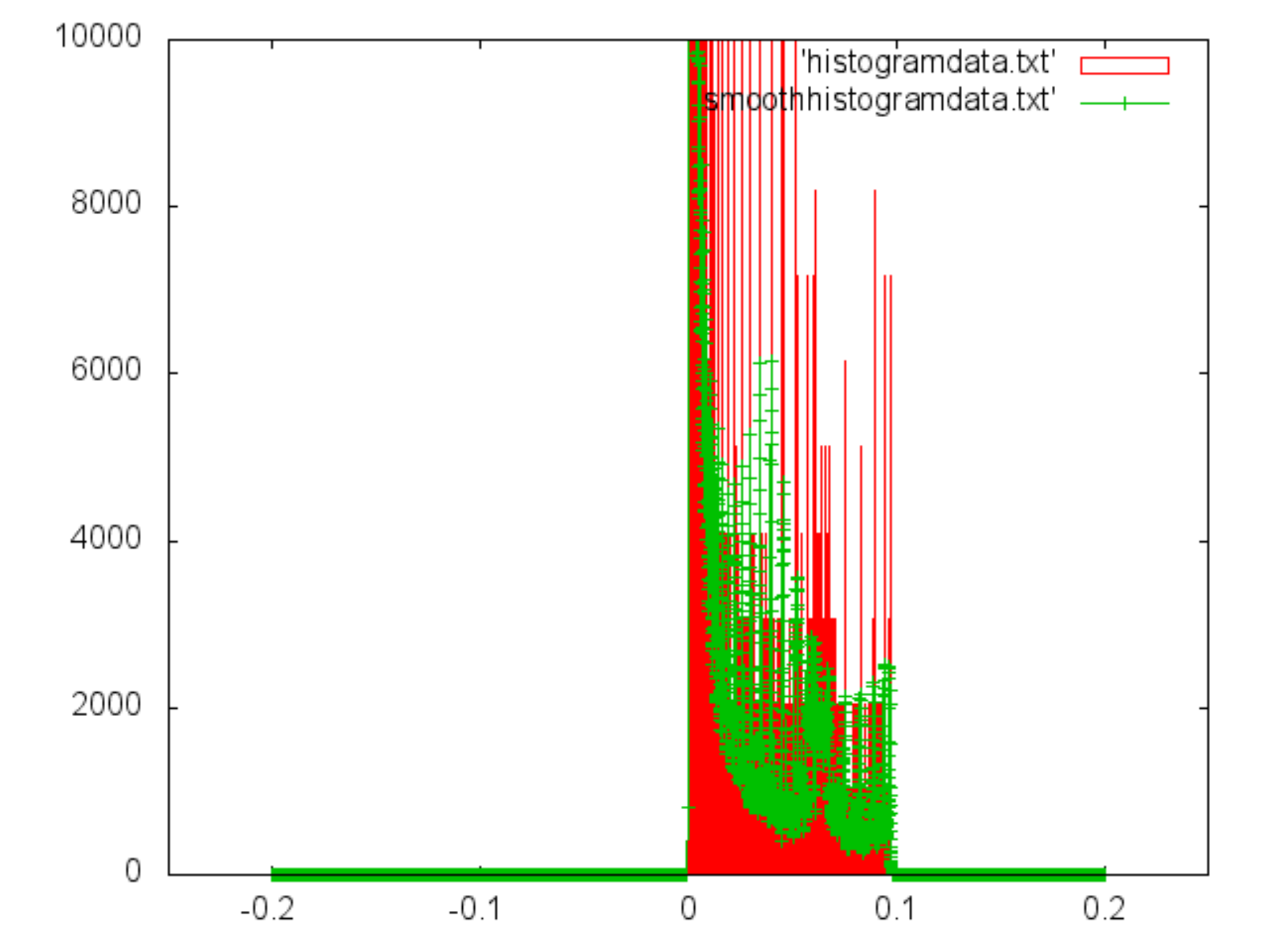}
\caption{These  figures show histograms of the gradient of the reversing
component of the large scale magnetic field in the direction normal to the
unperturbed current sheet, i.e. $\partial_yB_x$.  Upper left panel is for the
highest power simulation, P=1.  Upper right panel is for P=0.5. Lower left is
for P=1 but with no large scale magnetic field reversal, i.e. simply locally
driven strong turbulence.  Bins with twice the number of cells as  the
corresponding bin with the opposite sign of $\partial_yB_x$ are shown in green.
Lower right shows the first simulation in the absence of turbulent forcing.
From \cite{Vishniacetal12}. \label{current}}
\end{figure*}

The structure of the reconnection region was analyzed by Vishniac et al.
\cite{Vishniacetal12} based on the numerical work by Kowal et al.
\cite{Kowaletal09}.  While this paper only examined simulations with relatively
large forcing, the results seem to favour the latter picture, in which the
reconnection region is broad, the magnetic shear is more or less coincident with
the outflow zone, and the turbulence within it is broadly similar to turbulence
in a homogeneous system.  In particular, this analysis showed that peaks in the
current were distributed throughout the reconnection zone, and that
the width of these peaks were not a strong function of their strength.  The
single best illustration of the results is shown in Figure~\ref{current} which
shows histograms of magnetic field gradients in the simulations with strong
and moderate driving power, with no magnetic field reversal but with driven
turbulence, and with no driven turbulence at all, but a passive magnetic field
reversal (i.e. Sweet-Parker reconnection).  A few features stand out in this
figure.  First, all the simulations with driven turbulence have a roughly
gaussian distribution of magnetic field gradients.  In the case with no field
reversal (panel c) the peak is narrow and symmetric around zero.  In the
presence of a large scale field reversal the peak is slightly broadened, and
skewed.  (The simulation without reconnection was run at a lower resolution, so
the total number of cells is smaller by a factor of 8.)  Finally, the last
panel shows a very spiky distribution of points to the right of the origin.  The
spikiness is an artifact of the numerical grid.  In the absence of turbulence
the same values tend to repeat.  That occupied bins are all for positive
magnetic field gradients is a trivial consequence of the background solution and
the laminar nature of Sweet-Parker reconnection.

It is striking that turbulent reconnection does not produce any strong feature
corresponding to a preferred value of the magnetic field gradient.  Instead one
sees a systematic bias towards large positive values.  We conclude from the lack
of coherent features within the outflow zone, and the broad distribution of
values of the gradient of the magnetic field, that the second picture is best.
The current sheet and the outflow zone are roughly coincident and this volume is
filled with turbulent structures.

One weakness of this analysis is that it has been tested only for relatively
strong magnetic turbulence.  Although the driven turbulence in these simulations
was subalfvenic, they were not very weak.  We can expect that the skew in figure
\ref{current} will become stronger at as the turbulent velocities are turned
down.  At some point the mean gradient should begin to affect the turbulent
spectrum.

\subsection{Testing of magnetic Richardson diffusion}

As we discussed, the LV99 model is intrinsically related to the concept of
Richardson diffusion in magnetized fluids. Thus by testing the Richardson
diffusion of magnetic field, one also provides tests for the theory of turbulent
reconnection.

\begin{figure}[!ht]
\centering
\raisebox{-0.5\height}{\includegraphics[width=0.49\textwidth]{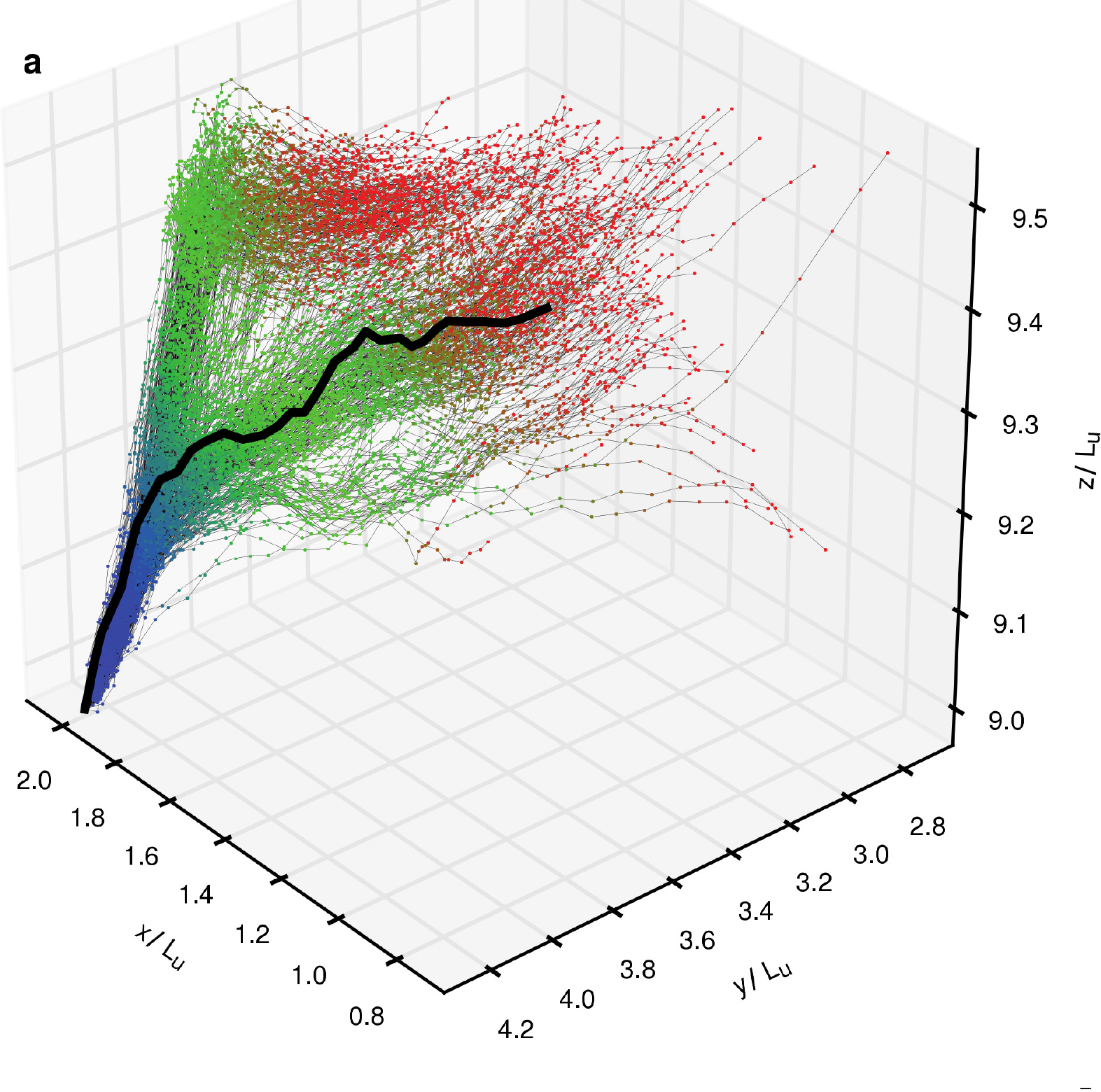}}
\raisebox{-0.5\height}{\includegraphics[width=0.49\textwidth]{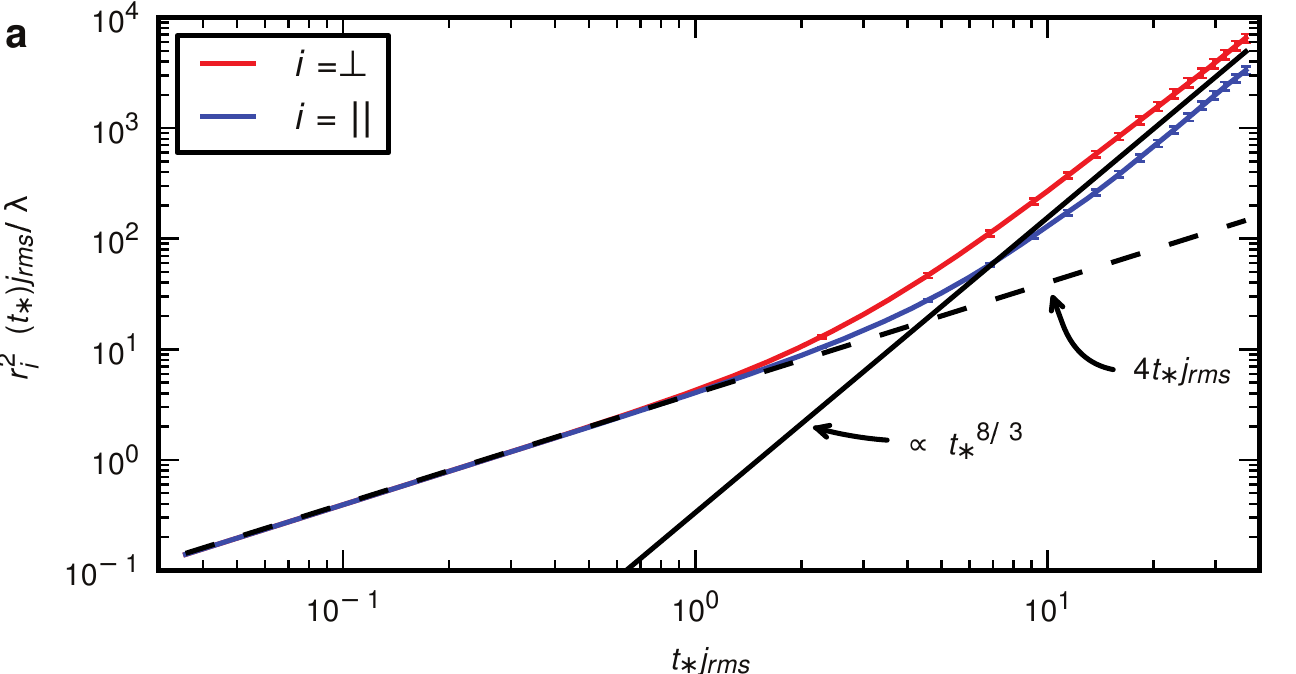}}
\caption{{\it Left panel}. Stochastic trajectories that arrive at a fixed point
in the archived MHD flow, color-coded \textcolor{red}{red},
\textcolor{green}{green}, and \textcolor{blue}{blue} from earlier to later
times.  From \cite{Eyinketal13}.
{\it Right panel.} Mean-square dispersion of field-lines backwards in time, with
\textcolor{red}{red} for direction parallel and \textcolor{blue}{blue} for
direction perpendicular to the local magnetic field.  From \cite{Eyinketal13}.
\label{figure10}}
\end{figure}

The first numerical tests of Richardson diffusion were related to magnetic field
wandering predicted in LV99 \cite{Maronetal04, Lazarianetal04,
Beresnyak13a}.  In Figure~\ref{figure10} we show the results obtained in
\cite{Lazarianetal04}.  There we clearly see different regimes of magnetic field
diffusion, including the $y\sim x^{3/2}$ regime.  This is a manifestation of the
spatial Richardson diffusion.

A direct testing of the temporal Richardson diffusion of magnetic field-lines
was performed recently in \cite{Eyinketal13}.  For this experiment,
stochastic fluid trajectories had to be tracked backward in time from a fixed
point in order to determine which field lines at earlier times would arrive to
that point and be resistively ``glued together''.  Hence, many time frames of an
MHD simulation were stored so that equations for the trajectories could be
integrated backward.  The results of this study are illustrated in
Figure~\ref{figure10}.  The left panel shows the trajectories of the
arriving magnetic field-lines, which are clearly widely dispersed backward in
time, more resembling a spreading plume of smoke than a single ``frozen-in''
line.  Quantitative results are presented in the right panel, which plots the
root-mean-square line dispersion in directions both parallel and perpendicular
to the local mean magnetic field.  Times are in units of the resistive time
$1/j_{rms}$ determined by the rms current value and distances in units of the
resistive length $\lambda/j_{rms}$.  The dashed line shows the standard
diffusive estimate $4\lambda t,$  while the solid line shows the Richardson-type
power-law $t^{8/3}$.  Note that this simulation exhibited a $k^{-3/2}$ energy
spectrum (or H\"older exponent 1/4) for the velocity and magnetic fields,
similar to other MHD simulations at comparable Reynolds numbers, and the
self-consistent Richardson scaling is with exponent 8/3 rather 3.  Although a
$t^{8/3}$ power-law holds both parallel and perpendicular to the local field
direction, the prefactor is greater in the parallel direction, due to
backreaction of the magnetic field on the flow via the Lorentz force.  The
implication of these results is that standard diffusive motion of field-lines
holds for only a very short time, of order of the resistive time, and is then
replaced by super-diffusive, explosive separation by turbulent relative
advection.  This same effect should occur not only in resistive MHD but whenever
there is a long power-law turbulent inertial range.  Whatever plasma mechanism
of line-slippage holds at scales below the ion gyroradius--- electron inertia,
pressure anisotropy, etc.---will be accelerated and effectively replaced by the
ideal MHD effect of Richardson dispersion.

\section{Observational consequences and tests}
\label{sec:obs_cons_tests}

Historically, studies of reconnection were motivated by observations of Solar
flares. There we deal with the collisionless turbulent plasmas and it is
important to establish whether plasma microphysics or LV99 turbulent dynamics
determine the observed solar reconnection.

Qualitatively, one can argue that there is observational evidence in favor
of the LV99 model.  For instance, observations of the thick reconnection current
outflow regions observed in the Solar flares \cite{CiaravellaRaymond08} were
predicted within LV99 model at the time when the competing plasma Hall term
models were predicting X-point localized reconnection.  However, as plasma
models have evolved to include tearing and formation of magnetic islands (see
\cite{Drakeetal10}) it is necessary to get to a quantitative level to compare
the predictions from the competing theories and observations.

To be quantitative one should relate the idealized model LV99 turbulence driving
to the turbulence driving within solar flares. In LV99 the turbulence driving
was assumed isotropic and homogeneous at a distinct length scale $L_{inj}.$ A
general difficulty with observational studies of turbulent reconnection is the
determination of $L_{inj}$.  One possible approach is based on the the relation
$\varepsilon \simeq u_L^4/V_AL_{inj}$ for the weak turbulence energy cascade
rate.  The mean energy dissipation rate $\varepsilon$ is a source of plasma
heating, which can be estimated from observations of electromagnetic radiation
(see more in ELV11). However, when the energy is injected from reconnection
itself, the cascade is strong and anisotropic from the very beginning.  If the
driving velocities are sub-Alfv\'{e}nic, turbulence in such a driving is
undergoing a transition from weak to strong at the scale $L M_A^2$ (see
\S\ref{ssec:richardson_diff}).  The scale of the transition corresponds to the
velocity $M_A^2 V_A$.  If turbulence is driven by magnetic reconnection, one can
expect substantial changes of the magnetic field direction corresponding to
strong turbulence.  Thus it is natural to identify the velocities measured
during the reconnection events with the strong MHD turbulence regime.  In other
words, one can use:
\begin{equation}
V_{rec}\approx U_{obs, turb} (L_{inj}/L_x)^{1/2},
\label{obs}
\end{equation}
where $U_{obs, turb}$ is the spectroscopically measured turbulent velocity
dispersion. Similarly, the thickness of the reconnection layer should be defined
as
\begin{equation}
\Delta\approx L_x (U_{obs, turb}/V_A) (L_{inj}/L_x)^{1/2}.
\label{delta_obs}
\end{equation}

Naturally, this is just a different way of presenting LV99 expressions, but
taking into account that the driving arises from reconnection and therefore
turbulence is strong from the very beginning (see more in \cite{Eyinketal13}.
The expressions given by Eqs.~(\ref{obs}) and (\ref{delta_obs}) can be compared
with observations in (\cite{CiaravellaRaymond08}).  There, the widths of the
reconnection regions were reported in the range from 0.08$L_x$ up to 0.16$L_x$
while the the observed Doppler velocities in the units of $V_A$ were of the
order of 0.1.  It is easy to see that these values are in a good agreement with
the predictions given by Eq.~(\ref{delta_obs}).  We note, that if we associate
the observed velocities with isotropic driving of turbulence, which is
unrealistic for the present situation, then a discrepancy with
Eq.~(\ref{delta_obs}) would appear.  Because of that \cite{CiaravellaRaymond08}
did not get quite as good quantitative agreement between observations and theory
as we did, but still within observational uncertainties. In \cite{Sychetal09},
authors explaining quasi-periodic pulsations in observed flaring energy releases
at an active region above the sunspot, proposed that the wave packets arising
from the sunspots can trigger such pulsations. This is exactly what is expected
within the LV99 model.


As we discussed in \S\ref{ssec:role_plasma_effects} the criterion for the
application of LV99 theory is that the outflow region is much larger than the
ion Larmor radius $\Delta \gg \rho_i$.  This is definitely satisfied for the
solar atmosphere where the ratio of $\Delta$ to $\rho_i$ can be larger than
$10^6$. Plasma effects can play a role for small scale reconnection events
within the layer, since the dissipation length based on Spitzer resistivity is
$\sim 1$ cm, whereas $\rho_i\sim 10^3$ cm (Table~\ref{tab:parameters}). However, as we
discussed earlier, this does not change the overall dynamics of turbulent
reconnection.

Reconnection throughout most of the heliosphere appears similar to that in the
Sun. For example, there are now extensive observations of reconnection jets
(outflows, exhausts) and strong current sheets in the solar wind
\cite{Gosling12}. The most intense current sheets observed in the solar wind are
very often not observed to be associated with strong (Alfv\'enic) outflows and
have widths at most a few tens of the proton inertial length $\delta_i$ or
proton gyroradius $\rho_i$ (whichever is larger). Small-scale current sheets of
this sort that do exhibit observable reconnection have exhausts with widths at
most a few hundreds of ion inertial lengths and frequently have small shear
angles (strong guide fields) \cite{Goslingetal07, GoslingSzabo08}.  Such
small-scale reconnection in the solar wind requires collisionless physics for
its description, but the observations are exactly what would be expected of
small-scale reconnection in MHD turbulence of a collisionless plasma
\cite{Vasquezetal07}. Indeed, LV99 predicted that the small-scale reconnection
in MHD turbulence should be similar to large-scale reconnection, but with nearly
parallel magnetic field lines and with ``outflows'' of the same order as the
local, shear-Alfv\'enic turbulent eddy motions. It is worth emphasizing that
reconnection in the sense of flux-freezing violation and disconnection of plasma
and magnetic fields is required at every point in a turbulent flow, not only
near the most intense current sheets. Otherwise fluid motions would be halted by
the turbulent tangling of frozen-in magnetic field lines. However, except at
sporadic strong current sheets, this ubiquitous small-scale turbulent
reconnection has none of the observable characteristics usually attributed to
reconnection, e.g. exhausts stronger than background velocities, and would be
overlooked in observational studies which focus on such features alone.

However, there is also a prevalence of very large-scale reconnection events in
the solar wind, quite often associated with interplanetary coronal mass
ejections and magnetic clouds or occasionally magnetic disconnection events at
the heliospheric current sheet  \cite{Phanetal09, Gosling12}.
These events have reconnection outflows with widths up to nearly $10^5$ of the
ion inertial length and appear to be in a prolonged, quasi-stationary regime
with reconnection lasting for several hours. Such large-scale reconnection is as
predicted by the LV99 theory when very large flux-structures with
oppositely-directed components of magnetic field impinge upon each other in the
turbulent environment of the solar wind. The ``current sheet'' producing such
large-scale reconnection in the LV99 theory contains itself many ion-scale,
intense current sheets embedded in a diffuse turbulent background of weaker (but
still substantial) current. Observational efforts addressed to
proving/disproving the LV99 theory should note that it is this broad zone of
more diffuse current, not the sporadic strong sheets, which is responsible for
large-scale turbulent reconnection. Note that the study \cite{Eyinketal13}
showed that standard magnetic flux-freezing is violated at general points in
turbulent  MHD, not just at the most intense, sparsely distributed sheets. Thus,
large-scale reconnection in the solar wind is a very promising area for LV99.
The situation for LV99 generally gets better with increasing distance from the
sun, because of the great increase in scales. For example, reconnecting flux
structures in the inner heliosheath could have sizes up to $\sim$100 AU, much
larger than the ion cyclotron radius $\sim10^3$ km \cite{LazarianOpher09}.

The magnetosphere is another example that is under active investigation by the
reconnection community.  The situation there is different, as $\Delta\sim
\rho_i$ is the general rule and we expect plasma effects to be dominant.
Turbulence of whistler waves, e.g. electron MHD (EMHD) turbulence may play its
role, however.  For instance, \cite{Huangetal12} reported a
magnetotail event in which they claim that turbulent electromotive force is
responsible for reconnection.  The turbulence at those scales is not MHD.  We
may speculate that the LV99 can be generalized for the case of EMHD and apply to
such events.  This should be the issue of further studies.

It may be worth noting that the possibility of in-situ measurements of
magnetospheric reconnection make it a very attractive subject for the
reconnection community.  Upcoming missions like the Magnetospheric Multiscale
Mission (MMS), set to launch in 2014, will provide detailed observations of
reconnection diffusion regions, energetic particle acceleration, and
micro-turbulence in the magnetospheric plasma.  In addition to the exciting
prospect of better understanding of the near-Earth space environment, the hope
has been expressed that this mission will provide insight into magnetic
reconnection in a very wide variety of astrophysical and terrestial plasmas.  We
believe that magnetospheric observations may indeed shed light on magnetic
reconnection in man-made settings such as fusion machines (tokamaks or
spheromaks) and laboratory reconnection experiments, which also involve
collisionless plasmas and overall small length scales.  However, magnetospheric
reconnection is a rather special, non-generic case in astrophysics, with
$\Delta$ of the order or less than $\rho_i$, while the larger scales involved in
most astrophysical processes imply that $\Delta\gg \rho_i$.  We claim that this
is the domain where turbulence and the broadening of $\Delta$ that it entails
must be accounted for.  Thus, magnetospheric reconnection, in the opinion of the
present reviewers, is a special case which will provide insight mainly into
micro-scale aspects of reconnection, which are of more limited interest in
general astrophysical environments.  Reconnection elsewhere in the solar system,
including the sun, its atmosphere, and the larger heliosphere (solar wind,
heliosheath, etc.) are better natural laboratories for observational study of
generic astrophysical reconnection in both collisionless and collisional
environments.

\section{Extending LV99 theory}
\label{sec:lv99_extension}

\subsection{Reconnection in partially ionized gas}
\label{ssec:ionized}

Turbulence in the partially ionized gas is different from that in fully ionized
plasmas.  One of the critical differences arises from the viscosity caused by
neutrals atoms.  This results in the media viscosity being substantially larger
than the media resistivity.  The ratio of the former to the latter is called the
Prandtl number and in what follows we consider high Prandtl number turbulence.
In reality, MHD turbulence in the partially ionized gas is more complicated as
decoupling of ions and neutrals and other complicated effects occur at
sufficiently small scale.  The discussion of these regimes is given in
\cite{Lazarianetal04}.  However, for the purposes of reconnection, we believe
that a simplified discussion below is adequate, as follows from the fact that we
discussed earlier, namely, that the LV99 reconnection is determined by the
dynamics of large scales of turbulent motions.

The high Prandtl number turbulence was studied numerically in \cite{Choetal02,
Choetal03, SchekochihinCowley04} and theoretically in \cite{Lazarianetal04}.  What is important for
our present discussion is that for scales larger than the viscous damping scale
the turbulence follows the usual GS95 scaling, while it develops a shallow power
law magnetic tail and steep velocity spectrum below the viscous damping scale
$\ell_{\perp, crit}$.  The existence of the GS95 scaling at sufficiently large
scales means that our considerations about Richardson diffusion and magnetic
reconnection that accompany it should be valid at these scales.  Thus, our goal
is to establish the scale of current sheets starting from where the Richardson
diffusion will induce the accelerated separation of magnetic field lines.

In high Prandtl number media the GS95-type turbulent motions decay at the scale
$l_{\bot, crit}$, which is much larger than the scale at which Ohmic
dissipation becomes important.  Thus over a range of scales less than $l_{\bot,
crit}$ to some much smaller scale magnetic field lines preserve their identity.
These magnetic field lines are being affected by the shear on the scale
$l_{\bot, crit}$, which induces a new regime of turbulence described in
\cite{Choetal02} and \cite{Lazarianetal04}.

To establish the range of scales at which magnetic fields perform Richardson
diffusion one can observe that the transition to the Richardson diffusion is
expected to happen when field lines get separated by the perpendicular scale of
the critically damped eddies $l_{\bot, crit}$.  The separation in the
perpendicular direction starts with the scale $r_{init}$ following the Lyapunov
exponential growth with the distance $l$ measured along the magnetic field
lines, i.e. $r_{init} \exp(l/l_{\|, crit})$, where $l_{\|, crit}$ corresponds to
critically damped eddies with $l_{\perp, crit}$.  It seems natural to associate
$r_{init}$ with the separation of the field lines arising from the action of
Ohmic resistivity on the scale of the critically damped eddies
\begin{equation}
r_{init}^2=\eta l_{\|, crit}/V_A,
\label{int}
\end{equation}
where $\eta$ is the Ohmic resistivity coefficient.

The problem of magnetic line separation in turbulent fluids was considered for
chaotic separation in smooth, laminar flows by Rechester \& Rosenbluth
\cite{RechesterRosenbluth78} and for superdiffusive separation in turbulent
plasmas by Lazarian \cite{Lazarian06}. Following the logic in the paper and
taking into account that the largest shear arises from the critically damped
eddies, it is possible to determine the distance to be covered along magnetic
field lines before the lines separate by the distance larger than the
perpendicular scale of viscously damped eddies is equal to
\begin{equation}
L_{RR}\approx l_{\|, crit} \ln (l_{\bot, crit}/r_{init})
\label{RR}
\end{equation}
Taking into account Eq. (\ref{int}) and that
\begin{equation}
l_{\bot, crit}^2=\nu l_{\|, crit}/V_A,
\end{equation}
where $\nu$ is the viscosity coefficient.  Thus Eq. (\ref{RR}) can be rewritten
\begin{equation}
L_{RR}\approx l_{\|, crit}\ln Pt
\label{RR2}
\end{equation}
where $Pt=\nu/\eta$ is the Prandtl number.

If the current sheets are much longer than $L_{RR}$, then magnetic field lines
undergo Richardson diffusion and according to \cite{Eyinketal11} the
reconnection follows the laws established in LV99.  In other words, on scales
significantly larger than the viscous damping scale LV99 reconnection is
applicable.  At the same time on scales less than $L_{RR}$ magnetic reconnection
may be slow\footnote{Incidentally, this can explain the formation of density
fluctuations on scales of thousands of Astronomical Units, that are observed in
the ISM.}.  This small scale reconnection regime requires further studies.  For
instance, results of laminar reconnection in the partially ionized gas obtained
analytically in \cite{VishniacLazarian99} and studied numerically by
\cite{HeitschZweibel03} can be applicable.  This approach has been recently used
by \cite{Leakeetal12} to explain chromospheric reconnection that takes place in
weakly ionized plasmas.  In this review we, however, are interested at
reconnection at large scales and therefore do not dwell on small scale
phenomena.

For the detailed structure of the reconnection region in the partially ionized
gas the study in \cite{Lazarianetal04} is relevant. There the magnetic
turbulence below the scale of the viscous dissipation is accounted for.
However, those magnetic structures on the small scales cannot change the overall
reconnection velocities.

\subsection{Development of turbulence due to magnetic reconnection}

Astrophysical fluids are generically turbulent.  However, even if the initial
magnetic field configuration is laminar, magnetic reconnection ought to induce
turbulence due to the outflow (LV99, \cite{LazarianVishniac09}).  This effect
was confirmed by observing the development of turbulence both in recent 3D
Particle in Cell (PIC) simulations (\cite{Karimabadietal13}) and 3D MHD
simulations (\cite{Beresnyak13b, Kowaletal13}).

Earlier on, the development of chaotic structures due to tearing was reported in
\cite{Loureiroetal09} as well as in subsequent publications
(see \cite{Bhattacharjeeetal09}).  However, we should stress
that there is a significant difference between turbulence development in 2D and
3D simulations.  As we discussed in \S\ref{ssec:lv99_model} the very nature of
turbulence is different in 2D and 3D.  In addition, the effect of the outflow is
very different in simulations with different dimentionality.  For instance, in
2D the development of the Kelvin-Hemholtz instability is suppressed by the field
that is inevitably directed parallel to the outflow.  On the contrary, the
outflow can induce this instability in the generic 3D configuration.  In
general, we do expect realistic 3D systems to be more unstable and therefore
prone to development of turbulence.  This corresponds well to the results of 3D
simulations that we refer to.

Beresnyak \cite{Beresnyak13b} studied the properties of reconnection-driven
turbulence and found its correspondence to those expected for MHD turbulence
(see \S\ref{ssec:lv99_model}).  The difference with isotropically driven
turbulence is that magnetic energy is observed to be dominant compared with
kinetic energy.  The periodic boundary conditions adopted in
\cite{Beresnyak13b} limits the time span over which magnetic
reconnection can be studied and therefore the simulations focus on the process
of establishing reconnection.  Nevertheless, as the simulations reveal a nice
turbulence power law behavior, one can apply the approach of turbulent
reconnection and closely connected to it, Richardson diffusion
(see \S\ref{ssec:richardson_diff}).

Beresnyak (2013, private communication) used LV99 approach and obtained expressions 
describing the evolution of the reconnection layer in the transient regime of turbulence
development that he observes. Below we provide our theoretical account of the results in \cite{Beresnyak13b} using
our understanding of turbulent reconnection also based on LV99 theory. However, we get expressions which differ from those by 
Beresnyak. 
 
The logic of our derivation is really simple. As the magnetic fluxes get into contact the width of the reconnection layer $\Delta$
is growing.  The rate at which this happens is limited by the mixing rate
induced by the eddies at the scale $\Delta$, i.e.
\begin{equation}
\frac{1}{\Delta} \frac{d \Delta}{dt}\approx g \frac{V_{\Delta}}{\Delta}
\label{Delta_start}
\end{equation}
with a factor $g$ which takes into account possible inefficiency in the
diffusion process. As $V_{\Delta}$ is a part of the turbulent cascade, i.e. the
mean value of $V_{\Delta}^2\approx \int \Phi(k_1) dk_1$, where
\begin{equation}
\Phi=C_k \epsilon^{2/3}k^{-5/3}_1,
\label{Phi1}
\end{equation}
and $C_k$ is a Kolmogorov constant, which for ordinary MHD turbulence is
calculated in \cite{Beresnyak12}, but in our special case may be different.  If
the energy dissipation rate $\varepsilon$ were time-independent, then the layer
width would be implied by Eqs.~(\ref{Delta_start}) and (\ref{Phi1}) to grow according to
Richardson's law $\Delta^2 \sim \varepsilon t^3.$ However, in the transient
regime considered, energy dissipation rate is evolving. If the y-component of
the magnetic field is reconnecting and the cascade is strong, then the mean
value of the dissipation rate $\epsilon$ is
\begin{equation}
\epsilon\approx \beta V_{Ay}^2/(\Delta/V_{\Delta}),
\label{eps}
\end{equation}
where $\beta$ is another coefficient measuring the efficiency of conversion of
mean magnetic energy into turbulent fluctuations. This coefficient can be obtained from
numerical simulations.

The ability of the cascade to be strong from the very beginning follows from the large
perturbations of the magnetic fields by magnetic reconnection, while magnetic energy
can still dominate the kinetic energy. The latter factor that can be experimentally measured
is given by a parameter $r_A$. 
With this factor and making use of Eqs.(\ref{Phi1})
and (\ref{eps}),  the expression for
$V_{\Delta}$ can be rewritten in the following way:
\begin{equation}
V_\Delta\approx C_k r_A (V_{Ay}^2 V_\Delta \beta)^{2/3}
\label{V_Delta}
\end{equation}
where the dependences on $k_1\sim 1/\Delta$ cancel out.

This provides the expression for the turbulent velocity at the injection scale $V_\Delta$
\begin{equation}
V_{\Delta}\approx (C_K r_A)^{3/4} V_{Ay} \beta^{1/2}
\label{V_Delta_final}
\end{equation}
as a function of the experimentally measurable parameters of the system. Thus
the growth of the turbulent reconnection zone is according to
Eq.(\ref{Delta_start})
\begin{equation}
\frac{d\Delta}{dt}\approx g \beta^{1/2} (C_K r_A)^{3/4} V_{Ay}
\label{growth}
\end{equation}
which predicts the nearly constant growth of the outflow region as seen in Fig.3
in \cite{Beresnyak13b}.

Using the values of the numerical constants provided to us by Beresnayk we get a fair correspondence with 
the results of simulations in \cite{Beresnyak13b}.  However, we feel that further testings are necessary.

\begin{figure}[!ht]
\centering
\includegraphics[width=\textwidth]{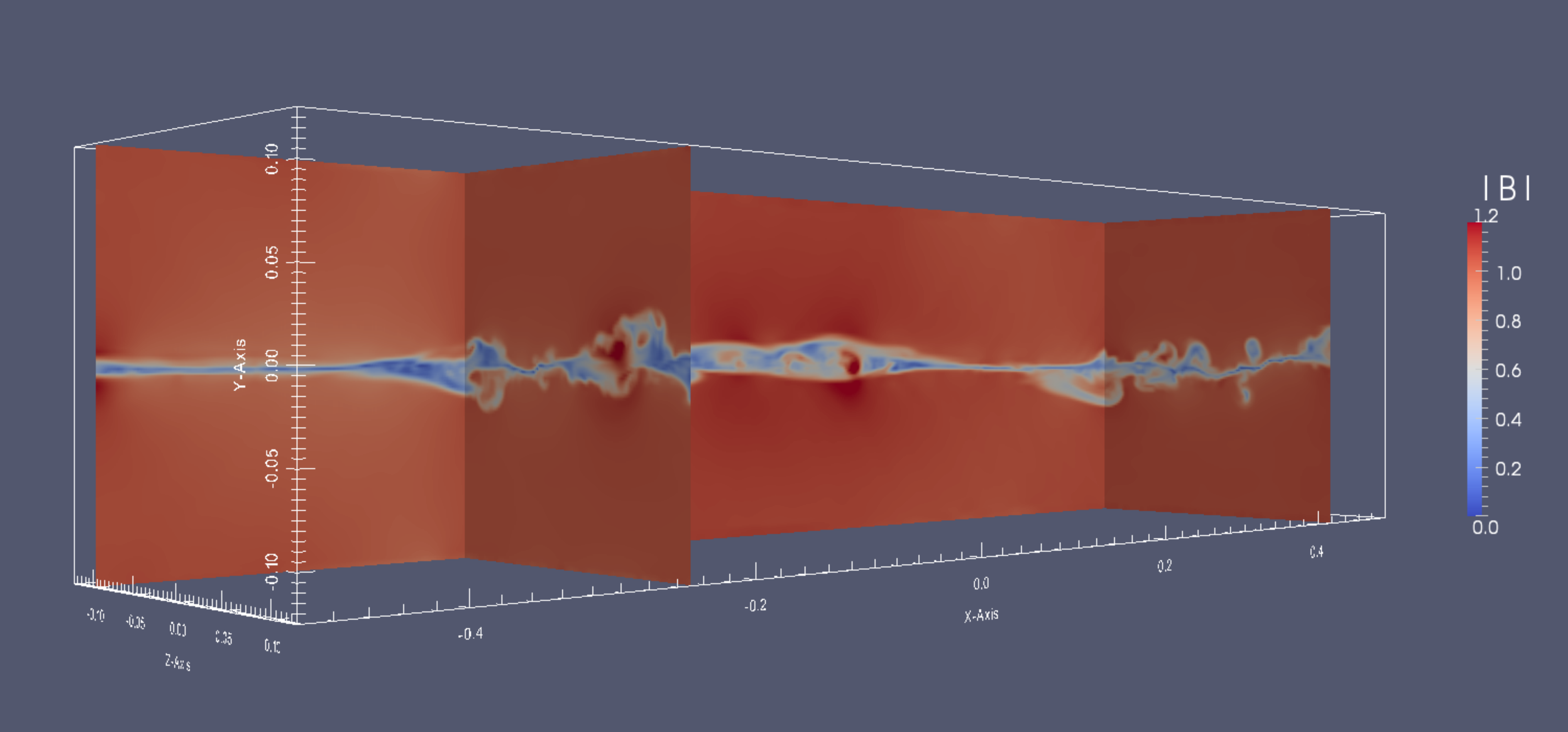}
\caption{Visualization of the model of turbulence generated by the seed
reconnection from \cite{Kowaletal13}. Three different cuts (one XY plane at
Z=-0.1 and two YZ-planes at X=-0.25 and X=0.42) through the computational domain
show the strength of magnetic field $|\vec{B}|$ at the evolution time $t=1.0$.
Kelvin-Helmholtz-type structures are well seen in the planes perpendicular to
the reconnecting magnetic component $B_x$.  In the Z direction, the
Kelvin-Helmholtz instability is slightly suppressed by the guide field of the
strength $B_z=0.1 B_x$ (with $B_x=1.0$ initially). The initial current sheet is
located along the XZ plane at Y=0.0. A weak ($E_{kin} \approx 10^{-4} E_{mag}$)
random velocity field was imposed initialy in order to seed the reconnection.
\label{fig:bmag_cuts}}
\end{figure}

As the reconnection gets into the steady state regime, one expects the outflow to play
an important role and therefore the reconnection rate gets modified. This regime cannot be studied in periodic box
simulations like those in \cite{Beresnyak13b} as they require studies for more than one Alfven crossing time. 
Studies with open boundary conditions are illustrated by Figure \ref{fig:bmag_cuts} from our new study.

The equations
for the reconnection rate were obtained in LV99 for the isotropic injection of energy.
For the case of anisotropic energy injection of turbulence we should apply the approach in
\S~\ref{sec:obs_cons_tests}. Using Eq. (\ref{delta_obs}) and identifying $V_\Delta$ with the total 
velocity dispersion, which is similar to the use 
of $U_{obs, turb}$ in Eq. (\ref{obs}) one can get
\begin{equation}
V_{rec}\approx V_\Delta (\Delta/L_x)^{1/2}
\label{prel}
\end{equation}
where the mass conservation condition provides the relation $V_{rec} L_x\approx V_{Ay} \Delta$.
Using the latter condition one gets
\begin{equation}
V_{rec} \approx V_{Ay} (C_K r_A)^{3/2} \beta
\label{rec_self}
\end{equation}
which somewhat slower than the rate at which the reconnection layer was growing
initially.
The latter behavior of reconnection is also present for the Sweet-Parker
reconnection, since the reconnection rate can be faster even before the
formation of steady state current sheet (see \cite{Kowaletal09}).

We are going to compare the prediction given by Eq. (\ref{ref_self}) against the results of
 recent simulations illustrated by
Figure~\ref{fig:bmag_cuts}.  The figure shows a few slices of the magnetic field
strength $|\vec{B}|$ through the three-dimensional computational domain with
dimensions $L_x=1.0$ and $L_y=L_z=0.25$.  The simulation was done with the
resolution $2048 \times 512 \times 512$.  Open boundary conditions along the X
and Y directions allowed studies of steady state turbulence.  At the presented
time $t=1.0$ the turbulence strength increased by two orders of magnitude from
its initial value of $E_{kin} \approx 10^{-4} E_{mag}$.  Initially, only the
seed velocity field at the smallest scales was imposed (a random velocity vector
was set for each cell).  We expect that most of the injected energy comes from
the Kelvin-Helmholtz instability induced by the local interactions between the
reconnection events, which dominates in the Z-direction, along which a weak
guide field is imposed ($B_z=0.1 B_x$).  As seen in the planes perpendicular to
$B_x$ in Figure~\ref{fig:bmag_cuts}, Kelvin-Helmholtz-like structures are
already well developed at time $t=1.0$. Turbulent structures are also observed
within the XY-plane, which probably are generated by the strong interations of
the ejected plasma from the neighboring reconnection events.  More detailed
analysis of the spectra of turbulence and efficiency of the Kelvin-Helmholtz
instability as the turbulent injection mechanism are presented in
\cite{Kowaletal13}.

\subsection{Effect of energy dissipation in the reconnection layer}

In the original LV99 paper it was argued that only a small fraction of the
energy stored in the magnetic field is lost during large-scale reconnection and
the magnetic energy is instead converted nearly losslessly to kinetic energy of
the outflow. This can only be true, however, when the Alfv\'enic Mach number
$M_A=u_L/V_A$ is small enough. If $M_A$ becomes too large, then it was argued in
ELV11 that energy dissipation in the reconnection layer becomes non-negligible
compared to the available magnetic energy and there is a consequent reduction of
the outflow velocity. Note that even if $M_A$ is initially small, reconnection
may drive stronger turbulence (see previous subsection) and increase the
fluctuation velocities $u_L$ in the reconnection layer. This scenario may be
relevant to post-CME reconnection, for example, where there is empirical
evidence that the energy required to heat the plasma in the reconnection layer
(``current sheet'') to the observed high temperatures is from energy cascade due
to turbulence generated by the reconnection itself \cite{Susinoetal13}. In
addition, $V_A$ within the reconnection layer will be smaller than the upstream
values, because of annihilation of the anti-parallel components, which will
further increase the Alfv\'enic Mach number. If $M_A$ rises to a sufficiently large
value, then the energy dissipated becomes large enough to cause a reduction in
the outflow velocity $v_{out}$ below the value $V_A$ assumed in LV99 and the
predictions of the theory must be modified. We consider here briefly the
modification proposed in ELV11 and some of its consequences.

The effect of turbulent dissipation can be estimated from steady-state energy
balance in the reconnection layer:
\begin{equation}
\frac{1}{2}v_{out}^3 \Delta = \frac{1}{2}V_A^2 v_{ren} L_x - \varepsilon L_x \Delta,  \label{Ebal}
\end{equation}
where kinetic energy carried away in the outflow is balanced against magnetic
energy transported into the layer minus the energy dissipated by turbulence.
Here we estimate the turbulent dissipation using the formula $\varepsilon =
u_L^4/V_A L_i$ for sub-Alfv\'enic turbulence \cite{Kraichnan65}. If one divides
(\ref{Ebal}) by $\Delta=L_x v_{rec}/v_{out}$, one gets
\begin{equation}
                  v_{out}^3 = V_A^2 v_{out} - 2 \frac{u_L^4}{V_A} \frac{L_x}{L_i},
\end{equation}
which is a cubic polynomial for $v_{out}$. The solutions are easiest to obtain
by introducing the ratios $f=v_{out}/V_A$ and $r= 2 M_A^4 (L_x/L_i)$ which
measure, respectively, the outflow speed as a fraction of $V_A$ and the energy
dissipated by turbulence in units of the available magnetic energy, giving
\begin{equation}
    r = f - f^3.
\label{cubic}
\end{equation}
When $r=0$, the only solution of (\ref{cubic}) with $f>0$ is $f=1,$ recovering
the LV99 estimate $v_{out}=V_A$ for $M_A\ll 1.$ For somewhat larger values of
$r,$ $f\simeq 1-(r/2)$, in agreement with the formula $f=(1-r)^{1/2}$ that
follows from eq.(65) in ELV11, implying a slight decrease in $v_{out}$ compared
with $V_A.$ Note that formula (\ref{cubic}) cannot be used to determine $f$ for
too large $r$, because it has then no positive, real solutions! This is easiest
to see by considering the graph of $r$ vs. $f$. The largest value of $r$ for
which a positive, real $f$ exists is $r_{max} = 2/\sqrt{27}\approx 0.385$ and
then $f$ takes on its minimum value $f_{min} = 1/\sqrt{3}\approx 0.577$. This
implies that the LV99 approach is limited to $M_A$ sufficiently small, because
of the energy dissipation inside the reconnection layer and the consequent
reduction of the outflow velocity.  This is not a very stringent limitation,
however, because $r$ is proportional to $M_A^4$. If one assumes $L_x \simeq
L_i$, one may consider values of $M_A$ up to $0.662$. Given the neglect of
constants of order unity in the above estimate, we may say only that the LV99
approach is limited to $M_A\lesssim 1.$ At the extreme limit of applicability of
LV99, $v_{out}$ is still a sizable fraction of $V_A$, i.e. 0.577, not a
drastically smaller value.

The effect of the reduced outflow velocity may be, somewhat paradoxically, to
{\it increase} the reconnection rate. The reason is that field-lines now spend a
time $L_x/v_{out}$ exiting from the reconnection layer, greater than assumed in
LV99 by a factor of $1/f.$ This implies a thicker reconnection layer $\Delta$
due to the longer time-interval of Richardson diffusion in the layer, greater
than LV99 by a factor of $(1/f)^{3/2}.$ The net reconnection speed
$v_{rec}=v_{out}\Delta/L_x$ is thus larger by a factor of $(1/f)^{1/2}.$ The
increased width $\Delta$ more than offsets the reduced outflow velocity
$v_{out}.$ However, this effect can give only a very slight increase, at most by
a factor of $3^{1/4}\simeq 1.31$ for $f_{min}=1/\sqrt{3}.$ We see that for the
entire regime $M_A\lesssim 1$ where LV99 theory is applicable, energy
dissipation in the reconnection layer implies only very modest corrections. It
is worth emphasizing that ``large-scale reconnection'' in super-Alfv\'enic
turbulence with $M_A>1$ is a very different phenomenon, because magnetic fields
are then so weak that they are easily bent and twisted by the turbulence. Any
large-scale flux tubes initially present will be diffused by the turbulence
through a process much different than that considered by LV99. For a discussion
of this regime, see \cite{KimDiamond01}.

\subsection{Relativistic reconnection}

Magnetic turbulence in a number of astrophysical highly magnetized objects,
accretion disks near black holes, pulsars, gamma ray bursts may be in the
relativistic regime when the Alfv\'{e}n velocity approaches that of light.  The
equations that govern magnetized fluid in this case look very different from the
ordinary MHD equations.  However, studies by \cite{Cho05} and
\cite{ChoLazarian13} show that for both balanced and
imbalanced turbulence, the turbulence spectrum and turbulence anisotropies are
quite similar in this regime and the non-relativistic one.  This suggests that
the Richardson diffusion and related processes of LV99-type magnetic
reconnection should cary on to the relativistic case (see Lazarian \& Yan 2012). This prediction was confirmed by the recent numerical simulations Makoto Takomoto (2014, private communication) who with his relativistic code adopted the approach in Kowal et al (2009) and showed that the rate of 3D relativistic magnetic reconnection gets independent of resistivity.  

The suggestion that LV99 is applicable to relativistic reconnection motivated the use of the model for explaining gamma ray
bursts in \cite{Lazarianetal03} and \cite{ZhangYan11} studies (see also
\S\ref{ssec:flares_bursts}) and in accretion disks around black holes and
pulsars studies \cite{deGouveiadalPinoLazarian05, Giannios13}. Now, as the
extension of the model to relativistic case has be confirmed these and other cases where the relativistic analog of LV99 process was discussed to be applicable (see Lyutikov \& Lazarian 2013) are given numerical support.   

 Naturally, more detailed
studies of both relativistic MHD turbulence and relativistic magnetic
reconnection are required (see also chapter by de Gouveia Dal Pino and Kowal in
this volume and references therein). It is evident that in
magnetically-dominated, low-viscous plasmas turbulence is a generic
ingredient and thus it must be taken into account for relativistic magnetic
reconnection. As we discuss elsewhere in the review the driving of turbulence may by external forcing or it can be driven by reconnection itself.

\section{Implications of fast reconnection in turbulent fluids}
\label{sec:implications}

\subsection{Flux freezing in astrophysical fluids}
\label{ssec:flux_freez}

Since the concept was first proposed by Hannes Alfv\'{e}n in 1942, the principle
of frozen-in field lines has provided a powerful heuristic which allows simple,
back-of-the-envelope estimates in place of full solutions (analytical or
numerical) of the MHD equations (\cite{Parker79}, \cite{Kulsrud05}).  As such,
the flux-freezing principle has been applied to gain insight into diverse
processes, such as star formation, stellar collapse, magnetic dynamo, solar wind
magnetospheric interactions, etc. However, it has long been understood that
magnetic flux-conservation, if strictly valid, would forbid magnetic
reconnection, because field-lines frozen into a continuous plasma flow cannot
change their topology.  Thus, the flux-freezing principle is in apparent
contradiction with numerous observations of fast reconnection in
high-conductivity plasmas.

Quite apart from these serious empirical difficulties, the flux-freezing
principle has recently been shaken by a fundamental theoretical problem.
Standard mathematical proofs of flux-freezing in MHD always assume, implicitly,
that velocity and magnetic fields remain smooth as $\eta \rightarrow 0$.
However, MHD solutions with small resistivities and viscosities (high magnetic
and kinetic Reynolds numbers) are generally turbulent.  These solutions exhibit
long ranges of power-law spectra corresponding to very non-smooth or ``rough''
magnetic and velocity fields.  Fluid particle (Lagrangian) trajectories in such
rough flows are known to be non-unique and stochastic (see \cite{Bernardetal98,
GawedzkiVergassola00, EEijnden00a, EEijnden00b, EEijnden01,
Chavesetal03}, and, for reviews, \cite{Kupiainen03} and \cite{Gawedzki08}).  In
fact, it is possible to show that, in the limit of infinite Reynolds number,
there is an infinite random ensemble of particle motions for the same initial
conditions!  This remarkable phenomenon has been called {\it spontaneous
stochasticity}.  It view of the above, it is immediately clear as a consequence
that standard flux-freezing cannot hold in turbulent plasma flows.  After all,
the usual idea is that magnetic field-lines at high conductivity are tied to the
plasma flow and follow the fluid motion. However, if the latter is non-unique
and stochastic, then which fluid element will the field-line follow?

The phenomenon of spontaneous stochasticity in magnetic field was shown to be
inseparably related to LV99 reconnection theory in ELV11.  It provides, however,
a new outlook on the problem of magnetic field in turbulent fluids.  The notion
of the violation of the flux conservation Alfv\'{e}n theorem is implicit in LV99
(but it is expressed explicitly in \cite{VishniacLazarian99}). At the moment we
can definitively assert that the domain of the Alfv\'{e}n theorem on flux
freezing is limited to laminar fluids only.

In view of the longstanding misconceptions about the general validity of
magnetic flux-conservation for high-conductivity MHD, it is worth making a few
more detailed remarks.  The standard textbook proofs of flux-conservation (e.g.
\cite{Chandrasekhar61}) all make implicit assumptions that are
violated in turbulent flow.  The proofs typically start with the ideal induction
equation $$ \partial_t {\bf B} = \nabla\times ({\bf u}\times {\bf B}) $$ and
consider a material surface $S(t)$ advected by velocity ${\bf u}$.  Then the
time-derivative of the flux integral becomes $$ \frac{d}{dt}\int_{S(t)} {\bf
B}(t)\cdot d{\bf A} = \int_{S(t)} \partial_t{\bf B}(t)\cdot d{\bf A} +
\int_{C(t)} {\bf B}(t)\cdot ({\bf u}\times d{\bf x}). $$  The first term from
the evolution of ${\bf B}$ and the second term from the motion of the surface
cancel identically, implying constant flux through the surface. Of course, in
reality, there is always a finite conductivity $\sigma$, however large, and the
induction equation is $$ \partial_t {\bf B} = \nabla\times ({\bf u}\times {\bf
B}) + \lambda \triangle{\bf B},$$ with $\lambda=c^2/4\pi\sigma$.  The last term
represents a diffusion of magnetic field lines in or out of the surface element,
so that flux is no longer exactly conserved.

For a laminar velocity field, this diffusion effect is small.  It is not hard to
see that a pair of field lines will attain a displacement ${\bf r}(t)$ apart
under the combined effect of advection and diffusion obeying $$
\frac{d}{dt}\langle r^2\rangle = 12\lambda + 2\langle {\bf r}\cdot \delta {\bf
u}({\bf r})\rangle $$ where $\delta {\bf u}({\bf r})$ is the relative advection
velocity at separation ${\bf r}$.  Thus, $$ \frac{d}{dt}\langle r^2\rangle \leq
12\lambda + 2\|\nabla {\bf u}\|\langle r^2\rangle, $$ where $\|\nabla {\bf u}\|$
is the maximum value of the velocity-gradient $\nabla {\bf u}$.  It follows that
two lines initially at the same point, by time $t$ can have separated at most
\begin{equation}
\langle r^2(t)\rangle \leq 6\lambda   \frac{e^{2\|\nabla{\bf u}\| t}-1}{\|\nabla{\bf u}\|}. \label{star}
\end{equation}
If we thus consider a smooth laminar flow with a fixed, finite value of
$\|\nabla{\bf u}\|$, then $\langle r^2(t)\rangle\rightarrow 0$ as
$\lambda\rightarrow 0$.  Under such an assumption, magnetic field lines do not
diffuse a far distance away from the solution of the deterministic ordinary differential equation $d{\bf
x}/dt={\bf u}({\bf x},t)$, and the magnetic line-diffusion becomes a negligible
effect.  In that case, magnetic flux is conserved better as $\lambda$ decreases.

However, in a turbulent flow, the above argument fails!  The inequality
(\ref{star}) still holds, of course, but it no longer restricts the dispersion
of field-lines under the joint action of resistivity and advection.  As is
well-known, a longer and longer inertial range of power-law spectrum $E(k)$
occurs as viscosity $\nu$ decreases and the maximum velocity gradient
$\|\nabla{\bf u}\|$ becomes larger and larger.  In fact, energy dissipation
$\varepsilon=\nu\|\nabla{\bf u}\|^2$ is observed to be non-vanishing as
$\nu\rightarrow 0$ in turbulent flow, requiring velocity gradients to grow
unboundedly.  Estimating $\|\nabla {\bf u}\|\sim (\varepsilon/\nu)^{1/2}$, the
upper bound (\ref{star}) becomes
\begin{equation}
\langle r^2(t)\rangle \leq 6\lambda (\nu/\varepsilon)^{1/2} [ \exp(2t (\varepsilon/\nu)^{1/2} ) - 1].
\label{star2}
\end{equation}
This bound allows unlimited diffusion of field-lines. Consider first the case
$\lambda=\nu$ or $Pt=1$, for simplicity, where Richardson's theory implies that
\begin{equation}
\langle r^2(t)\rangle  \sim 12\lambda t +\varepsilon t^3.
\end{equation}
The rigorous
upper bound always lies strictly above Richardson's prediction and, in fact,
goes to infinity as $\nu=\lambda\rightarrow 0$!  The case of large Prandtl
number is just slightly more complicated, as previously discussed in \S\ref{ssec:ionized}.  When $Pt\gg 1,$ the inequality (\ref{star2}) holds as
an equality for times $t\ll t_{trans}$ with
\begin{equation}
 t_{trans}= \frac{\ln(Pt)}{2(\varepsilon/\nu)}.
 \end{equation}
   This is then followed by a Richardson
diffusion regime
\begin{equation}
\langle r^2(t)\rangle  \sim 6(\nu^3/\varepsilon)^{1/2} +
\varepsilon (t - t_{trans})^3, \,\,\,\,\,\,\,\,\,\,\,\,\,\,\,\, t\gg
t_{trans},
\end{equation}
assuming that the kinetic Reynolds number is also large and a
Kolmogorov inertial range exists at scales greater than the Kolmogorov length
$(\nu^3/\varepsilon)^{1/4}.$  Once again, the upper bound (\ref{star2}) is much
larger than Richardson's prediction and, at times longer than $t_{trans},$ the
dispersion of field lines is independent of resistivity.

The textbook proofs of magnetic flux-freezing for ideal MHD are therefore based
on unstated assumptions which are explicitly violated in turbulent flows.  They
are mathematically valid derivations with appropriate assumptions, but
physically inapplicable in typical astrophysical systems with plasma turbulence
at MHD scales.  It is worth emphasizing that any attempt to obtain fast
reconnection (independent of resistivity) within a similar hydromagnetic
description must likewise account for flux-freezing violation.  For example, it
has been conjectured \cite{Mandtetal94,ShayDrake98} that reconnection rates are
independent of resistivity in Hall MHD X-point reconnection.  This proposal
faces the same {\it a priori} theoretical difficulty as MHD-based theories,
since magnetic field-lines remain frozen-in to the electron fluid in ideal Hall
MHD.  The conjectured failure of flux-freezing in Hall MHD X-point reconnection
even as $\lambda\rightarrow 0$ must therefore be explained.  Analytical studies
of Hall reconnection indicate that the mechanism may be mathematically similar
to the turbulent LV99 case, in that gradients of the electron fluid velocity
${\bf u}^e$ in the direction of the outgoing reconnection jets are predicted to
diverge proportional to $S,$ the Lundquist number
\cite{Malyshkin08,Shivamoggi11}.

The Hall effects discussed above, as well as other microscopic plasma effects,
are not expected to modify the Richardson diffusion of magnetic field lines at
length scales much greater than the ion Larmor radius (see Appendix B of ELV11
and section 3.5 of this review). However, one may worry that additional {\it
hydrodynamic} effects at large scales may fundamentally alter Richardson
diffusion. For example, in the Kraichnan-Kazantsev turbulence model \cite{Kraichnan65}, where
``spontaneous stochasticity'' was first predicted, it was shown that a
sufficiently high degree of compressibility may eliminate Richardson dispersion
entirely and replace it with instead a coalescence of fluid particles
\cite{GawedzkiVergassola00,EvandenEijnden01}. If such effects were found in
compressible MHD turbulence, then they could strongly alter the quantitative
predictions of LV99, at the very least. This is a particular source of concern
because most astrophysical plasmas are compressible, with Mach numbers ranging
from a bit less than unity (subsonic) to very large (hypersonic). Note that the
numerical tests of Richardson dispersion reported in section 4.6 were for
incompressible MHD turbulence. Could compressible MHD turbulence be
fundamentally different?

There is at this time no complete theory of Richardson dispersion for MHD
turbulence (or, for that matter, for hydrodynamic turbulence), but there are
several reasons to believe that compressibility effects will be minimal on the
turbulent motion of field lines relevant to reconnection. First, very high
degrees of compressibility are required in the Kraichnan model \cite{Kraichnan65} to eliminate
spontaneous stochasticity. In 3D the kinetic energy in the potential part of the
flow must be 10 times greater than in the solenoidal part! Such extreme
compressibility is rare in astrophysics. Of course, the Kraichnan model velocity
is Gaussian and contains no compressible coherent structures like shocks, which
may magnify the compressibility effects. It is well-known that the simple
Burgers model, which is entirely potential flow, exhibits no spontaneous
stochasticity but instead coalescence of particles in shocks
\cite{BauerBernard99}.  However, Burgers differs in another crucial respect from
the Kraichnan model in that it is time-irreversible. As discussed in
\cite{Eyink11} and ELV11, it is the Richardson dispersion of magnetic field
lines {\it backward in time} which is relevant to breakdown of flux-freezing. As
shown in \cite{Eyinketal13}, the Burgers model does exhibit spontaneous
stochasticity backward in time and field lines will thus not be ``frozen-in''
for vanishing resistivities. This is completely unlike the Kraichnan model for
pure potential flow in which fluid particles coalesce backward in time as well
as forward. In the Burgers model, therefore, magnetic field lines which arrive
together at the shock become glued together to produce a resultant magnetic
field at the shock. This is the same thing that happens at each point in
incompressible MHD turbulence! Our second argument is thus that micro-scale
shocklets in compressible MHD turbulence will probably glue field lines together
in a manner almost indistinguishable from the surrounding ``sea'' of rough
turbulence with continuous velocities. Finally, we note that the compressible
MHD wave modes (slow and fast magnetosonic waves) are found in numerical
simulations to decouple dynamically from the solenoidal shear-Alfv\'en modes,
which exhibit turbulence characteristics very similar to those of incompressible
MHD \cite{ChoLazarian02, ChoLazarian03}. Since shear-Alfv\'en waves produce the
dominant fluid motions normal to the direction of the mean magnetic field, they
will be the principal drivers of magnetic field-line diffusion across a
turbulent reconnection layer. While more research into compressible MHD
turbulence is desirable, the above facts support the view that compressibility
effects will not strongly alter turbulent magnetic reconnection.

\subsection{Solar flares and gamma ray bursts}
\label{ssec:flares_bursts}

Preexisting turbulence is a rule for astrophysical systems.  However, for
sufficiently low $M_A$ the LV99 reconnection rates may be quite small.  Would
this mean that the reconnection will stay slow?  LV99 model predicts {\it
reconnection instability} that can drive reconnection in a bursty fashion in low
$\beta$ plasmas.  If initially $M_A$ is very small, the magnetic field wandering
is small and therefore the reconnection is going to proceed at a slow pace.  However, the system of two highly
magnetized flux tubes being in contact is unstable to the development of
turbulence arising from magnetic reconnection.  Indeed, if the outflow gets
turbulent, turbulence should, first of all, increase the magnetic field
wandering thus increasing the width of the outflow $\Delta$.  Second, the
increase of $\Delta$ increases the energy injection in the system via the
increase of $V_{rec}$.  Both factors drive up the level of turbulence in the
system\footnote{For instance, the increase of $\Delta$ increases the Reynolds
number of the outflow, making the outflow more turbulent.} inducing a positive
feedback which in magnetically dominated media will lead to explosive
reconnection.

A characteristic feature of this reconnection instability is that it has a
threshold and therefore it can allow the accumulation of the flux prior to
reconnection.  In other words, as remarked before, LV99 model predicts that the reconnection can be
both fast and slow, which is the necessary requirement of bursty reconnection
frequently observed in nature, e.g. in solar flares.  This process may be
related to the bursts of reconnection observed in simulations in
\cite{Lapenta08}.  In addition, LV99 predicted the process of
{\it triggered reconnection} when reconnection in one part of the volume sends
perturbations that initiate reconnection in adjacent volumes.  Such process was,
as we mentioned earlier, also reported recently in the observations of
\cite{Sychetal09}.

The value of the threshold for initiating the burst depends on the system.  For
instance, tearing instability associated with magnetic reconnection (see
\cite{Loureiroetal09, Bhattacharjeeetal09}) in 3D should create turbulence in
agreement with the numerical simulations that we discussed in
\S\ref{sec:numerical_testing}.  This shows how the tearing and turbulent mechanisms
may be complementary, with tearing triggering turbulent reconnection.  Note
that, once turbulence develops, the LV99 mechanism can provide much faster
reconnection compared to tearing and tearing becomes a subdominant process.
Depending on the value of the Reynolds number of the outflow, the emerging
turbulence may completely supplant the tearing instability as the driver of
reconnection.  We believe that such flares of turbulent reconnection  can
explain a wide variety of astrophysical processes ranging from solar flares to
gamma ray bursts as well as bursty reconnection in the pulsar winds (eg. \cite{deGouveiadalPinoLazarian05}).

A simple quantitative model of flares was presented in
\cite{LazarianVishniac09}.  There it is assumed that since stochastic
reconnection is expected to proceed unevenly, with large variations in the
thickness of the current sheet, one can expect that some unknown fraction of
this energy will be deposited inhomogeneously, generating waves and adding
energy to the local turbulent cascade.

For the sake of simplicity, the plasma density is assumed to be uniform so that the
Alfv\'{e}n speed and the magnetic field strength are interchangeable.  The
nonlinear dissipation rate for waves is
\begin{equation}
\tau_{nonlinear}^{-1}\sim\max\left[ {k_\perp^2 v_{wave}^2\over k_\|V_A},k_\perp^2 VL\right],
\end{equation}
where the first rate is the self-interaction rate for the waves and the second
is the dissipation rate induced by the ambient turbulence (see
\cite{BeresnyakLazarian08}).  The important point here is that
$k_\perp$ for the waves falls somewhere in the inertial range of the strong
turbulence.  Eddies at that wavenumber will disrupt the waves in one eddy
turnover time, which is necessarily less than $L/V_A$.  Therefore, the bulk of
the wave energy will go into the turbulent cascade before escaping from the
reconnection zone.

An additional simplification is achieved by assuming that some fraction
$\epsilon$ of the energy liberated by stochastic reconnection is fed into the
local turbulent cascade.  The evolution of the  turbulent energy density per
area is
\begin{equation}
{d\over dt}\left(\Delta V^2\right)=\epsilon V_A^2 V_{rec}-V^2\Delta {V_A\over L_x},
\end{equation}
where the loss term covers both the local dissipation of turbulent energy, and
its advection out of the reconnection zone.  Since $V_{rec}\sim v_{turb}$  and
$\Delta\sim L_x(V/V_A)$,  it is possible to rewrite this by defining
$\tau\equiv tV_A/L_x$ so that
\begin{equation}
{d\over d\tau}M_A^3\approx \epsilon M_A-M^3_A.
\end{equation}
If $\epsilon$ is a constant then
\begin{equation}
V\approx V_A\epsilon^{1/2}(1-e^{-2\tau/3})^{1/2}.
\end{equation}
This implies that the time during which reconnection rate rises to
$\epsilon^{1/2}V_A$ is comparable to the ejection time from the reconnection
region ($\sim L_x/V_A$).

Within this toy model $\epsilon$ is not defined.  Its value can be constrained
through observations.  Given that reconnection events in the solar corona seem
to be episodic, with longer periods of quiescence, this is suggestive that
$\epsilon$ is very small, for example, depends strongly on the ratio of
the  thickness of the current sheet to $L_x$.  In particular, if it scales as
$M_A$ to some power greater than two then initial conditions dominate the early
time evolution.

Another route by which magnetic reconnection might be self-sustaining via
turbulence injection would be in the context of a series of topological knots in
the magnetic field, each of which is undergoing reconnection.  For simplicity,
one can assume that as each knot undergoes reconnection it releases a
characteristic energy into a volume which has the same linear dimension as the
distance to the next knot.  The density of the energy input into this volume is
roughly $\epsilon V_A^2 V/L_x$, where here $\epsilon$ is defined as the
efficiency with which the magnetic energy is transformed into turbulent energy.
Thus one gets
\begin{equation}
\epsilon {V_A^2V\over L_x}\sim {v'^3\over L_k},
\end{equation}
where $L_k$ is the distance between knots and $v'$ is the turbulent velocity
created by the reconnection of the first knot.  This process will proceed
explosively if $v'>V$ or
\begin{equation}
V_A^2 L_k\epsilon> V^2 L_x.
\end{equation}
The condition above is easy to fulfill.  The bulk motions created by
reconnection can  generate turbulence as they interact with their surrounding,
so $\epsilon$ should be of order unity.  Moreover the length of any current
sheet should be at most comparable to the distance to the nearest distinct
magnetic knot.  The implication is that each magnetic reconnection event will
set off its neighbors, boosting their reconnection rates from $V_L$, set by the
environment, to $\epsilon^{1/2}V_A(L_k/L_x)^{1/2}$ (as long as this is less than
$V_A$).  The process will take a time comparable to the crossing time $L_x/V_L$
to begin, but once initiated will propagate through the medium with a speed
comparable to speed of reconnection in the individual knots.  The net effect can
be a kind of modified sandpile model for magnetic reconnection in the solar
corona and chromosphere.  As the density of knots increases, and the energy
available through magnetic reconnection increases, the chance of a successfully
propagating reconnection front will increase.

This picture is broadly supported by current observations and numerical
simulations of solar flares and CME's.  For example, simulations by
\cite{Lynchetal08} of the ``breakout model'' of CME initiation show that an
extremely complex magnetic line structure develops in the ejecta during
and after the initial breakout reconnection phase, even under the severe
numerical resolution constraints of such simulations.  In the very high
Lundquist-number solar environment, this complex field must correspond to a
strongly turbulent state, within which the subsequent ``anti-breakout
reconnection'' and post-CME current sheet occur.  Direct observations of such
current sheets \cite{CiaravellaRaymond08,Bemporad08} verify the presence of
strong turbulence and greatly thickened reconnection zones, consistent with the
LV99 model.  In the numerical simulations, the ``trigger'' of the initial
breakout reconnection is numerical resistivity and there is no evidence of
turbulence or complex field-structure during the  eruptive flare onset.  This is
very likely to be a result of the limitations on resolution, however, and we
expect that developing turbulence will accelerate reconnection in this phase of
the flare as well.

While the details of the physical processes discussed above can be altered, it
is clear that LV99 reconnection induces bursts in highly magnetized plasmas.
This can be applicable not only to the solar environment but also to more exotic
environments, e.g. to gamma ray bursts.  The model of gamma ray bursts based on
LV99 reconnection was suggested in \cite{Lazarianetal03}.  It was elaborated and
compared with observations in \cite{ZhangYan11}.  Currently, the latter model is
considered promising and it attracts a lot of attention of researchers.  Flares
of reconnection that we described above can also be important for compact sources, like pulsars and black holes in microquasars and AGNs \cite{deGouveiadalPinoLazarian05}.  They
seem like a more natural way of explaining the observed phenomenon compared to
e.g. individual plasmoids (cf. \cite{Giannios13}).

\subsection{Reconnection diffusion and star formation}

Star formation theory was formulated several decades ago with an explicit
assumption that the fully ionized gas and magnetic field are coupled to very
high degree.  Therefore, the source of the decoupling was identified with the
presence of neutral atoms which do not directly feel magnetic fields, but
interact with ions that tend to follow magnetic field lines.  The slippage of
matter in respect to magnetic field was called ambipolar diffusion and became
the textbook explanation for the processes of star formation in magnetized gas (see more details in the Chapters of E. Zweibel and of Lizano and Galli in this volume).

Naturally, fast magnetic reconnection changes the situation dramatically.  It is
clear that in turbulent astrophysical media the dynamics of matter and gas are
different from the idealized picture above and this presents a serious shift in
the conventional paradigm of star formation.

The process of moving of matter in respect to magnetic field was identified in
\cite{Lazarian05} (see also \cite{LazarianVishniac09}) and successfully tested
in the subsequent publications for the case of molecular clouds and protostellar
disks, e.g. \cite{SantosLimaetal10, SantosLimaetal12, SantosLimaetal13, 
degouv12, leao13}.  
The theory of transporting matter in turbulent magnetized medium 
is discussed at length in \cite{Lazarian11a} and \cite{Lazarianetal12} and we 
refer our reader to these publications.  
The process was termed ``reconnection diffusion'' to stress the importance of 
reconnection in the the diffusive transport.

\begin{figure}
\centering
\includegraphics[width=0.95\textwidth]{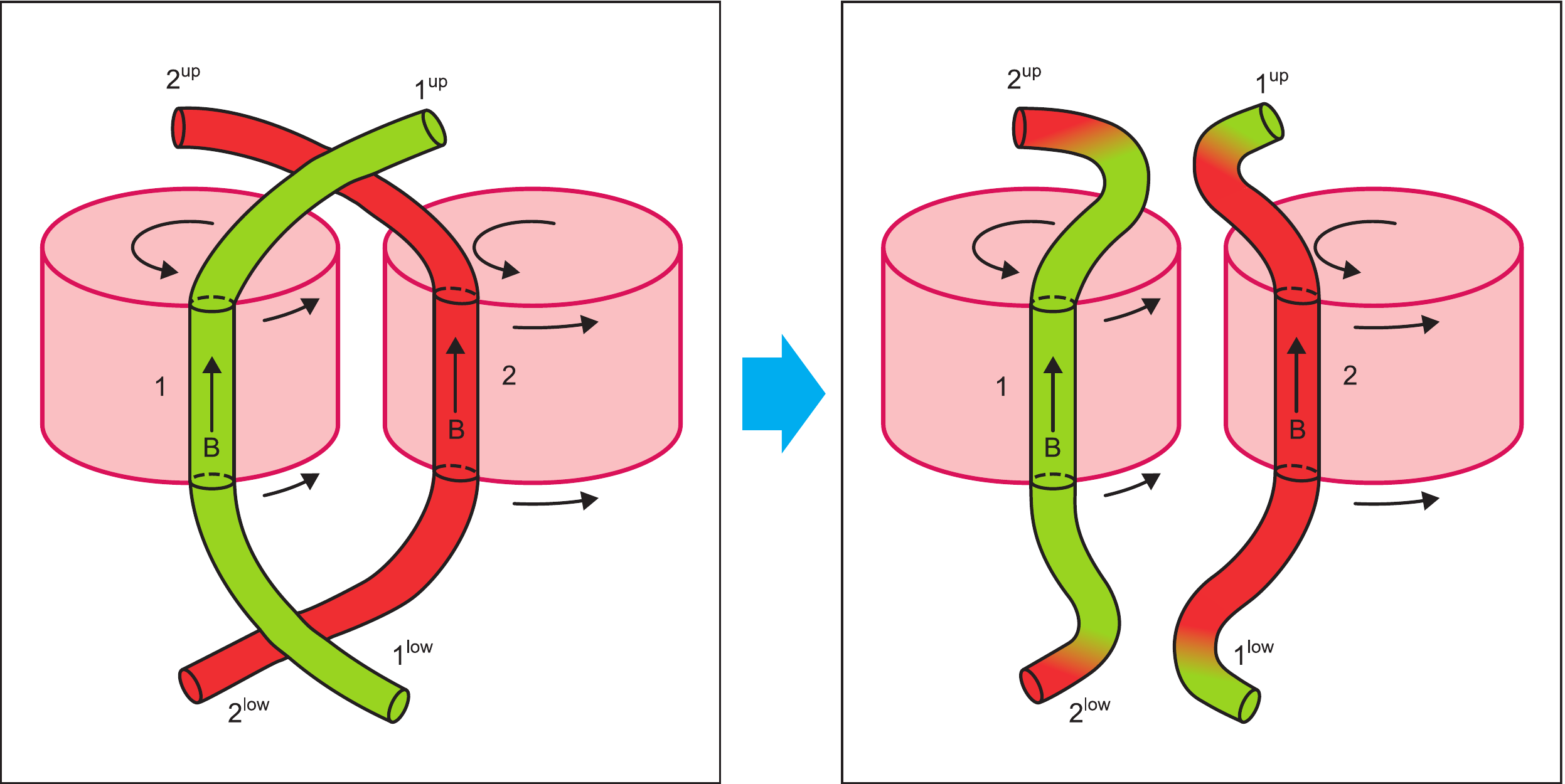}
\caption{Reconnection diffusion: exchange of flux with entrained matter.
Illustration of the mixing of matter and magnetic fields due to reconnection as
two flux tubes of different eddies interact. Only one scale of turbulent motions
is shown. In real turbulent cascade such interactions proceed at every scale of
turbulent motions. From \cite{Lazarian11a}. \label{mix}}
\end{figure}

The peculiarity of reconnection diffusion is that it requires nearly parallel
magnetic field lines to reconnect, while the textbook description of
reconnection is usually associated with anti-parallel description of magnetic
field lines.  One should understand that the configuration shown in
Figure~\ref{recon} is just a cross section of the magnetic fluxes
depicting the anti-parallel components of magnetic field.  Generically, in 3D
reconnection configurations the sheared component of magnetic field is present.
The process of reconnection diffusion is closely connected with the reconnection
between adjacent Alfv\'{e}nic eddies (see Figure~\ref{mix}).  As a result,
adjacent flux tubes exchange their segments with entrained plasmas and flux
tubes of different eddies get connected.  This process involves eddies of all
the sizes along the cascade and ensures fast diffusion which has similarities
with turbulent diffusion in ordinary hydrodynamic flows.

Finally, a number of studies attempted to understand the role of joint action of
turbulence and ambipolar diffusion.  For instance, \cite{Heitschetal04}
(henceforth HX04) performed 2.5D simulations of turbulence with two-fluid code
and examined the decorrelation of neutrals and magnetic field in the presence of
turbulence (see also the Chapter by Zweibel in this volume).  The study reported an enhancement of diffusion rate compared to the
ambipolar diffusion in a laminar fluid.  HX04 correctly associated the
enhancement with turbulence creating density gradients that are being dissolved
by ambipolar diffusion (see also \cite{Zweibel02}). However, in 2.5D simulations
of HX04 the numerical set-up precluded reconnection from taking
place as magnetic field was perpendicular to the plane of 2D mixing and
therefore magnetic field lines were absolutely parallel to each other.  This
will not happen in realistic astrophysical situations where reconnection will
be an essential part of the physical picture. Therefore, we claim that a
treatment of ``turbulent ambipolar diffusion'' without addressing the
reconnection issue is of academic interest.

Incidentally, the authors of HX04 reported an enhanced rate that is equal to the
turbulent diffusion rate $L V_L$.  The fact that
ambipolar diffusion rate does not enter the result in HX04 suggests that
ambipolar diffusion is irrelevant for the diffusion of matter in the presence of
turbulence.  This is another reason not to call the observed process ``turbulent
ambipolar diffusion'' \footnote{A similar process takes place in the case of
molecular diffusivity in turbulent hydrodynamic flows.  The result for the
latter flows is well known: in the turbulent regime, molecular diffusivity is
irrelevant for the turbulent transport.  The process is called therefore
``turbulent diffusivity'' without adding the superfluous and inappropriate word
``molecular''.}.

Therefore we believe that HX04 captured in their simulations a special
degenerate case of 2.5D turbulent diffusion where due to a special set up the
reconnection is avoided and magnetic field lines do not intersect.  We also note
that, in the presence of turbulence, the independence of the gravitational
collapse from the ambipolar diffusion rate was reported in numerical simulations
by \cite{Balsaraetal01}, although further higher resolution studies are still missing..

A comprehensive review dealing with reconnection diffusion is presented in
\cite{Lazarian13}.

\subsection{Heat and cosmic ray transport in the presence of reconnection}

Magnetic reconnection is a very fundamental basic process that happens in all
magnetized fluids.  As we discussed in \S\ref{sec:reconnection} magnetic
reconnection is closely related to the turnover processes of magnetic eddies as
well as magnetic field wandering.  The former is essential for the heat
advection via turbulent mixing of magnetized gas.  The process was invoked by
\cite{Choetal03} to explain the suppression of cooling flows
in galaxy clusters.  Fast LV99 magnetic reconnection was invoked to justify the
existence of magnetic eddies for the very high Lundquist number plasmas (see
more \cite{Lazarian09, Lazarian11b}).

Heat transport is also possible in magnetized plasma if electrons are streaming
along meandering magnetic field lines.  In \cite{Lazarian06} the
heat transfer by electron streaming was compared with that induced by turbulent
eddies and it was concluded that in typical clusters of galaxies the latter
dominates.

Transport of cosmic rays along meandering magnetic field was invoked to solve
the problem of perpendicular diffusion in Milky Way in classical studies
\cite{Jokipii73}.  For the propagation of cosmic rays the dynamics of turbulent
magnetized plasmas is not important as $c/V_A$ is usually large.  However, the
formation of the complicated web of the wandering magnetic field lines that is
consistent with the Kolmogorov-type scaling of turbulence statistics does
necessarily require fast magnetic reconnection.

\subsection{Reconnection and First-order Fermi acceleration}

The process of LV99 reconnection invokes shrinking magnetic loops.  It is clear
from Figure 1 in the Chapter by de Gouveia Dal Pino and Kowal in this volume 
that particles entrained on such a loop will experience acceleration.  
This process that naturally follows from the LV99 model was invoked by
\cite{deGouveiadalPinoLazarian05} to predict efficient First-Order Fermi
acceleration of cosmic rays in the reconnection regions (see also
\cite{Lazarian05}).  The latter are traditionally associated with the
acceleration of particles in shocks\footnote{The First-Order Fermi acceleration
is a process in which the energy gain is proportional to the first order of the
ratio of the shock velocity to that of light.  It should be distinguished from
the stochastic Second-Order Fermi acceleration which is proportional to the
square of this ratio. (see more details in de Gouveia Dal Pino and Kowal's
chapter in this volume).}.  Later research revealed the high promise of the
process for explaining various physical processes.  Recently, the acceleration
of cosmic rays in reconnection has been invoked to explain results on the
anomalous cosmic rays obtained by Voyager spacecrats (\cite{LazarianOpher09,
Drakeetal10}), the local anisotropy of cosmic rays (\cite{LazarianDesiati10})
and the acceleration of cosmic rays in clusters of galaxies
(\cite{LazarianBrunetti11}), as well as in the surrounds of  compact sources and
black holes \cite{deGouveiadalPinoLazarian05} and relativistic jets
\cite{Giannios13}. Naturally, the process of acceleration is much more
widespread and not limited to the explored examples.


In addition to the acceleration of cosmic rays parallel to magnetic field,
acceleration perpendicular to the magnetic field is also possible, as discussed
in \cite{Kowaletal12b,Lazarianetal12b}.  The advantage of such a perpendicular
acceleration is that the gain of energy is taking place every Larmor period of
the cosmic ray.  The efficiency of perpendicular acceleration was observed in
simulations of \cite{Kowaletal12b}, where the simulations of turbulent
reconnection were used to study the acceleration of cosmic rays (see more
details in de Gouveia Dal Pino and Kowal's chapter in this volume).

\subsection{Dissipation of turbulence in current sheets}

MHD turbulence cascade does not depend on the details of microphysics.  However,
at sufficiently small scales current sheets are formed and those may dissipate a
substantial part of the turbulent cascade.  As we discussed in
\S\ref{sec:reconnection} within LV99 model small scale reconnection events may
happen due to ordinary Ohmic or plasma effects.  In particular, the small scale
current sheets can be in the collisionless regime.  Therefore it is not easy to
distinguish the nature of magnetic reconnection by studying the processes of
electron and proton heating.

\section{Discussion}
\label{sec:discussion}

\subsection{Interrelation of LV99 reconnection and modern understanding of MHD turbulence}

MHD turbulence has advanced considerably in the last 20 years.  It is easy to
understand that strong Alfv\'{e}nic turbulence that induces Richardson diffusion
does require fast reconnection.  Indeed, eddy type motions that are produced by
such turbulence can happen only if the magnetic field of the eddies relaxes on
the time scale of eddy turnover.  Calculations in LV99 showed that
the GS95 theory \cite{GoldreichSridhar95} is self-consistent when the small-scale magnetic
reconnection between adjacent turbulent eddies happens with the LV99 predicted
rate\footnote{Indeed, within the GS95 picture the reconnection happens with nearly parallel lines with magnetic pressure
gradient $V_A^2/l_{\|}$ being reduced by a factor $l_{\bot}^2/l_{\|}^2$, since only reversing component is available for
driving the outflow. At the same time the length of the contracted magnetic field lines is also reduced from $l_{\bot}$ but $l_{\bot}^2/l_{\||}$.
Therefore the acceleration is $\tau_{eject}^{-2} l_{\bot}^2/l_{\||}$. As a result, the Newtons' law gives 
$V_A^2 l_{\bot}^2/l_{\|}^3 \approx \tau_{eject}^{-2} l_{\bot}^2/l_{\||}$. This provides the result for the ejection rate $\tau_{eject}^{-1}\approx V_A/l_{\|}$.
The length over which the magnetic eddies intersect is $l_{\bot}$ and the rate of reconnection is $V_{rec}/l_{\bot}$. For the stationary reconnection
this gives $V_{rec}\approx V_{A} l_{\bot}/l_{\}}$, which provides the reconnection rate $V_A/l_{\|}$, which is exactly the rate of the eddy turnovers
in GS95 turbulence.}. This result also follows from the Richardson diffusion that we discussed in the chapter.

by a factor . This rate varies from $\sim V_A$ for largest eddies in
transAlfv\'{e}nic turbulence to a small fraction of $V_A$ for the smallest
eddies.  Obviously, no mechanism that produces a fixed reconnection rate, e.g.
the rate of $0.1 V_A$ that for decades was a sort of Holy Grail rate for the
researchers attempting to explain Solar flares, can make modern theories of MHD
turbulence, both the GS95 and its existing
modifications, self-consistent.  At the same time, ELV11 showed that the
Lagrangian dynamics of turbulent fluids do require fast magnetic reconnection.
Or, reversing the role of cause and effect, the Lagrangian phenomenon of
Richardson dispersion produces a breakdown in the standard form of flux-freezing
for a turbulent MHD flow. The reconnection rates that are dictated by the
well-established process of Richardson diffusion coincide with those predicted
by LV99.

In other words, LV99 reconnection is an intrinsic and inseparable element of MHD
turbulence.  There can be other types of magnetic reconnection, that are
important in particular circumstances, but in turbulent fluids the LV99 type
seems inevitable.

\subsection{Suggestive evidence on fast reconnection}

A study of tearing instability of current sheets in the presence of background
2D turbulence that observed the formation of large-scale islands was performed
in \cite{Politanoetal89}.  While one can argue that observed long-lived islands
are the artifact of adopted 2D geometry, the authors present evidence for {\it
fast energy dissipation} in 2D MHD turbulence and show that this result does not
change as they change the resolution.  A more recent work of
\cite{MininniPouquet09} provides evidence for {\it fast dissipation} also in 3D
MHD turbulence.  This phenomenon is consistent with the idea of fast
reconnection, but cannot be treated as a direct evidence of the process.
Although related, fast dissipation and fast magnetic reconnection are rather
different physical processes, dealing with decrease of energy on the one hand
and decrease of magnetic flux on the other.

Works by Galsgaard and Nordlund, in particular \cite{GalsgaardNordlund97b},
could also be interpreted as an indirect support for fast reconnection.  The
authors showed that in their simulations they could not produce highly twisted
magnetic fields.  One possible interpretation of this result could be the fast
relaxation of magnetic field via reconnection.  In this case, these observations
could be related to the numerical finding of \cite{LapentaBettarini11} which
shows that reconnecting magnetic configurations spontaneously get chaotic and
dissipate, which, as discussed in \cite{LapentaLazarian12}, may
be related to the LV99 model.  However, in view of many uncertainties of the
numerical studies, this relation is unclear.  The highest resolution simulations
of \cite{GalsgaardNordlund97b} were only $136^3$ and with Reynolds number so
small that they could not allow a turbulent inertial-range.

\subsection{Convergence of different approaches to fast reconnection}

The LV99 model of magnetic reconnection in the presence of weakly stochastic
magnetic fields was proposed more than a decade ago.  In fact, LV99 and the idea
of collisionless X-point reconnection mediated by the Hall effect are
essentially coeval.  At the same time,  due to a few objective factors, it met
less enthusiasm in the community than the X-point collisionless reconnection.
One can speculate what were the factors responsible for this slow start.  For
one thing, the collisionless X-point reconnection was initiated and supported by
numerical simulations, while the numerical testing of LV99 became possible only
recently.  In addition, the acceptance of the idea of astrophysical fluids
generically being in turbulent state was only taking roots in 1999 (but see
\cite{Chandrasekhar49}!) and at that time it had much less observational
support.  By now we have much more evidence which justifies the claim that
models ignoring pre-existent turbulence have little relevance to astrophysics.
Finally, the analytical solutions of LV99 were based on the use and extension of
the GS95 model of turbulence.  However, the GS95 theory was far from being
universally accepted at the time LV99 was published\footnote{In fact, this
unsatisfactory situation with the theory of turbulence motivated some of us to
work seriously on testing turbulence models (see \cite{ChoVishniac00, Choetal02, ChoLazarian02, ChoLazarian03})}.

The situation has changed substantially by now.  With GS95, as we discussed
earlier, being widely accepted, with more observational evidence of ubiquitous
turbulence in astrophysical environments and with the successful testing of the
LV99 model, it is more difficult to argue against the importance of turbulence for
astrophysical reconnection.  Moreover, the LV99 model has received more support
from solar observations \S\ref{sec:obs_cons_tests}, which both showed that
magnetic reconnection can be fast in collisional media, where the aforementioned
collisionless reconnection does not work.  Solar observations also confirmed
LV99 predictions on the thickness of reconnection regions and on triggering
reconnecttion by the neighboring reconnection events.  Last, but not the least,
a very important development took place, namely, the LV99 model was connected to
the modern developments in the Lagrangian description of magnetized fluids and
the equivalence of the approach in LV99 and that based on spontaneous
stochasticity was established (see \S\ref{sec:reconnection} and \S\ref{sec:numerical_testing}).

One can argue that we have observed the convergence of LV99 with other
directions of reconnection research.  In particular, recent models of
collisionless reconnection have acquired several features in common with the
LV99 model.  In particular, they have moved to consideration of volume-filling
reconnection (see \cite{Drakeetal06}).  While much of the
discussion may still be centered around 2D magnetic islands produced by
reconnection, in three dimensions these islands are expected to evolve into
contracting 3D loops or ropes \cite{Daughtonetal08}, which is broadly similar to
what is depicted in Figure~\ref{fig:bmag_cuts}, at least in the sense of
introducing stochasticity to the reconnection zone.  Moreover, it is more and
more realized that the 3D geometry of reconnection is essential and that the 2D
physics is not adequate and may be misleading. This essentially means the end of
the epoch of the dominance of collisionless X-point reconnection.  The interest
of the models alternative to LV99 shifted to chaotically broadened extended
Y-shaped outflow regions, which were advocated in LV99 and confirmed by
observations.

The departure from the concept of laminar reconnection and the introduction of
magnetic stochasticity is also apparent in a number of recent papers appealing
to the tearing mode instability to drive fast reconnection (see
\cite{Loureiroetal09}, \cite{Bhattacharjeeetal09}). These studies showed that
tearing modes do not require collisionless environments and thus collisionality
is not a necessary ingredient of fast reconnection\footnote{The largest-scale
Hall MHD simulations performed to date \cite{Huangetal11} do show somewhat
higher reconnection rates for laminar X-point solutions than for plasmoid
unstable regimes, but the X-point solutions lose stability and seem to have
lower reconnection rates with decreasing ratios $\delta_i/L_x.$}.  Finally, the
reported development of turbulence in 3D numerical simulations clearly testifies
that the reconnection induces turbulence even if the initial reconnection
conditions are laminar. Naturally, one should expect that turbulence modifies
tearing instability and induces its own laws for reconnection thus making for
many situations the tearing modes only the trigger to self-supported turbulent
reconnection.  If this is the case, the final non-linear stage of the
reconnection should allow a theoretical description based on the LV99 model.

All in all, in the last decade, the models competing with LV99 have undergone a
substantial evolution, from 2D collisionless X-point reconnection based mostly
on Hall effect to 3D reconnection where the collisionless condition is no more
required, Hall effect is not employed, but magnetic stochasticity and turbulence
play an important role in the thick Y-shaped reconnection regions.  In other
words, a remarkable convergence has taken place.

Saying all the above, we want to stress that collisionless X-point reconnection
may nevertheless be suitable for the description of reconnection when the reconnecting
flux-structures are comparable with the ion gyro scale, which is the case of the
reconnection studied {\it situ} in the magnetosphere (see Table~\ref{tab:parameters}).  However, this is a special case of magnetic
reconnection with, we argue, atypical features compared with most astrophysical
reconnection.

\subsection{Recent attempts to relate turbulence and reconnection}

Gue et al. \cite{Guoetal12} proposed a model based on the earlier idea of mean field
approach suggested initially in \cite{KimDiamond01}.  In the latter paper the
author concluded that the reconnection rate should be always slow in the
presence of turbulence. On the contrary, models in \cite{Guoetal12} invoke
hyperresistivity and get fast reconnection rates. Similarly, invoking the mean
field approach \cite{HigashimoriHoshino12} presented their model of turbulent
reconnection.

The mean field approach invoked in the aforementioned studies was critically analyzed
by \cite{Eyink11}, and below we briefly present some arguments from
that study. The principal difficulty is with the justification of using the mean
field approaches to explain fast magnetic reconnection. In such an approach
effects of turbulence are described using parameters such as anisotropic
turbulent magnetic diffusivity and hyper-resistivity experienced by the fields
once averaged over ensembles. The problem is that it is the lines of the full
magnetic field that must be rapidly reconnected, not just the lines of the mean
field.  ELV11 stress that the former implies the latter, but not conversely. No
mean-field approach can claim to have explained the observed rapid pace of
magnetic reconnection unless it is shown that the reconnection rates obtained in
the theory are strictly independent of the length and timescales of the
averaging. More detailed discussion of the conceptual problems of the
hyper-resistivity concept and mean field approach to magnetic reconnection is
presented in \cite{Lazarianetal04} and ELV11.

\subsection{Reconnection and numerical simulations}

As discussed in section \S\ref{ssec:numerical_approach}, a brute force 
numerical approach to astrophysical reconnection is impossible. 
Therefore our numerical studies of reconnection diffusion in 
\cite{SantosLimaetal10, SantosLimaetal12, SantosLimaetal13, leao13} deal with a
different domain of Lundquist numbers and the theoretical justification why 
{\it for the given problem} the Lundquist number regime is not essential. 
For the case of reconnection diffusion simulations, LV99 theory
predicts that the dynamics of reconnection is independent from the Lundquist
number and therefore the reconnection in the computer simulations {\it in the
presence of turbulence} adequately represents the astrophysical process.

The above numerical results explored the consequences of
reconnection diffusion.  Similarly, as numerical studies of ambipolar diffusion
do not ``prove'' the very concept of ambipolar diffusion, our studies were not
intended to ``prove'' the idea of reconnection diffusion.  Our goal was to
demonstrate that, {\it in agreement with the theoretical expectations}, the
process of reconnection diffusion is important for a number of astrophysical
set-ups relevant to star formation.


\subsection{Plasma physics and reconnection}

We have been primarily interested in this review in reconnection phenomena
at scales much larger than the ion gyro-radius $\rho_i.$ We have also made the claim---
which may appear paradoxical to some---that these phenomena can be explained
by hydrodynamical processes in turbulent MHD regimes. Microscopic plasma processes
do play a role, however, which should be briefly explained. Consider a collisionless
turbulent plasma, such as the solar wind, in which the MHD description of the cascade
terminates at the ion gyro radius.   At scales smaller than $\rho_i$
but larger than $\rho_e$, the plasma is described by an ion kinetic equation and
a system of ``electron reduced MHD'' (ERMHD) equations for kinetic Alfv\'{e}n
waves \cite{Schekochihinetal07,Schekochihinetal09}.  This system exhibits the
``Hall effect'', with distinct ion and electron mean flow velocities and
magnetic field-lines frozen-in to the electron fluid.  The ERMHD equations (or
the more general ``electron MHD'' or EMHD equations) produce the typical
features of ``Hall reconnection'' such as quadrupolar magnetic fields in the
reconnection zone \cite{UzdenskyKuslrud06}\footnote{Because the Hall MHD
equations have played a prominent role in magnetic reconnection research of the
past decade \cite{Shayetal98,
Shayetal99,Wangetal00,Birnetal01,Drake01,Malakitetal09,Cassaketal10}, it is
worth remarking that those equations are essentially never applicable in
astrophysical environments.  A derivation of Hall MHD based on collisionality
requires that the ion skin-depth $\delta_i$ must satisfy the conditions
$\delta_i\gg L\gg \ell_{mfp,i}$.  The second inequality is needed so that a
two-fluid description is valid at the scales $L$ of interest, while the first
inequality is needed so that the Hall term remains significant at those scales.
However, substituting $\delta_i=\rho_i/\sqrt{\beta_i}$ into (\ref{lmfp-rho})
yields the result $$\frac{\ell_{mfp,i}}{\delta_i}\propto
\frac{\Lambda}{\ln\Lambda}\frac{v_{th,i}}{c}. $$  The ratio $v_{th,i}/c$ is
generally small in astrophysical plasmas, but the plasma parameter $\Lambda$ is
usually large by even much, much more (see Table~\ref{tab:parameters}).  Thus, it
is usually the case that $\ell_{mfp,i}\gg \delta_i,$ unless the ion temperature
is extremely low.  A collisionless derivation of Hall MHD from gyrokinetics
requires also a restrictive condition of cold ions (\cite{Schekochihinetal09},
Appendix E).  Thus, Hall MHD is literally valid only for cold, dense plasmas
like those produced in some laboratory experiments, such as the MRX reconnection
experiment \cite{Yamada99,Yamadaetal10}.}.  At length scales smaller than
$\rho_e,$ kinetic equations are required to describe both the ions and the
electrons.  It is at these scales that the magnetic flux finally ``unfreezes''
from the electron fluid, due to effects such as Ohmic resistivity, electron
inertia, finite electron gyroradius, etc.  However, as we have discussed at
length in this review, these weak effects are vastly accelerated by turbulent
advection and manifested, in surprising ways, at far larger length scales.

\begin{acknowledgement}
A.L. research is supported by the NSF grant AST~1212096, Vilas Associate Award
as well as the support 1098 from the NSF Center for Magnetic Self-Organization.
The research is supported by the Center for Magnetic Self-Organization in
Laboratory and Astrophysical Plasmas.  Stimulating environment provided by
Humboldt Award at the Universities of Cologne and Bochum, as well as a
Fellowship at the International  Institute of Physics (Brazil) is acknowledged.
G.K. acknowledges support from FAPESP (projects no. 2013/04073-2 and
2013/18815-0). Part of the computations were performed using supercomputer
RANGER (Teragrid AST080005N, TACC, USA,
\url{https://www.xsede.org/tg-archives/}), supercomputer GALERA (ACK TASK,
Poland, \url{http://www.task.gda.pl/}), and supercomputer ALPHACRUCIS (LAi,
IAG-USP, Brazil, \url{http://lai.iag.usp.br/}).  We thank Andrey Beresnyak for
useful discussions of the generation of turbulence in the process of magnetic
reconnection.
\end{acknowledgement}


\begin{thebibliography}{99.}
\bibitem{Alfven42} Alfv\'{e}n, H.\ 1942, Ark. Mat., Astron. o. Fys., 29B, 1
\bibitem{Parker79} Parker, E.~N.\ 1979, Oxford, Clarendon Press; New York, Oxford University Press, 1979, 858 p.,
\bibitem{Parker70} Parker, E.~N.\ 1970, The Astrophysical Journal, 162, 665
\bibitem{Lovelace76} Lovelace, R.~V.~E.\ 1976, Nature, 262, 649
\bibitem{PriestForbes02} Priest, E.~R., \& Forbes, T.~G.\ 2002, Astronomy \& Astrophysics Reviews, 10, 313
\bibitem{Innesetal97} Innes, D.~E., Inhester, B., Axford, W.~I., \& Wilhelm, K.\ 1997, Nature, 386, 811
\bibitem{YokoyamaShibata95} Yokoyama, T., \& Shibata, K.\ 1995, Nature, 375, 42
\bibitem{Masudaetal94} Masuda, S., Kosugi, T., Hara, H., Tsuneta, S. \& Ogawara, Y.\ 1994, Nature, 371, 495
\bibitem{Shayetal98} Shay,~M.~A., Drake,~J.~F., Denton,~R.~E., \& Biskamp,~D.\ 1998, Journal of Geophysical Research, 103, 9165
\bibitem{Drake01} Drake, J.~F.\ 2001, Nature, 410, 525
\bibitem{Drakeetal06} Drake, J.~F., Swisdak, M., Che, H., \& Shay, M.~A.\ 2006, Nature, 443, 553
\bibitem{Daughtonetal06} Daughton, W., Scudder, J., \& Karimabadi, H.\ 2006, Physics of Plasmas, 13, 072101
\bibitem{Parker93} Parker, E.~N.\ 1993, The Astrophysical Journal, 408, 707
\bibitem{Ossendrijver03} Ossendrijver, M.\ 2003, Astronomy \& Astrophysics Reviews, \textbf{11}, 287
\bibitem{Sturrock66} Sturrock, P.~A.\ 1966, Nature, 211, 695
\bibitem{ZhangYan11} Zhang, B., \& Yan, H.\ 2011, The Astrophysical Journal, 726, 90
\bibitem{Lazarianetal04} Lazarian, A., Vishniac, E.~T., \& Cho, J.\ 2004, The Astrophysical Journal, 603, 180
\bibitem{Foxetal05} Fox, D.~B., et al.\ 2005, Nature, 437, 845
\bibitem{Galamaetal98} Galama, T.~J., et al.\ 1998, Nature, 395, 670
\bibitem{ShibataMagara11} Shibata,~K. \& Magara,~T.\ 2001, Living Reviews in Solar Physics, 8, 6
\bibitem{BrowningLazarian13} Browning, P., \& Lazarian, A., 2013, Space Science Reviews, 178, 325
\bibitem{KarimabadiLazarian13} Karimabadi, H. \& Lazarian, A. 2013, Physics of Plasmas, in press
\bibitem{Kowaletal09} Kowal, G., Lazarian, A., Vishniac, E.~T., \& Otmianowska-Mazur, K.\ 2009, The Astrophysical Journal, 700, 63
\bibitem{Parker57} Parker,~E.~N.\ 1957, Journal of Geophysical Research, 62, 509
\bibitem{Sweet58} Sweet,~P.~A.\ 1958, Proceedings from IAU Symposium no. 6. edited by Bo Lehnert, Cambridge University Press, p.~123
\bibitem{Petschek64} Petschek,~H.~E.\ 1964, The Physics of Solar Flares, AAS-NASA Symposium (NASA SP-50),
ed. W. H. Hess (Greenbelt, MD: NASA), 425
\bibitem{Biskamp96} Biskamp,~D.\ 1996, Astrophysics and Space Science, 242, 165
\bibitem{Shayetal04} Shay,~M.~A., Drake,~J.~F., \& Swisdak,~M.~M.\ 2004, Physics of Plasmas, 11, 2199
\bibitem{Bhattacharjeeetal03} Bhattacharjee,~A., Ma,~Z.~W., \&  Wang,~X.\ 2003, Lecture Notes in Physics, 614, 351
\bibitem{Wangetal01} Wang,~X., Bhattacharjee,~A., \& Ma,~Z.~W.\ 2001, Physical Review Letters, 87, 265003
\bibitem{Smithetal04} Smith,~D., Ghosh,~S., Dmitruk,~P., \& Matthaeus,~W.~H.\ 2004, Geophysysical Research Letters, 31, L02805
\bibitem{Fitzpatrick04} Fitzpatrick,~R.\ 2004, Physics of Plasmas, 11, 937
\bibitem{Malyshkin08} Malyshkin, L.~M.\ 2008, Physical Review Letters, 101, 225001
\bibitem{Shivamoggi11} B. K. Shivamoggi,  2011, Phys. Plasmas 18, 052304
\bibitem{Yamada07} Yamada,~M.\ 2007, Physics of Plasmas, 14, 058102
\bibitem{NormanFerrara96} Norman, C.~A., \& Ferrara, A.\ 1996, The Astrophysical Journal, 467, 280
\bibitem{Ferriere01} Ferri{\`e}re, K.~M.\ 2001, Reviews of Modern Physics, 73, 1031
\bibitem{Subramanianetal06} Subramanian, K., Shukurov, A., \& Haugen, N.~E.~L.\ 2006, Monthly Notices of Royal Astronomical Society, 366, 1437
\bibitem{EnsslinVogt06} En{\ss}lin, T.~A., \& Vogt, C.\ 2006, Astronomy \& Astrophysics, 453, 44
\bibitem{Chandran05} Chandran, B.~D.~G.\ 2005, The Astrophysical Journal, 632, 809
\bibitem{BalbusHawley98} Balbus, S.~A., \& Hawley, J.~F.\ 1998, Reviews of Modern Physics, 70, 1
\bibitem{GalsgaardNordlund97a} Galsgaard, K., \& Nordlund, {\AA}.\ 1997, Journal of Geophysical Research, 102, 219
\bibitem{GerrardHood03} Gerrard, C.~L., \& Hood, A.~W.\ 2003, Solar Physics, 214, 151
\bibitem{Leamonetal98} Leamon, R.~J., Smith, C.~W., Ness, N.~F., Matthaeus, W.~H., \& Wong, H.~K.\ 1998, Journal of Geophysical Research, 103, 4775
\bibitem{Baleetal05} Bale, S.~D., Kellogg, P.~J., Mozer, F.~S., Horbury, T.~S., \& Reme, H.\ 2005, Physical Review Letters, 94, 215002
\bibitem{Schueckeretal04} Schuecker, P., Finoguenov, A., Miniati, F., B{\"o}hringer, H., \& Briel, U.~G.\ 2004, Astronomy \& Astrophysics, 426, 387
\bibitem{VogtEnsslin05} Vogt, C., \& En{\ss}lin, T.~A.\ 2005, Astronomy \& Astrophysics, 434, 67
\bibitem{Armstrongetal95} Armstrong,~J.~W., Rickett,~B.~J., \& Spangler,~S.~R.\ 1995, The Astrophysical Journal, 443, 209
\bibitem{ChepurnovLazarian10} Chepurnov, A. \& Lazarian, A.\ 2010, The Astrophysical Journal, 710, 853
\bibitem{Eyink08} Eyink, G.~L.\ 2008, Physica D Nonlinear Phenomena, 237, 1956
\bibitem{LazarianPogosyan00} Lazarian, A. \& Pogosyan, D.\ 2000, The Astrophysical Journal, 537, 720
\bibitem{LazarianPogosyan04} Lazarian, A. \& Pogosyan, D.\ 2004, The Astrophysical Journal, 616, 943
\bibitem{LazarianPogosyan06} Lazarian, A. \& Pogosyan, D.\ 2006, The Astrophysical Journal, 652, 1348
\bibitem{LazarianPogosyan08} Lazarian, A. \& Pogosyan, D.\ 2008, The Astrophysical Journal, 686, 350
\bibitem{Lazarian09} Lazarian,~A.\ 2009, Space Science Reviews, 143, 357
\bibitem{Vekshteinetal70} Vekshtein, G.~E., Ryutov, D.~D., \& Sagdeev, R.~Z.\ 1970, Soviet Journal of Experimental and Theoretical Physics Letters, 12, 291
\bibitem{Kulsrud83} Kulsrud, R.\ 1983, Handbook of Plasma Physics, eds. M. N. Rosenbluth \& R. Z. Sagdeev (North Holland, New York)
\bibitem{Eyinketal11} Eyink,~G.~L., Lazarian,~A., \& Vishniac,~E.~T.\ 2011, The Astrophysical Journal, 743, 51
\bibitem{Braginsky65} Braginsky, S.~I.\ 1965. Rev. Plasma Phys. 1, 205.
\bibitem{Fitzpatrick11} Fitzpatrick, R.\ 2011, ``Introduction to Plasma Physics'', online lecture notes, URL: {\tt http://farside.ph.utexas.edu/teaching/plasma/plasma.html}
\bibitem{Bemporad08} Bemporad, A.\ 2008, The Astrophysical Journal, 689, 572
\bibitem{Zimbardoetal10} Zimbardo, G., Greco, A., Sorriso-Valvo, L., Perri, S., V{\"o}r{\"o}s, Z., Aburjania, G.,
Chargazia, K., \& Alexandrova, O.\ 2010, Space Science Reviews, 156, 89
\bibitem{Kowaletal11} Kowal, G., Falceta-Gon\c{c}alves, D.~A., \& Lazarian, A.\ 2011, New Journal of Physics, 13, 053001
\bibitem{SantosLimaetal13} Santos-Lima,~R., de Gouveia Dal Pino,~E.~M., Kowal,~G., Falceta-Gon\c{c}alves,~D.~A., Lazarian,~A., \& Nakwacki,~M.~S.\ 2013, The Astrophysical Journal, in press, arXiv:1305.5654
\bibitem{SchekochihinCowley04} Schekochihin, A.~A., Cowley, S.~C., Maron, J. L., \& McWilliams, J. C.\ 2004, The Astrophysical Journal, 612, 276
\bibitem{SchekochihinCowley06} Schekochihin, A.~A., \& Cowley, S.~C.\ 2006, Physics of Plasmas, 13, 056501
\bibitem{LazarianBeresnyak06} Lazarian, A., \& Beresnyak, A.\ 2006, Monthly Notices of the Royal Astronomical Society, 373, 1195
\bibitem{Schekochihinetal07} Schekochihin, A.~A., Cowley, S.~C., \& Dorland, W.\ 2007, Plasma Physics and Controlled
Fusion, 49, 195
\bibitem{Schekochihinetal09} Schekochihin, A.~A., Cowley, S.~C., Dorland, W., Hammett, G.~W., Howes, G.~G., Quataert,
E., \& Tatsuno, T.\ 2009, The Astrophysical Journals, 182, 310
\bibitem{ChoLazarian02} Cho, J., \& Lazarian, A.\ 2002, Physical Review Letters, 88, 245001
\bibitem{ChoLazarian03} Cho, J., \& Lazarian, A.\ 2003, Monthly Notices of Royal Astronomical Society, 345, 325
\bibitem{KowalLazarian10} Kowal, G., \& Lazarian, A.\ 2010, The Astrophysical Journal, 720, 742
\bibitem{GoldreichSridhar95} Goldreich, P. \& Sridhar, S.\ 1995, The Astrophysical Journal, 438, 763
\bibitem{LithwickGoldreich01} Lithwick, Y., \& Goldreich, P.\ 2001, The Astrophysical Journal, 562, 279
\bibitem{Biskamp03} Biskamp, D.\ 2003, Magnetohydrodynamic Turbulence, by Dieter Biskamp, pp.~310.~ISBN 0521810116.~Cambridge, UK: Cambridge University Press, September 2003.,
\bibitem{Shebalinetal83} Shebalin, J.~V., Matthaeus, W.~H., \& Montgomery, D.\ 1983, Journal of Plasma Physics, 29, 525
\bibitem{Higdon84} Higdon, J.~C.\ 1984, The Astrophysical Journal, 285, 109
\bibitem{LazarianVishniac99} Lazarian, A. \& Vishniac, E.~T.\ 1999, The Astrophysical Journal, 517, 700
\bibitem{ChoVishniac00} Cho, J., \& Vishniac, E.~T.\ 2000, The Astrophysical Journal, 539, 273
\bibitem{MaronGoldreich01} Maron, J., \& Goldreich, P.\ 2001, The Astrophysical Journal, 554, 1175
\bibitem{Choetal02} Cho, J., Lazarian, A., \& Vishniac, E.~T.\ 2002, The Astrophysical Journal, 564, 291
\bibitem{Podesta10} Podesta, J.~J.\ 2010, Twelfth International Solar Wind Conference, 1216, 128
\bibitem{Wicksetal10} Wicks, R.~T., Horbury, T.~S., Chen, C.~H.~K. \&  Schekochihin, A.~A.\ 2010, Monthly Notices of Royal Astronomical Society, 407, L31
\bibitem{Wicksetal11} Wicks, R.~T., Horbury, T.~S., Chen, C.~H.~K. \&  Schekochihin, A.~A.\  2011, Physical Review Letters, 106, 045001
\bibitem{MontgomeryMatthaeus95} Montgomery, D. \& Matthaeus, W.~H.\ 1995, The Astrophysical Journal, 447, 706
\bibitem{Galtieretal00} Galtier, S., Nazarenko, S. V., Newell, A. C. \& Pouquet, A.\ 2000, J. Plasma Phys., 63, 447
\bibitem{NgBhattacharjee96} Ng, C.~S. \& Bhattacharjee, A.\ 1996, The Astrophysical Journal, 465, 845
\bibitem{Lazarian06} Lazarian, A.\ 2006, The Astrophysical Journal Letters, 645, L25
\bibitem{Choetal03} Cho, J., Lazarian, A., \& Vishniac, E.~T.\ 2003, Turbulence and Magnetic Fields in Astrophysics, 614, 56
\bibitem{Boldyrev06} Boldyrev, S.  2006, Phys. Rev. Lett., 96, 115002
\bibitem{BeresnyakLazarian06} Beresnyak, A., \& Lazarian, A.\ 2006, The Astrophysical Journal Letters, 640, L175
\bibitem{BeresnyakLazarian09} Beresnyak, A., \& Lazarian, A.\ 2009, The Astrophysical Journal, 702, 1190
\bibitem{Gogoberidze07} Gogoberidze, G.\ 2007, Physics of Plasmas, 14, 022304
\bibitem{BeresnyakLazarian10} Beresnyak, A., \& Lazarian, A.\ 2010, The Astrophysical Journal Letters, 722, L110
\bibitem{Beresnyak11} Beresnyak, A.\ 2011, Physical Review Letters, 106,  075001
\bibitem{Beresnyak12} Beresnyak, A.\ 2012, Monthly Notices of the Royal Astronomical Society, 422, 3495
\bibitem{BrandenburgLazarian13} Brandenburg, A. \& Lazarian, A., 2013, Space Science Reviews, in press
\bibitem{Zeldovich57} Ya B Zeldovich, 1957, JETP Phys 4, 460
\bibitem{Speiser70} Speiser, T.~W.\ 1970, Planetary Space Science, 18, 613
\bibitem{Jacobson84} Jacobson, A.~R., \& Moses, R.~W.\ 1984, Physical Review A, 29, 3335
\bibitem{Strauss86} Strauss, H.~R.\ 1986, Physics of Fluids, 29, 3668
\bibitem{BhattacharjeeHameiri86} Bhattacharjee, A., \& Hameiri, E.\ 1986, Physical Review Letters, 57, 206
\bibitem{HameiriBhattacharjee87} Hameiri, E., \& Bhattacharjee, A.\ 1987, Physics of Fluids, 30, 1743
\bibitem{DiamondMalkov03} Diamond, P.~H., \& Malkov, M.\ 2003, Physics of Plasmas, 10, 2322
\bibitem{Strauss88} Strauss, H.~R.\ 1988, The Astrophysical Journal, 326, 412
\bibitem{ShibataTanuma01} Shibata, K.  \& Tanuma, S. 2001, Earth Planets Space, 53, 473
\bibitem{Waelbroeck89} Waelbroeck, F.~L.\ 1989, Physics of Fluids B, 1, 2372
\bibitem{Loureiroetal09} Loureiro, N. F.,  Uzdensky, D. A. , Schekochihin, A. A., Cowley, S. C.  \& Yousef, T. A.\ 2009, Mon. Not. R. Astron. Soc., 399, L146
\bibitem{Bhattacharjeeetal09} Bhattacharjee, A.,  Huang, Y.-M.,  Yang, H.  \& Rogers, B.\ 2009, Phys. Plasmas, 16, 112102
\bibitem{Karimabadietal13} Karimabadi, H., Roytershteyn, V., Wan, M., et al.\ 2013, Physics of Plasmas, 20, 012303
\bibitem{Beresnyak13b} Beresnyak, A.\ 2013, arXiv:1301.7424
\bibitem{Dahlburgetal92} Dahlburg, R.~B., Antiochos, S.~K., \& Zang, T.~A.\ 1992, Physics of Fluids B, 4, 3902
\bibitem{DahlburgKarpen94} Dahlburg, R.~B., \& Karpen, J.~T.\ 1994, Space Science Reviews, 70, 93
\bibitem{Dahlburg97} Dahlburg, R.~B.\ 1997, Journal of Plasma Physics, 57, 35
\bibitem{FerraroRogers04} Ferraro, N.~M., \& Rogers, B.~N.\ 2004, Physics of Plasmas, 11, 4382
\bibitem{MatthaeusLamkin85} Matthaeus, W.~H. \& Lamkin, S.~L.\ 1985, Physics of Fluids, 28, 303
\bibitem{MatthaeusLamkin86} Matthaeus, W.~H. \& Lamkin, S.~L.\ 1986, Physics of Fluids, 29, 2513
\bibitem{Watsonetal07} Watson, P.~G., Oughton, S. \& Craig, I.~J.~D.\ 2007, Physic. Plasmas, 14, 032301
\bibitem{EyinkBenveniste13} Eyink, G.~L., \& Benveniste, D.\ 2013, Phys. Rev. E, 88, 041001
\bibitem{Kupiainen03} Kupiainen, A.\ 2003, Ann. Henri Poincar\'{e}, 4, Suppl. 2, S713
\bibitem{Yamadaetal06} Yamada, M., Ren, Y., Ji, H., Breslau, J., Gerhardt, S., Kulsrud, R., \& Kuritsyn, A.\ 2006, Physics of Plasmas, 13, 052119
\bibitem{Kowaletal12} Kowal, G., Lazarian, A., Vishniac, E.~T., \& Otmianowska-Mazur, K.\ 2012, Nonlinear Processes in Geophysics, 19, 297
\bibitem{Vishniacetal12} Vishniac, E.~T., Pillsworth, S., Eyink, G.~L., Kowal, G., Lazarian, A., Murray, S.\ 2012, Nonlinear Processes in Geophysics, 19, 605
\bibitem{Maronetal04} Maron, J., Chandran, B.~D., \& Blackman, E.\ 2004, Physical Review Letters, 92, 045001
\bibitem{Beresnyak13a} Beresnyak, A.\ 2013, The Astrophysical Journal Letters, 767, L39
\bibitem{Eyinketal13} Eyink,~G.~L., Vishniac,~E.~T., Lalescu,~C., Aluie,~H., Kanov,~K., B\"urger,~K., Burns,~R., Meneveau,~C.
and Szalay,~A. \ 2013. Nature, 497, 466
\bibitem{CiaravellaRaymond08} Ciaravella, A., \& Raymond, J.~C.\ 2008, The Astrophysical Journal, 686, 1372
\bibitem{Drakeetal10} Drake, J.~F., Opher, M., Swisdak, M., \& Chamoun, J.~N.\ 2010, The Astrophysical Journal, 709, 963
\bibitem{Sychetal09} Sych, R., Nakariakov, V.~M., Karlicky, M., \& Anfinogentov, S.\ 2009, Astronomy \& Astrophysics, 505, 791
\bibitem{Gosling12} Gosling, J.~T., 2012, Space Science Reviews, 172, 187
\bibitem{Goslingetal07} J. T. Gosling,T. D. Phan, R. P. Lin, and A. Szabo\ 2007, Geophys. Res. Lett. 34, L15110
\bibitem{GoslingSzabo08} J. T. Gosling and A. Szabo\ 2008, J. Geophys. Res. 113, A10103
\bibitem{Vasquezetal07} B. J. Vasquez, V. I. Abramenko, D. K. Haggerty, and C. W. Smith\ 2007, J. Geophys. Res. 112, A11102
\bibitem{Phanetal09} T. D. Phan, J. T. Gosling, and M. S. Davis\ 2009, Geophys. Res. Lett. 36, L09108
\bibitem{LazarianOpher09} Lazarian, A. and Opher, M.\ 2009, The Astrophysical Journal, 703, 8
\bibitem{Huangetal12} S.-Y. Huang et al.\  2012, Geophys. Res. Lett., 39, L11104
\bibitem{RechesterRosenbluth78} Rechester, A.~B., \& Rosenbluth, M.~N.\ 1978, Physical Review Letters, 40, 38
\bibitem{VishniacLazarian99} Vishniac, E. \& Lazarian, A.\ 1999, in: Plasma Turbulence and Energetic Particles in Astrophysics; Proceedings of the International Conference, Cracow, Poland, 5-10 September, 1999 eds. M. Ostrowski \& R. Schlickeiser. Obserwatorium Astronomiczne, Uniwersytet Jagiello\'{n}ski, Krak\'{o}w.
\bibitem{HeitschZweibel03} Heitsch, F., \& Zweibel, E.~G.\ 2003, The Astrophysical Journal, 583, 229
\bibitem{Leakeetal12} Leake, J.~E., Lukin, V.~S., Linton, M.~G., \& Meier, E.~T.\ 2012, The Astrophysical Journal, 760, 109
\bibitem{LazarianVishniac09} Lazarian, A., \& Vishniac, E.~T.\ 2009, Revista Mexicana de Astronomia y Astrofisica Conference Series, 36, 81
\bibitem{Kowaletal13} Kowal, G., Lazarian, A., Falceta-Gon\c{c}alves, D.~A., \& Vishniac, E.~T., in preparation
\bibitem{Susinoetal13} R. Susino, A. Bemporad, and S. Kruker, arXiv:1310.2853v1 [astro-ph.SR]
\bibitem{Kraichnan65} Kraichnan, R. H.\ 1965, Phys. Fluids, 8, 1385
\bibitem{KimDiamond01} Kim, E.-j., \& Diamond, P.~H.\ 2001, The Astrophysical Journal, 556, 1052
\bibitem{Cho05} Cho, J., 2005, The Astrophysical Journal, 621, 324
\bibitem{ChoLazarian13} Cho, J. \& Lazarian, A. 2013, The Astrophysical Journal Letters, in press
\bibitem{Lazarianetal03} Lazarian, A., Petrosian, V., Yan, H., \& Cho, J.\ 2003, arXiv:astro-ph/0301181
\bibitem{Kulsrud05} Kulsrud, R.\ 2005 Princeton University Press,  Princeton, NJ
\bibitem{Bernardetal98} Bernard, D., Gaw\c{e}dzki, K.  \& Kupiainen,  A.\ 1998, J. Stat. Phys., 90, 519
\bibitem{GawedzkiVergassola00} Gaw\c{e}dzki, K. \& Vergassola, M.\ 2000, Physica D, 138, 63
\bibitem{EvandenEijnden01} E, W. \& vanden-Eijnden, E.\ 2001, Physica D, 152-153, 636
\bibitem{EEijnden00a} E, W., \& Vanden-Eijnden, E.\ 2000a, Physics of Fluids, 12, 149
\bibitem{EEijnden00b} E, W. \& Vanden-Eijnden, E.\ 2000b, Proceedings of National Academy of Sciences of the United States of America, 97,  8200
\bibitem{EEijnden01} E, W. \& Vanden-Eijnden, E.\ 2001, Physica D: Nonlinear Phenomena, 152, 636
\bibitem{Chavesetal03} Chaves, M., Gaw\c{e}dzki, K., Horvai, P., Kupiainen, A.  \& Vergassola, M. 2003, J. Stat. Phys., 113, 643
\bibitem{Gawedzki08} Gaw\c{e}dzki, K.\ 2008, arXiv:0806.1949
\bibitem{Chandrasekhar61} S. Chandrasekhar, 1961, Hydrodynamic and Hydromagnetic Stability, Oxford University Press
\bibitem{Mandtetal94} M. E. Mandt, R. E. Denton, and J. F. Drake, 1994, Geophys. Res. Lett. 21, 73 (1994)
\bibitem{ShayDrake98} Shay, M.~A., \& Drake, J.~F.\ 1998, Geophysical Review Letters, 25, 3759
\bibitem{BauerBernard99} M. Bauer \& D. Bernard\ 1999, J. Phys. A: Math. Gen., 32, 5179
\bibitem{Eyink11} Eyink,~G.~L.\ 2011, Physical Review E, 83, 056405
\bibitem{Lapenta08} Lapenta, G.\ 2008, Phys. Rev. Lett., 100, 235001
\bibitem{BeresnyakLazarian08} Beresnyak, A., \& Lazarian, A.\ 2008, The Astrophysical Journal, 682, 1070
\bibitem{Lynchetal08} B. J. Lynch et al, 2008, Astrophys. J. 683: 1192
\bibitem{Giannios13} Giannios, D.\ 2013, Monthly Notices of the Royal Astronomical Society, 431, 355
\bibitem{Lazarian05} Lazarian, A.\ 2005, Magnetic Fields in the Universe: From Laboratory and Stars to Primordial Structures., 784, 42
\bibitem{SantosLimaetal10} Santos-Lima, R., Lazarian, A., de Gouveia Dal Pino, E.~M., \& Cho, J.\ 2010, The Astrophysical Journal, 714, 442
\bibitem{SantosLimaetal12} Santos-Lima, R., de Gouveia Dal Pino, E.~M., \& Lazarian, A.\ 2012, The Astrophysical Journal, 747, 21
\bibitem{Lazarian11a} Lazarian, A.\ 2011a, arXiv:1108.2280
\bibitem{Lazarianetal12} Lazarian, A., Esquivel, A., \& Crutcher, R.\ 2012, The Astrophysical Journal, 757, 154
\bibitem{Heitschetal04} Heitsch, F., Zweibel, E.~G., Slyz, A.~D., \& Devriendt, J.~E.~G.\ 2004, The Astrophysical Journal, 603, 165
\bibitem{Zweibel02} Zweibel, E.~G.\ 2002, The Astrophysical Journal, 567, 962
\bibitem{Balsaraetal01} Balsara, D.~S., Crutcher, R.~M., \& Pouquet, A.\ 2001, The Astrophysical Journal, 557, 451
\bibitem{Lazarian13} Lazarian, 2013, Space Science Reviews, accepted
\bibitem{Lazarian11b} Lazarian, A.\ 2011b, arXiv:1111.0694
\bibitem{Jokipii73} Jokipii, J.~R.\ 1973, The Astrophysical Journal, 183, 1029.
\bibitem{deGouveiadalPinoLazarian05} de Gouveia dal Pino, E.~M., \& Lazarian, A.\ 2005, Astronomy \& Astrophysics, 441, 845
\bibitem{LazarianDesiati10} Lazarian, A., \& Desiati, P.\ 2010, The Astrophysical Journal, 722, 188
\bibitem{LazarianBrunetti11} Lazarian, A., \& Brunetti, G.\ 2011, Memorie della Societa Astronomica Italiana, 82, 636
\bibitem{Lazarianetal11} Lazarian, A., Kowal, G., Vishniac, E., \& de Gouveia Dal Pino, E.\ 2011, Planetary and Space Science, 59, 537
\bibitem{Lazarianetal12b} Lazarian, A., Vlahos, L., Kowal, G., Yan, H., Beresnyak, A., de Gouveia Dal Pino, E.~M.\ 2012, Space Science Reviews, 173, 557
\bibitem{Kowaletal12b} Kowal, G., de Gouveia Dal Pino, E.~M., \& Lazarian, A.\ 2012, Physical Review Letters, 108, 241102
\bibitem{Politanoetal89} Politano, H., Pouquet, A. \& Sulem, P.~L.\ 1989, Physics of Fluids B, 1, 2330
\bibitem{degouv12} de Gouveia Dal Pino, E.~M., Le{\~a}o, M.~R.~M., Santos-Lima, R., Guerrero, G., \& Lazarian, A. \ 2012, Phys. Scripta, 86, 018401
\bibitem{leao13} Le{\~a}o, M.~R.~M., de Gouveia Dal Pino, E.~M., Santos-Lima, R. \& Lazarian, A.\ 2013, The Astrophysical Journal, 777, 46L
\bibitem{MininniPouquet09} Mininni, P.~D. \& Pouquet, A.\ 2009, Physical Review E, 80, 025401
\bibitem{GalsgaardNordlund97b} Galsgaard, K. \& Nordlund, \AA.\ 1997, Journal of Geophysical Research, 102, 231
\bibitem{LapentaBettarini11} Lapenta, G., \& Bettarini, L.\ 2011, arXiv:1102.4791
\bibitem{LapentaLazarian12} Lapenta, G., \& Lazarian, A.\ 2012, Nonlinear Processes in Geophysics, 19, 251
\bibitem{Chandrasekhar49} S. Chandrasekhar, 1949, Astrophys. J. 110, 329
\bibitem{Daughtonetal08} Daughton, W., Roytershteyn, V., Albright, B.~J., Bowers, K., Yin, L., \& Karimabadi, H.\ 2008, AGU Fall Meeting Abstracts, A1705
\bibitem{Huangetal11} Y.-M. Huang et al., 2011, Phys. Plasmas 18, 072109
\bibitem{Guoetal12} Guo, Z. B., Diamond, P. H., \& Wang,  X. G., 2012, The Astrophysical Journal, 757, 173
\bibitem{HigashimoriHoshino12} Higashimori, K., \& Hoshino, M., 2012, Journal of Geophysical Research (Space Physics), 117, 1220
\bibitem{UzdenskyKuslrud06} Uzdensky, D.~A. \& Kulsrud, R.~M.\ 2006,  Physics of Plasmas, 13, 062305
\bibitem{Shayetal99} Shay, M.~A., Drake, J.~F., Rogers, B.~N., \& Denton, R.~E.\ 1999, Geophysical Review Letters, 26, 2163
\bibitem{Wangetal00} Wang, X., Bhattacharjee, A., \& Ma, Z.~W.\ 2000, Journal of Geophysical Research, 105, 27
633
\bibitem{Birnetal01} Birn, J., et al.\ 2001, Journal of Geophysical Research, 106, 3715
\bibitem{Malakitetal09} Malakit, K., Cassak, P.~A., Shay, M.~A., \& Drake, J.~F.\ 2009, Geophysical Review Letters, 36, 7107
\bibitem{Cassaketal10} Cassak, P.~A., Shay, M.~A., \& Drake, J.~F.\ 2010, Physics of Plasmas, 17, 062105
\bibitem{Yamada99} Yamada, M.\ 1999, Journal of Geophysical Research, 104, 14529
\bibitem{Yamadaetal10} Yamada, M., Kulsrud, R., \& Ji, H.\ 2010, Reviews of Modern Physics, 82, 603
\end{thebibliography}
\end{document}